\documentclass[%
 reprint,
 amsmath,amssymb,
 aps,
]{revtex4-2}

\usepackage{graphicx}% Include figure files
\usepackage{dcolumn}% Align table columns on decimal point
\usepackage{bm}% bold math
\usepackage{natbib}
\usepackage{graphicx}
\usepackage{float}
\usepackage{subfig}
\usepackage{amsopn}
\usepackage{amsmath}
\usepackage{hyperref}
\usepackage{cleveref}
\usepackage{color,xcolor}

\newcommand{\dudx}[2]{\frac{\partial{#1}}{\partial{#2}}}

\newcommand{\tpr}{\boldsymbol{\otimes}}   % tensor product
\newcommand{\dpr}{\boldsymbol{\cdot}}     % dot product
\newcommand{\vc}[1]{\boldsymbol{#1}}      % bold for vector values

\begin{document}

\preprint{APS/123-QED}

\title{On the computation of thermo-relaxing multi-component flows with the Baer-Nunziato model}% Force line breaks with \\

\author{Chao Zhang}
 %\altaffiliation[Also at ]{Institute of Applied Physics and Computational Mathematics, Beijing, China.}
 \email{zhang-c@mail.ru}
\author{Lifeng Wang}%
 \altaffiliation[Also at ]{Center for Applied Physics and Technology, HEDPS, Peking University, Beijing, China.}
 \email{wang_lifeng@iapcm.ac.cn}
\affiliation{%
 Institute of Applied Physics and Computational Mathematics, Beijing, China.
}%

\date{\today}% It is always \today, today,

\begin{abstract}
In inertial confinement fusion (ICF) implosions, mixing the ablator into the fuel and the hot spot is one of the most adverse factors that lead to ignition degradation. Recent experiments in the Marble campaign at the Omega laser facility and the National Ignition Facility (NIF) demonstrate the significance of the temperature separation in heterogeneous mixing flows  [Haines B.M., et al. Nature Communications, 2020].
In the present work we provide an approach to deal with thermally disequilibrium multi-component flows with the ultimate aim to investigate the temperature separation impact on mixing and fusion burn. The present work is two-fold: (a) we derive a model governing the multi-component flows in thermal disequilibrium with transport terms, (b) we use the derived model to study the Rayleigh–Taylor (RT) instability in thermally relaxing multi-component systems. The model is reduced from the full disequilibrium multi-phase Baer-Nunziato model in the limit of small Knudsen number $Kn << 1$. Velocity disequilibrium is closed with the diffusion laws and only one mass-weighted velocity is retained formally.  Thus, the complex wave structure of the original Baer-Nunziato model is simplified to a large extent and the obtained model is much more computationally affordable. Moreover, the capability to deal with finite temperature relaxation is kept. Efficient numerical methods for solving the proposed model are also presented.
Equipped with the proposed model and numerical methods, we further investigate the impact of thermal relaxation on the RT instability development at the inertial confinement fusion (ICF) deceleration stage.
On the basis of numerical simulations, we have found that for the RT instability at an interface between the high-density low-temperature component and the low-density high-temperature component, the thermal relaxation significantly suppresses the development of the instability.
\end{abstract}

\maketitle

%\tableofcontents

\section{Introduction}
\label{sec:intro}
%Compressible multicomponent flows are of significance to many applications, such as the inertial confinement fusion (ICF),  the explosion of core-collapse supernova, underwater explosion (UNDEX) and so forth. These physical processes include Rayleigh-Taylor (RT) and Richtmyer-Meshkov (RM) hydrodynamic instabilities that rapidly develop in the presence of small initial perturbations. Up to now, understanding the development of these nonlinear instabilities  still heavily relies on numerical simulations. The present research is motivated by the need to simulate the  laser-driven plasma instability developed at the interface between dissimilar materials within ICF capsules. In such applications, the transport phenomena (of mass, momentum and energy) accompanying the hydrodynamic process play significant role. First of all, the laser energy deposited in the plasma is transported through the heat conduction process. Moreover, at small spatial scales the effect of viscous dissipation and the mass diffusion begin to impact the instability growth \cite{robey2004effects}. Numerically, the transport process is vital for achieving a grid-converged DNS (Direct Numerical
%Simulation) \cite{vold2021plasma}.

Mixing of the ablator into the fuel and the hot spot is considered to be one of the most adverse factors that lead to ignition degradation in inertial confinement fusion (ICF) implosions. Mixing takes place at different scales driven by different mechanisms.
At the macroscopic scale where $Kn<<1$, the hydrodynamic instabilities (such as the Rayleigh–Taylor instability (RTI), Richtmyer–Meshkov instability (RMI) and Kelvin–Helmholtz instability (KHI))
play predominate roles in causing mixing. With the continuous development of the hydrodynamic instabilities, the flows transition into turbulence where mixing happens at very different scales. At the small scale where $Kn=\mathcal{O}(1)$ the mass diffusion is proceeding all the time as result of random molecular motions (i.e., the kinetic effect).
These two mechanisms have fundamental difference and scaling laws. For example, the mixing length caused by RTI and further induced turbulence can be scaled as $L_{mix} = \alpha Agt^2$, where $\alpha$ is a constant, $A$ is the Atwood number, $g$ is the acceleration, and $t$ the time.
In contrast, the mixing length caused by mass diffusion is expressed as $L_{mix} = \beta \sqrt{D t}$, where $D$ is the mass diffusivity and $\beta$ is a constant. Recent works demonstrate that the the mass diffusion maybe the leading mixing mechanism for the implosion experiments under moderate temperature and convergence\cite{zylstra2018diffusion}. 

\begin{figure}[htbp]
{\includegraphics[width=0.5\textwidth]{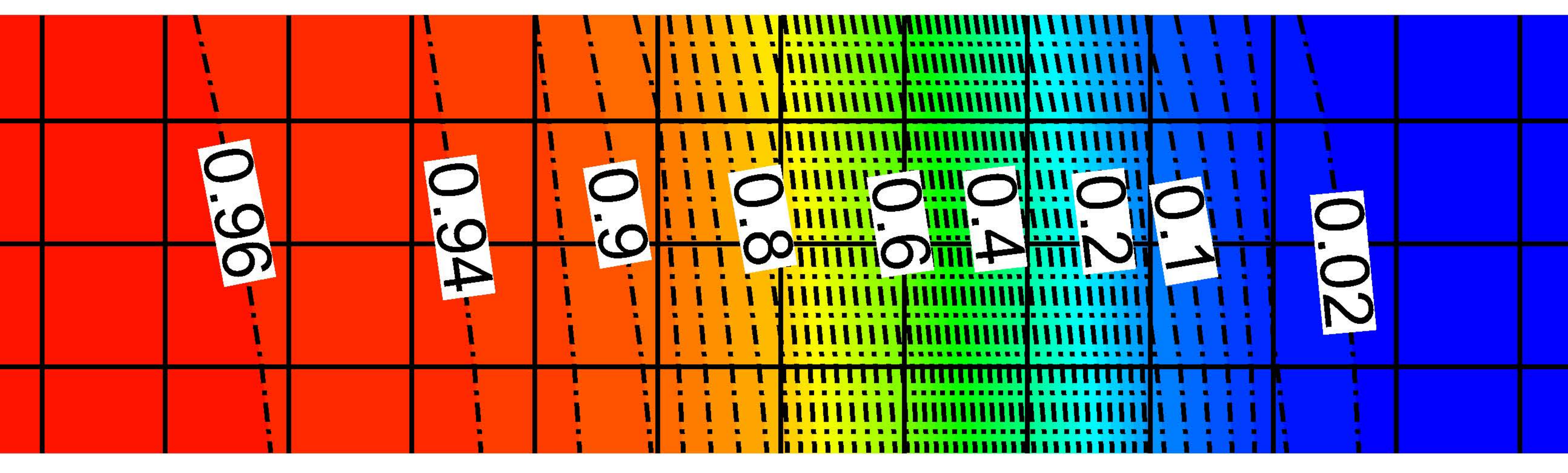}}
\caption{The typical mass fraction distribution on the Eulerian grid in the case of heterogeneous mixing.}
\label{fig:mix_on_grid_a}
\end{figure}

The ICF mixing is usually categorized into two types, i.e., the atomic mixing and the chunk mixing. In the former the mixing of the components takes place on the atomic scale, while in the latter the constituents are separated from each other as either at a rippled interface or in discrete clumps \cite{Wilson2011}.
The ICF mixing is intrinsically heterogeneous as a combination of atomic mixing and non-atomic mixing. In the course of hydrodynamic development towards turbulence, the chunk mixing dominate the early stage and the atomic mixing takes over at later times.  Meanwhile the mass diffusion is always producing atomic mixing.
From the perspective of numerical modeling, to discriminate different types of mixing with direct numerical simulation is a formidable task due to their very different characteristic scales. Therefore, one has to rely on mixing models that allow the coarse-grained description of the atomic mixing on an affordable grid (\Cref{fig:mix_on_grid_a}).
The grid does not resolve the small scale mixing topology and the component mass fraction span over several computational cells in the macroscopic description. In the mixing cells some closure relation is needed to make the governing equations solvable. One of the most frequently used closure is the temperature equilibrium,
which is assumed in the ICF-relevant mixing models such as the k-L model \cite{Dimonte2006,schilling2021self} and the BHR model \cite{Besnard1992,grinstein2021coarse}. These models have been successfully used to simulate some experiments on the Omega laser facility and the National Ignition Facility (NIF). However, they give numerical results that deviate from experimental measurements in the case of temperature separation.
The root of such failure is attributed to  the thermal equilibrium assumption, which has been proved in subsequent experiments at the Marble platform at NIF \cite{olson2020development,murphy2021results} (referred to as ``Marble'' henceforth).

% \begin{figure}
% \end{figure}

As noted by Haines et al. \cite{haines2020nc,olson2020development}, strong ion temperature separation arises during the shock flash. Mechanisms  that drive the component temperatures towards equilibrium include electron thermal conduction and local collision in atomic mixing. They happen in a timescale comparable to that of the ICF implosion (at least 1.6ns for the Marble campaign on the Omega facility \cite{haines2020nc}).
Thus, the ion temperatures of each species are not fully equilibrated in the course of ICF implosions. This means that in numerical simulation the components should have their own temperature in a mixing cell.

\begin{figure*}
\centering
{\includegraphics[width=0.51\textwidth]{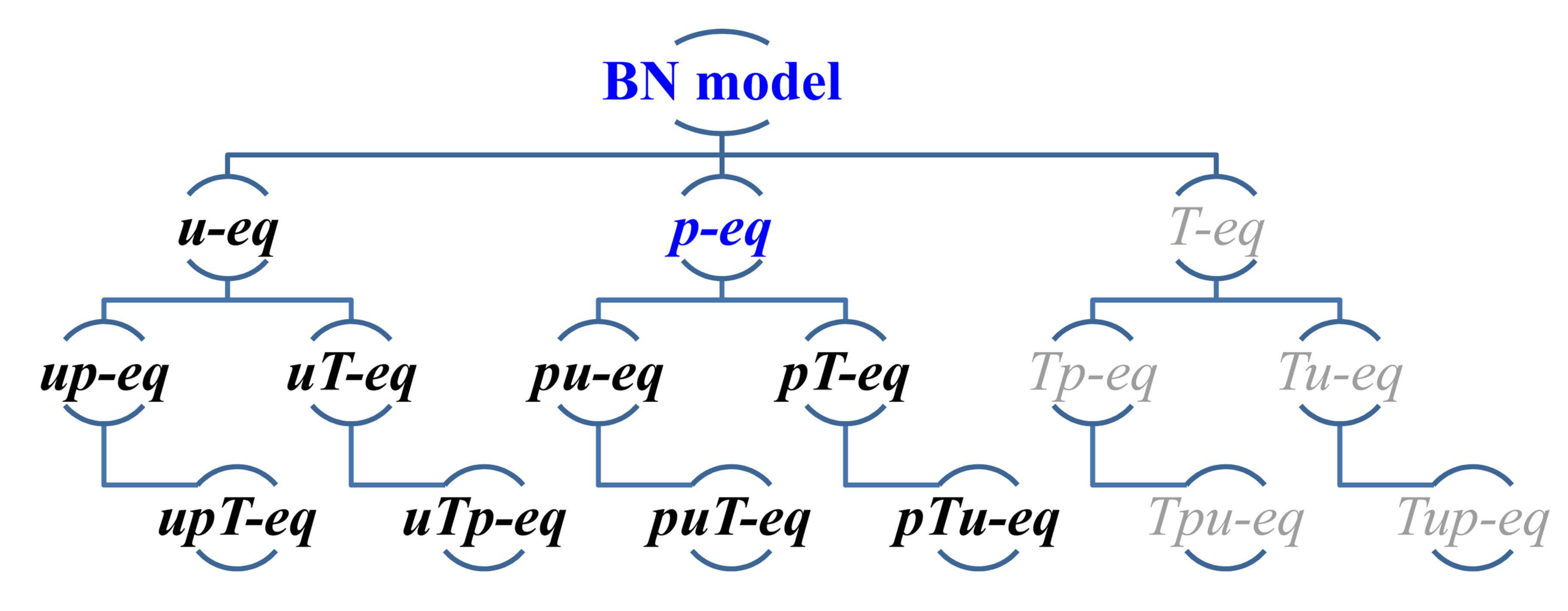}}
\caption{The hierarchy of the reduced models of the Baer-Nunziato model.}
\label{fig:BN_Hier}
\end{figure*}

{\color{black}We aim to develop a multicomponent model to describe the temperature separation in mixing. Note that the phenomenlohgical two-field turbulence models \cite{llor2005statistical,llor2003new,youngs1994numerical,saurel2012modelling} has the potential to deal with turbulence mixing with finite temperature relaxation. However, here we adopt a more direct and strict derivation from a fully disequilibrium model - the BN model\cite{BAER1986861}.} 
Due to the complexity and relaxation stiffness of the original BN model, a hierarchy of reduced models has been established (\Cref{fig:BN_Hier}).
Reduced models are derived in the limit of instantaneous relaxation of corresponding variable (chemical potential, pressure, velocity, and temperature). For example, in the case where the phase velocities relax instantaneously, one can derive  the $u$-eq model via the asymptotic analysis.
Starting from the  $u$-eq model, one can further derive the $up$-eq model (i.e., the Kapila's five equation model \cite{kapila2001two}) on the basis of the instantaneous phase pressure relaxation. Following such procedure, a full hierarchy of reduced models can be obtained. A similar hierarchy has been described by Lund \cite{lund2012hierarchy}, however, the velocity disequilibrium that is vital for modeling mass diffusion is neglected there.

The applicability of a particular reduced model is determined by the corresponding assumption on relaxation rates. In most applications the thermal relaxation time is smaller than the mechanical (pressure and velocity) relaxation times, thus models in the right subsidiary are scarcely used.
In term of the current work, we need a model that retains the thermal disequilibrium. Moreover, since mass diffusion is related to the species velocity difference, the velocity disequilibrium should also be maintained.
Thus, the possible models to satisfy these requirements are the original BN model and the $p$-eq model. Here we focus on reformulation the former by invoking the diffusion laws to close the velocity difference and abandoning the terms of order $\mathcal{O}(\text{Kn}^2)$.
Formally the derived model has one velocity, i.e., the mass weighted velocity. In fact, hyperbolic sub-system of the model coincides with the six equation model presented in \cite{Saurel2009}, which is more robust than the Kapila's five equation model in numerical implementation.  This reduction significantly simplifies the wave structure of the hyperbolic subsystem of the model,
thus improving the computational efficiency.

We propose second-order methods for solving the hyperbolic-parabolic-relaxation system. The numerical methods are validated against some benchmark problems. Being equipped with the model and numerical methods, we then investigate the RTI in the thermally relaxing multi-component flows, especially the dependence
of the mixing length on the thermal relaxation rate and the initial temperature separation.

The rest of the article is organized as follows. In \Cref{sec:model} we derive the reduced temperature-disequilibrium model with diffusions. In \Cref{sec:num_med} the numerical methods for solving the proposed model are briefly described.
In \Cref{sec:num_res} some numerical results for validating the numerical methods are presented. Analysis of the RTI problem is performed in particular detail to investigate the impact of the thermal relaxation.

\section{Model formulation}
\label{sec:model}
\subsection{The BN-type seven-equation model}
The starting point of the following model formulation is the complete  BN-type seven-equation model \cite{BAER1986861,saurel1999multiphase,petitpas2014,perigaud2005compressible}. It reads:
\begin{subequations} \label{eq:bn}
\begin{align}
\label{eq:bn:mass}
&\dudx{\alpha_k\rho_k}{t} + \nabla\dpr(\alpha_k\rho_k\vc{u}_k) = 0, \\
  \label{eq:bn:mom}
 & \dudx{\alpha_k\rho_k\vc{u}_k}{t} + \nabla\dpr\left(\alpha_k\rho_k\vc{u}_k\tpr\vc{u}_k - \alpha_k \overline{\overline{T}}_k \right) = \\ & -  \overline{\overline{T}}_I \dpr \nabla {\alpha_k}  + \mathcal{M}_k,\nonumber\\
  \label{eq:bn:en}
  &\dudx{\alpha_k \rho_k E_k}{t} +
  \nabla\dpr\left(
\alpha_k \rho_k E_k \vc{u}_k   - \alpha_k \overline{\overline{T}}_k \dpr \vc{u}_k
  \right)
  = \\ 
  &- \vc{u}_{I} \dpr \left( \overline{\overline{T}}_I \dpr \nabla {\alpha_k} \right)
    + \vc{u}_I \mathcal{M}_k
- p_I \mathcal{F}_k + \mathcal{Q}_k + q_k + \mathcal{I}_k, \nonumber\\
   \label{eq:bn:vol}
 & \dudx{\alpha_k }{t} +
  \vc{u}_I \dpr \nabla\alpha_k = \mathcal{F}_k,
\end{align}
\end{subequations}
where the notations used are standard: $\alpha_k, \; \rho_k, \; \vc{u}_k, \; p_k, \; \overline{\overline{T}}_k, \; E_k$ are the volume fraction, phase density, velocity, pressure, stress tensor, and total energy of phase $k$. 

{\color{black}The BN-type formulation can also be derived by averaging the one-phase governing equations\cite{drew1983mathematical,CHINNAYYA2004490,ABGRALL2003361}. Within such framework the volume fraction $\alpha_k$ is the spatially average characteristic function $X_k$, which takes the value of 1(or 0) in the presence (or absence) of the $k$-th component, i.e.,
\begin{align*}
\begin{split}
X_{k}(\vc{r},t)= \left \{
\begin{array}{ll}
    1,     & \text{the\;point\;} \vc{r} \text{\;in\;the\;material\;$k$},\\
    0,     & \text{otherwise.}
\end{array}
\right. 
\end{split} 
\end{align*}

Then, we have
\begin{equation*}
    \alpha_k = \frac{1}{V} \int_V X_k \text{d}V=\frac{V_k}{V}.
\end{equation*}
The volume fraction $\alpha_k$ can be perceived as a macroscopic measure of spatial mixing extent when the microscopic interfaces can not be accurately resolved. Moreover, the BN model can be used for describing both miscible and immiscible flows according to different interpretations of $\alpha_k$ and other variables \cite{hantke2021news}.
}

For the sake of clarity we constrict our discussions within the scope of two-phase flows, $k=1,2$. The phase density $\rho_k$ is defined as the mass per unit volume occupied by $k$-th phase. The mixture density $\rho$ is the sum of the partial densities $\alpha_k \rho_k$, i.e., $\rho = \sum {\alpha_k \rho_k}$.
The last equation \cref{eq:bn:vol} is written for only one component thanks to the saturation constraint for volume fractions
$\sum_{k=1}^2 \alpha_k  = 1$.
%the last equations are written only for only $N-1$ volume fractions, where $N$ is the number of components.
The total energy is $E_k  = e_k + \mathcal{K}_k$ where $e_k$, and $\mathcal{K}_k = \frac{1}{2}\vc{u}_k \dpr \vc{u}_k$ are the  internal energy and kinetic energy, respectively.
%$\mathcal{I}_k = \alpha_k I_k$ with $I_k$ being the external energy source released in the phase $k$.

The variables with the subscript ``I'' represent the variables at interfaces, for which there are several possible definitions \cite{saurel1999multiphase,perigaud2005compressible,saurel2018diffuse}.
%Whatever the definitions we choose, $\lim_{\eta\to\infty} p_{I} = \lim_{\eta\to\infty}p_{k} = p$, $\lim_{\vartheta\to\infty}\vc{u}_{I} = \lim_{\vartheta\to\infty}\vc{u}_{k} = \vc{u}$, and $\lim_{\vartheta\to\infty} {\overline{\overline{\tau}}}_{I} = \lim_{\vartheta\to\infty}{\overline{\overline{\tau}}}_{k} = {\overline{\overline{\tau}}}$.
Here we choose the following
\begin{align}
&\vc{u}_I = {\vc{u}} = {\sum y_k \vc{u}_k }, \quad p_I = \sum \alpha_k p_k, \nonumber\\
&\quad {\overline{\overline{\tau}}}_I = \sum \alpha_k {\overline{\overline{\tau}}}_k, \quad \overline{\overline{T}}_{I} = - p_I \overline{\overline{I}} +  \overline{\overline{\tau}}_I,\nonumber
\end{align}
where $y_k$ denotes the mass fraction $y_k  = \alpha_k \rho_k / \rho$, and ${\vc{u}}$ is the mass-fraction weighted mean velocity. {\color{black}Here, we have used the mass weighted velocity $\vc{u}$ to approximate the interface velocity $\vc{u}_I$.} {\color{black}Other possible definitions for $\vc{u}_I$ as convex combination of component velocities will result in the same reduced model since $(\vc{u}_I - \vc{u}) = \mathcal{O}(\text{Kn})$, as to be shown in \Cref{sec:reduceBN}.}

The interfacial stress $\overline{\overline{\tau}}_I$ is defined in such way that the thermodynamical laws are respected. {\color{black}In the present work we do not expand the issue on defining the interfacial variables since they are included in the terms of order $\mathcal{\text{Kn}}^2$ that are to be omitted.} 

The inter-phase exchange terms include the velocity relaxation $\mathcal{M}_k$, the pressure relaxation $\mathcal{F}_k$, and the temperature relaxation $\mathcal{Q}_k$. They are as follows:{\color{black}
\begin{align}
&\mathcal{M}_k=\vartheta\left( \vc{u}_{k^*}-\vc{u}_k \right), \nonumber\\ &\mathcal{F}_k=\varsigma\left( {p}_{k} - {p}_{k^*} \right), \nonumber\\ &\mathcal{Q}_k=\eta\left( T_{k^*} - T_{k} \right),\label{eq:relaxations}
\end{align}}
where $k^{*}$ denotes the conjugate component of the $k$-th component,  i.e., $k=1,\; k^{*} =2$ or $k=2,\; k^{*} =1$. The relaxation velocities are all positive $\vartheta > 0, \; \eta > 0, \; \varsigma > 0$.

The phase stress tensor, $\overline{\overline{T}}_k$, can be written as
\begin{equation}
 \overline{\overline{T}}_k  = - p_k \overline{\overline{I}} +  \overline{\overline{\tau}}_k.
\end{equation}

For the viscous part we use the  Newtonian approximation
\begin{equation}\label{eq:newton_vis}
\overline{\overline{\tau}}_k = 2\mu_k \overline{\overline{D}}_k + \left(\mu_{b,k} - \frac{2}{3}\mu_k \right) \nabla \dpr \vc{u}_k,
\end{equation}
where $\mu_k > 0$ is the coefficient of shear viscosity and  $\mu_{b,k} > 0$ is the coefficient of bulk viscosity.
The tensor $\overline{\overline{D}}_k$ is the deformation rate, which takes the following form
\[
\overline{\overline{D}}_k = \frac{1}{2}\left(  \nabla\vc{u}_k + \left( \nabla\vc{u}_k \right)^{\text{T}} \right).
\]

{\color{black}The heat conduction is represented by $q_k = \nabla \dpr \left( \lambda_k \nabla T_k \right)$. The term $\mathcal{I}_k$ is a volume energy source term, which can be used to model the laser energy adsorption in direct drive ICF.}

Performing straightforward calculus similar to that of Murrone \cite{murrone2005five}, one can derive the following equation for the internal energy
\begin{align}
&\dudx{\alpha_k \rho_k e_k}{t} + \nabla \dpr \left( \alpha_k \rho_k e_k \vc{u}_k \right) = - \alpha_k p_k \nabla \dpr \vc{u}_k \nonumber
\\ &- p_I \mathcal{F}_k + \left( \vc{u} - \vc{u}_k \right) \dpr \mathcal{M}_k   \nonumber
\\ &+ \left( \vc{u}_k - \vc{u} \right) \dpr \left( \overline{\overline{T}}_I \dpr \nabla \alpha_k \right) + \mathcal{G}_k,\label{eq:bn:int_en}
\end{align}
where $\mathcal{G}_k = q_k + \mathcal{I}_k + \mathcal{S}_k + \mathcal{Q}_k, \;\; \mathcal{S}_k = \alpha_k \overline{\overline{\tau}}_k : \overline{\overline{D}}_k$.

\subsection{Splitting diffusion-related terms}
In continuum mechanics the mass diffusion strength is described through the difference between the phase velocity and the velocity of the mass center.  We introduce the following definition of the diffusion velocity
\begin{equation}\label{eq:w_k}
\vc{w}_k = \vc{u}_k - \vc{u}.
\end{equation}

For future use, we gather the diffusion related terms separately.
To do this, we replace $\vc{u}_k$ by $\vc{u} + \vc{w}_k$ to obtain the following reformulation of \cref{eq:bn}:
\begin{subequations}
\begin{align}
&\dudx{\alpha_k\rho_k}{t} + \nabla\dpr(\alpha_k\rho_k {\vc{u}}) = - \nabla\dpr \vc{J}_k,\label{eq:avbn:mass}\\
&\dudx{\alpha_k\rho_k \vc{u}}{t} + \nabla\dpr\left( \alpha_k \rho_k \vc{u} \vc{u} - \alpha_k \overline{\overline{T}}_{ak} \right) = -  \overline{\overline{T}}_I \dpr \nabla \alpha_k \nonumber
\\&+ \mathcal{M}_k + \dudx{\alpha_k \rho_k \vc{w}_k}{t} + \nabla \dpr \left( 2\alpha_k \rho_k \vc{u} \vc{w}_k +  \alpha_k \rho_k \vc{w}_k \vc{w}_k\right) \nonumber
\\&+ \nabla \dpr \left( \alpha_k \overline{\overline{T}}_{wk} \right), \label{eq:avbn:mom}\\
&  \dudx{\alpha_k \rho_k {{E}}_{ak}}{t} +
  \nabla\dpr\left(
\alpha_k \rho_k {{E}}_{ak} \vc{u}   - \alpha_k \overline{\overline{T}}_{ak} \dpr \vc{u}
  \right)
  = \nonumber
  \\& - \vc{u}_{I} \dpr \left( \overline{\overline{T}}_I \dpr \nabla {\alpha_k} \right)
    + \vc{u} \mathcal{M}_k 
- p_I \mathcal{F}_k + \mathcal{Q}_k + q_k + \mathcal{I}_k \nonumber\\&
{\color{black}-} \dudx{\alpha_k \rho_k {E}_{wk} }{t} - \nabla \dpr \left( \alpha_k \rho_k  \left(  \vc{u} E_{wk} + {E}_{ak} \vc{w}_k + E_{wk} \vc{w}_k \right) \right) \nonumber
\\&+ \nabla \dpr \left( \alpha_k \overline{\overline{T}}_{ak} \dpr \vc{w}_k + \alpha_k \overline{\overline{T}}_{wk} \dpr \vc{u} + \alpha_k \overline{\overline{T}}_{wk} \dpr \vc{w}_k \right) \label{eq:avbn:en},\\
&\dudx{\alpha_k }{t} + \vc{u} \dpr \nabla\alpha_k = \mathcal{F}_k \label{eq:avbn:vol},
\end{align}
\end{subequations}
where the diffusion flux \[ \vc{J}_k = \alpha_k \rho_k \vc{w}_k.\]

We have used the following decomposition
\[
E_k = E_{ak} + E_{wk}, \quad \overline{\overline{T}}_k = \overline{\overline{T}}_{ak} + \overline{\overline{T}}_{wk},
\]
The quantities with subscript ``$ak$'' contains only the mass centered velocity $\vc{u}$, i.e.,
\[E_{ak} = e_k + \frac{1}{2}\vc{u} \dpr \vc{u},\]
\[\overline{\overline{T}}_{ak} =  - p_k \overline{\overline{I}} + 2\mu_k \overline{\overline{D}}_{ak} + \left( \mu_{b,k} - \frac{2}{3} \mu_k \right) \nabla\dpr \vc{u},\]
\[\overline{\overline{D}}_{k} = \overline{\overline{D}}_{ak} + \overline{\overline{D}}_{wk}  , \quad  \overline{\overline{D}}_{ak} = \frac{1}{2} \left(  \nabla \vc{u} + \left( \nabla\vc{u} \right)^{\text{T}} \right).\]

The terms with $\vc{w}_k$ can be regarded as diffusion-induced. Similarly, \cref{eq:bn:int_en} can be reformulated as
 \begin{align}\label{eq:bn:int_en1}
&\dudx{\alpha_k \rho_k e_k}{t} + \nabla \dpr \left( \alpha_k \rho_k e_k \vc{u} \right) = - \alpha_k p_k \nabla \dpr \vc{u} - p_I \mathcal{F}_k  \nonumber
\\&+ \vc{w}_k \dpr \mathcal{M}_k  + \vc{w}_k \dpr \left( \overline{\overline{T}}_I \dpr \nabla \alpha_k \right) - \nabla \dpr \left( \alpha_k \rho_k e_k \vc{w}_k \right) \nonumber
\\&- \alpha_k p_k \nabla \dpr \vc{w}_k + \mathcal{G}_k.
\end{align}

\subsection{Reduction of the BN model}\label{sec:reduceBN}
Continuum assumption is usually accepted in the case of a small Knudsen number, i.e.,
\begin{equation}
Kn = \frac{\lambda}{\Delta} \leq 0.001 << 1.
\end{equation}

For moderate Mach number and collision of ions with comparable masses, it can be shown that\cite{kagan2014}
\begin{equation}
\frac{|\vc{w}_k|}{u_{shock}} \approx \frac{\lambda}{\Delta} << 1.
\end{equation}

With such scale estimation, in the following we will drop the terms of order  $\mathcal{O}(|\vc{w_k}^2|)$.
The velocity relaxation $\mathcal{M}_k$ in the model (\ref{eq:bn}) is related to the ion friction $\mu_{kk^{*}}\nu_{i}n_k \left( \vc{w}_k - \vc{w}_{k^{*}} \right)$, where $\mu_{kk^{*}}$ is the reduced mass, $\nu_{i}$ is the ion collision frequency, and $n_k$ is the number density. This term is finite under the concerned scenario. Thus, the velocity relaxation rate is estimated to be $\mathcal{O}\left( 1 \right)$.
The general idea is to reduce the Baer-Nunziato model (\ref{eq:bn}) via order analysis. More concretely, we will drop $\mathcal{O}(|\vc{w_k}^2|)$ terms and close $\mathcal{O}(|\vc{w_k}|)$ terms with the established diffusion laws.

The diffusion velocity is defined with the diffusion laws such as the Fick's law, resulting in the redundance of the model (\ref{eq:avbn:mom}). Thus we can retain only one momentum equation, i.e., the sum of \cref{eq:avbn:mom}
\begin{equation}\label{eq:sum_mom}
    \dudx{\rho \vc{u}}{t} + \nabla\dpr\left( \rho \vc{u} \vc{u} - \sum \alpha_k \overline{\overline{T}}_{ak} \right) =   \sum \nabla \dpr \left( \alpha_k \overline{\overline{T}}_{wk} \right).
\end{equation}

With such evaluation, the equation for the internal energy \cref{eq:bn:int_en1} is reduced to
\begin{align}\label{eq:avbn:inten1}
    &\dudx{\alpha_k \rho_k e_k}{t} + \nabla \dpr \left( \alpha_k \rho_k e_k \vc{u} \right) = - \alpha_k p_k \nabla \dpr \vc{u} - p_I \mathcal{F}_k  \nonumber
    \\&- \nabla \dpr \left( \alpha_k \rho_k e_k \vc{w}_k \right) - \alpha_k p_k \nabla \dpr \vc{w}_k + \mathcal{G}_k,
\end{align}
%\begin{equation}\label{eq:avbn:inten1}
%    \dudx{\alpha_k \rho_k e_k}{t} + \nabla \dpr \left( \alpha_k \rho_k e_k \vc{u} \right) = - \alpha_k p_k \nabla \dpr \vc{u} - p_I \mathcal{F}_k  + \vc{w}_k \dpr \left( \sum \alpha_k \overline{\overline{T}}_{ak} \dpr \nabla \alpha_k \right) - \nabla \dpr \left( \alpha_k \rho_k e_k \vc{w}_k \right) - \alpha_k p_k \nabla \dpr \vc{w}_k
%\end{equation}

By summing  \cref{eq:avbn:en} and abandoning terms of $\mathcal{O}(|\vc{w}_k|^2)$, one can obtain
\begin{align}
  &\dudx{ \rho {{E}}_{a}}{t} +
  \nabla\dpr\left(\rho {{E}}_{a} \vc{u}   - \overline{\overline{T}}_{a} \dpr \vc{u}
  \right)
  =
  \sum\left( \mathcal{Q}_k + q_k + \mathcal{I}_k \right)  \nonumber
  \\&- \nabla\dpr \left(  \sum \alpha_k \rho_k e_k \vc{w}_k \right) \nonumber
  \\&+ \nabla \dpr \left(  \sum \alpha_k \overline{\overline{T}}_{ak}\dpr\vc{w}_k +   \sum \alpha_k \overline{\overline{T}}_{wk}\dpr\vc{u} \right)  \label{eq:avbn:en1},
\end{align}
where \[ E_a = \sum y_k E_{ak} .\]

Note that the enthalpy diffusion flux comes from the second and third terms on the right hand side of \cref{eq:avbn:en1}. 

After the above reduction we can have the following closed system consisting of the phase mass equations (\ref{eq:avbn:mass}), the mixture momentum equation (\ref{eq:sum_mom}), the phase internal energy equation (\ref{eq:avbn:inten1}) and the  volume fraction equation (\ref{eq:bn:vol}).  For clarity, we present the obtained model as follows
%\begin{subequations}\label{eq:final_model}
%\begin{align}
%&\dudx{\alpha_k\rho_k}{t} + \nabla\dpr(\alpha_k\rho_k {\vc{u}}) = - \nabla\dpr \vc{J}_k,\label{eq:avbn:mass}\\
% &\dudx{\rho \vc{u}}{t} + \nabla\dpr\left( \rho \vc{u} \vc{u} - \sum \alpha_k \overline{\overline{T}}_{ak} \right) =   \sum \nabla \dpr \left( \alpha_k \overline{\overline{T}}_{wk} \right), \\
%&\dudx{\alpha_k \rho_k e_k}{t} + \nabla \dpr \left( \alpha_k \rho_k e_k \vc{u} \right) = - \alpha_k p_k \nabla \dpr \vc{u}   \nonumber
%\\&- p_I \mathcal{F}_k + \vc{w}_k \dpr \left( \sum \alpha_k \overline{\overline{T}}_{ak} \dpr \nabla \alpha_k \right) \nonumber
%\\&- \nabla \dpr \left( \alpha_k \rho_k e_k \vc{w}_k \right) - \alpha_k p_k \nabla \dpr \vc{w}_k, \\
%&\dudx{\alpha_k }{t} +
%  \vc{u} \dpr \nabla\alpha_k = \mathcal{F}_k.
%\end{align}
%\end{subequations}
\begin{subequations}\label{eq:final_model}
\begin{align}
&\dudx{\alpha_k\rho_k}{t} + \nabla\dpr(\alpha_k\rho_k {\vc{u}}) = - \nabla\dpr \vc{J}_k,\\
 &\dudx{\rho \vc{u}}{t} + \nabla\dpr\left( \rho \vc{u} \vc{u} - \sum \alpha_k \overline{\overline{T}}_{ak} \right) =   \sum \nabla \dpr \left( \alpha_k \overline{\overline{T}}_{wk} \right), \\
&\dudx{\alpha_k \rho_k e_k}{t} + \nabla \dpr \left( \alpha_k \rho_k e_k \vc{u} \right) = - \alpha_k p_k \nabla \dpr \vc{u}   \nonumber
\\&- p_I \mathcal{F}_k - \nabla \dpr \left( \alpha_k \rho_k e_k \vc{w}_k \right) - \alpha_k p_k \nabla \dpr \vc{w}_k + \mathcal{G}_k , 
\\&\dudx{\alpha_k }{t} +
  \vc{u} \dpr \nabla\alpha_k = \mathcal{F}_k.\label{eq:final_modeld}
\end{align}
\end{subequations}

{\color{black}Note that the energy flux caused by mass diffusion is represented by $\alpha_k \rho_k e_k \vc{w}_k$. Energy diffusion effects of heat conduction and viscous dissipation is included in the term $\mathcal{G}_k$ (see the definition in \cref{eq:bn:int_en}).}

{\color{black}The absence of a mass diffusion term in the volume fraction equation (\ref{eq:final_modeld}) is due to the particular choice of interfacial velocity $\vc{u}_I = \vc{u}$. In more general case we can reformulate \cref{eq:bn:vol} as follows
\begin{equation*}
    \dudx{\alpha_k}{t} + \vc{u}\dpr \nabla \alpha_k = (\vc{u}-\vc{u}_I)\dpr\nabla\alpha_k + \mathcal{F}_k,
\end{equation*}
Here, one can see that the first term on the right hand side is actually of order $\mathcal{O}(\text{Kn}^2)$ as long as the $\vc{u}_I$ is of the same order as $\vc{u}_k$. This means that even in this general situation \cref{eq:final_modeld} holds in the context our approximation.}

In the framework of the fractional step method, the hydrodynamic subsystem  of (\ref{eq:final_model}) is non-conservative due to the equation for internal energy and the volume fraction. To {\color{black}alleviate} the non-conservativeness, we use the mixture energy equation  (\ref{eq:avbn:en1}) as an auxiliary equation {\color{black}in solving the hyperbolic sub-system} as the six equation model \cite{saurel2009simple}. {\color{black}An instantaneous pressure relaxation follow after solving the hyperbolic sub-system, thus resulting in a pressure-equilibrium model. Such approach ensures robustness in numerical implementation with marginal sacrifice of computation efficiency.}

{\color{black}Note that \cref{eq:avbn:en1} is not a strict consequence of \cref{eq:final_model} since the latter loses a momentum equation after reduction. However, \cref{eq:avbn:en1} is only invoked in solving the hyperbolic part where the consistency remains.}

{\color{black}We now check the entropy dissipative property of the model. The full entropy equation before reducetion can be derived in a way similar to \cite{zhang2022a,murrone2005five}, which read
%\begin{align}
%&\alpha_1 \rho_1 T_1 \frac{\mathrm{D}_1 s_1}{\mathrm{D} t}= \left(p_1-p_I\right) \mathcal{F}_1+\left(p_I-p_1\right)\left(\boldsymbol{u}_I-\boldsymbol{u}_1\right) \cdot \nabla \alpha_1 \nonumber\\ 
%&+\left(\boldsymbol{u}_I-\boldsymbol{u}_1\right) \cdot \mathcal{M}_1+\mathcal{G}_1,
%\end{align}
\begin{align}
&\alpha_k \rho_k T_k \frac{\mathrm{D}_k s_k}{\mathrm{D} t}= \left(p_k-p_I\right) \mathcal{F}_k + \mathcal{G}_k  \\ &+ p_k \vc{w}_k \cdot \nabla \alpha_k + \vc{w}_k \dpr \overline{\overline{T}}_I \dpr \nabla \alpha_k + \left(\boldsymbol{u}-\boldsymbol{u}_k\right) \cdot \mathcal{M}_k , \nonumber
\end{align}
The last three terms are evaluated to be of $\mathcal{O}(\text{Kn}^2)$ and therefore abandoned. Thus, the entropy equation is reduced to the following
\begin{align}\label{eq:reduce_entropy}
&\alpha_k \rho_k T_k \frac{\mathrm{D}_k s_k}{\mathrm{D} t}= \left(p_k-p_I\right) \mathcal{F}_k + \mathcal{G}_k.
\end{align}
The model (\ref{eq:final_model}) keeps the equations for the internal energy, and abandoning a momentum equation has no impact on the entropy equation. Therefore, the above reduced entropy equation can also be derived from the reduced model (\ref{eq:final_model}). With \cref{eq:reduce_entropy}, one can readily prove the entropy inequality {\color{black}in the absence of external heat flux and energy source}
%\begin{align}\label{eq:entropy_ineq}
%\sum_{k=1}^2 \alpha_k \rho_k  \frac{\mathrm{D}_k s_k}{\mathrm{D} t}= \sum_{k=1}^2 \frac{\left(p_k-p_I\right) \mathcal{F}_k + \mathcal{G}_k}{T_k} \geq 0.
%\end{align}
\begin{align}\label{eq:entropy_ineq}
\sum_{k=1}^2 \alpha_k \rho_k  \frac{\mathrm{D}_k s_k}{\mathrm{D} t} + \sum_{k=1}^2 \nabla\dpr\left( \frac{\vc{q}_k}{T_k}\right) - \sum_{k=1}^2 \frac{\mathcal{I}_k}{T_k} \geq 0.
\end{align}
The proof can be performed in a totally similar manner to that in our previous publication\cite{zhang2022a}, and we provide a brief proof in the appendix.
}

In the model (\ref{eq:final_model}) all the mass diffusion effects are gathered into the terms containing $\vc{w}_k$. In comparision with the model in literature\cite{Cook2009Enthalpy},  our model contains the mass diffusion contribution to the viscous stress. Moreover, as the original Baer-Nunziato model, this model includes all the disequilibrium effects in phase pressure and temperature, which is driven towards equilibrium with the corresponding relaxation terms.

The derived model  has certain  advantages over the widely used model of Cook \cite{Cook2009Enthalpy} in the following aspects:
(a) It retains the pressure/temperature disequilibria between phases through the relaxation terms, thus allowing us to consider the disequilibrium effects between components. Such disequilibrium is significant in some phenomena of plasma physics.
(b) It frees the hydrodynamic step of the temperature equilibrium constraint, which is the root of spurious oscillations for diffuse interface problems\cite{allaire2002five}. {\color{black}Some special numerical cures exist in literature\cite{JOHNSEN20125705,williams2019fully}.}
(c) As to the constitutive law for viscous stress, the viscous stress of each component is determined with their own velocity (derived by virture of the diffusion law), which is compatible with thermodynamic relations. In the model of Cook \cite{Cook2009Enthalpy}, only the mass weighted velocity is used for the evaluation of viscous stress, which is decoupled from the diffusion effect. These two approaches may lead to {\color{black}noticeable} difference in simulation. A numerical test is to be considered in \Cref{sec:num_res} to demonstrate this issue.

\subsection{Diffusion models}
To make the model solvable, some closure relations are needed to relate the diffusion terms to basic variables (including density, temperature, and pressure).  The diffusion coefficients are usually derived from kinetic theories or from experimental measurements. For ordinary neural flows without plasma, there are some well accepted equations to describe the diffusion processes, for example, the Fourier's law for heat flux, the Newton's law for viscous stress, and the Fick's law for mass diffusion. However, for plasma flows, the diffusion laws are far more complicated. The electron-ion disequilibrium may also play an important role. For simplicity, in the present paper we concentrate on the case of  electron-ion equilibrium and show the feasibility of our approach to consider multi-component plasma flows with diffusions. This assumption  should not limit the present model's applicability to the disequilibrium ion-electron temperature flows. Further work on the ion-electron disequilibrium plasma flows will follow in our next publication. Moreover, one can also regard our model as the limit of instantaneous ion-electron relaxation time of some ion/electron temperature disequilibrium model.

Based on the above discussions, we use the ion-electron equilibrium Spitzer-Harm model\cite{Spitzer1953} for the heat flux. The commonly used models to calculate the plasma viscosity and mass diffusion include the Clerouin's model\cite{clerouin1998viscosity}  for viscosity and the Paquette's model\cite{paquette1986diffusion} for mass diffusion, respectively.  In recent years, driven by the need to evaluate species mixing/separation in thermonuclear inertial confinement fusion plasmas, a series of transport models \cite{kagan2018,simakov2016a,molvig2014} for multicomponent plasmas are derived. In these models the mass diffusion is driven by the gradients of field variables such as species concentration, ion/electron pressure (baro-diffusion) and ion/electron temperature (thermal diffusion). For Simakov\cite{simakov2016hydrodynamic}, the closure law for the diffusion velocity is written  as follows:
\begin{equation}\label{eq:wk_simakov}
    \vc{w}_k = - \sum D_{kj} \vc{d}_j + D_{k}^T \nabla\left( \text{ln} T_i \right),
\end{equation}
and
\begin{align}\label{eq:dk_simakov}
    &\vc{d}_k = \nabla x_k + \left( x_k - y_k \right) \nabla \left( \text{ln} p \right) + \left( z_k - y_k \right) \frac{\nabla p_e}{p} \nonumber
    \\&+ \left( \frac{Z_k n_k}{n_e} - \frac{Z_k^2 n_k}{\sum_{j} Z_j^2 n_j} \right) \frac{n_e}{n} \frac{\beta_0 \nabla T_e}{T},
\end{align}
where $x_k$, $y_k$, $z_k$, $Z_k$ and $n_k$ are the number fraction, the mass fraction, the charge fraction, the charge number and the number density  of component $k$, respectively.

The parameter $\beta_0$ is a function of effective charge number,
\begin{equation}\label{eq:beta0}
    \beta_0 \left( Z_{eff} \right) = \frac{30Z_{eff}\left( 11Z_{eff} + 15\sqrt{2} \right)}{217Z_{eff}^2 + 604\sqrt{2}Z_{eff} + 288}.
\end{equation}

In the case of two components, the above representation is identical to that of Kagan \& Tang \cite{kagan2014}:
\begin{equation}\label{eq:Kagan}
    \vc{w}_k = - \frac{D}{y_k} \left( \nabla y_k + D_{pk} \nabla \text{log} p  \right),
\end{equation}
where we have omitted the electro- and thermo-diffusions. The baro-diffusion coefficient $D_{pk}$ is determined as follows
\begin{equation}\label{eq:Dpk}
    D_{pk} = y_k y_{k*} \left(M_{k*} - M_{k}\right) \left(  \frac{y_1}{M_1} + \frac{y_2}{M_2} \right).
\end{equation}
where $M_{k}$ is the ion mass of the $k$ component.

\section{Numerical method}
\label{sec:num_med}
The model (\ref{eq:final_model}) is solved by using the splitting procedure. According to physical processes, the model is split into three sub-systems: the hyperbolic sub-system, the diffusion sub-system and the the relaxation sub-system. These sub-systems are solved within each time step in order.
The solution of one sub-system serves as the initial condition for the next.
The numerical methods for the hyperbolic and parabolic part are similar to those described in our previous works \cite{zhang2022a}, thus are omitted here.  
Particularities here consist in the temperature relaxation and mass diffusion, which are to be described in this section.
%The particularities here lies in that the constitutive law for each component is dependent on its own velocity $\vc{w}_k$, which can be obtained with the aid of \cref{eq:wk_simakov,eq:w_k}.

\subsection{Thermal relaxation}\label{subsec:thermal_relax}
Here we describe some details for the solution of the thermal relaxation sub-system, which reads
\begin{equation}\label{eq:sub_relax}
%\dudx{\alpha_k\rho_ke_k}{t} = \mathcal{Q}_k = \eta \left( T_k^{*} - T_k \right)
    \dudx{\alpha_k\rho_ke_k}{t} = \mathcal{Q}_k - p_I \dudx{\alpha_k}{t}.
\end{equation}

To account for the interface motion due to arising pressure disequilibrium, we adopt the following assumption
\begin{equation}\label{eq:sub_relax_alpk}
\dudx{\alpha_k}{t} = \Lambda \mathcal{Q}_k
\end{equation}
where the exchanged energy $\mathcal{Q}_k  = \eta \left( T_k^{*} - T_k \right)$ and the interfacial pressure is approximated as $p_I = \sum{\alpha_k p_k}$.

The volume fraction varies in such a way that the phasic pressure equilibrium is maintained, implicitly meaning that the pressure relaxation rate is much larger than the thermal relaxation rate.  By using the pressure equilibrium condition \[\dudx{p_1}{t} = \dudx{p_2}{t},\] one can derive the evolution equation
for $\alpha_k$ in the case of polytropic gas EOS as follows:
\begin{equation}\label{eq:kappa}
   \Lambda =  \frac{1}{p_I + \frac{\sum{p_k/\alpha_k}}{\sum{\left( \gamma_k - 1 \right)/\alpha_k}}}.
\end{equation}

Note that partial densities and momentum remain unchanged during the thermal relaxation stage.
Reformulation of \cref{eq:sub_relax,eq:sub_relax_alpk,eq:kappa} gives
\begin{equation}\label{eq:dTkdt}
    \dudx{T_k}{t} = \widehat{\eta}_k \left( T_k^{*} - T_k \right),
\end{equation}
where \[
\widehat{\eta}_k = \frac{   \frac{\sum{p_k/\alpha_k}}{\sum{\left( \gamma_k - 1 \right)/\alpha_k}}    }{  \left(  \frac{\sum{p_k/\alpha_k}}{\sum{\left( \gamma_k - 1 \right)/\alpha_k}} + p_I \right) m_k C_{vk}} \eta.
\]

For the solution of the {\color{black}nonlinear \cref{eq:dTkdt,eq:sub_relax_alpk}}, iterative methods should be used, where for each iteration one solves the following linearized ODEs {\color{black}in each time step $\Delta t$}
\begin{equation}
    \dudx{T_k^{(s+1)}}{t} = \widehat{\eta}_k^{(s)} \left( T_k^{*(s+1)} - T_k^{(s+1)} \right),
\end{equation}
where $(s)$ denotes the iteration index. The analytical solution for the above ODEs are as follows
%\[
%T_1^{(s+1)} = A + B \widehat{\eta}_1^{(s)} , \;\; T_2^{(s+1)} = A - B \widehat{\eta}_2^{(s)}
%\]
%\[A = \frac{C T_{10}\widehat{\eta}_2^{(s)} + T_{20} \widehat{\eta}_1^{(s)}}{C \widehat{\eta}_2^{(s)} + \widehat{\eta}_1^{(s)}}, \]
%\[B = \frac{\left(T_{10}-T_{20}\right)  C^{t} }{C \widehat{\eta}_2^{(s)} + \widehat{\eta}_1^{(s)}}, \]
%\[C = e^{ -\widehat{\eta}_1^{(s)} - \widehat{\eta}_2^{(s)}}, \]
{\color{black}
\begin{align*}
&T_1^{(s+1)} = A + B \widehat{\eta}_1^{(s)} , \;\; T_2^{(s+1)} = A - B \widehat{\eta}_2^{(s)} \\
&A = \frac{ T_{10}\widehat{\eta}_2^{(s)} + T_{20} \widehat{\eta}_1^{(s)}}{ \widehat{\eta}_2^{(s)} + \widehat{\eta}_1^{(s)}}, \;\; B = \frac{\left(T_{10}-T_{20}\right)  C }{ \widehat{\eta}_2^{(s)} + \widehat{\eta}_1^{(s)}}, \\
&C = e^{ - \left( \widehat{\eta}_1^{(s)} + \widehat{\eta}_2^{(s)} \right) \Delta t },
\end{align*}}
where $T_{10}$ and $T_{20}$ are the initial component temperature at the beginning of the temperature relaxation stage.

The iterations are performed until the convergence condition $|T_k^{(s+1)}-T_k^{(s)}|<\epsilon$ is satisfied.

\subsection{Mass diffusion}
The sub-system for the mass diffusion reads:
\begin{eqnarray*}
  \dudx{\alpha_k\rho_k}{t} &=& -\nabla\dpr\left( \alpha_k \rho_k \vc{w}_k \right) := \mathcal{C}_k,\\
  \dudx{\alpha_k\rho_k e_k}{t} &=& - \alpha_k \rho_k \vc{w}_k \dpr \nabla e_k  \\&-& \alpha_k p_k \nabla\dpr \vc{w}_k - p_I \dudx{\alpha_k}{t} := \mathcal{E}_k - p_I \dudx{\alpha_k}{t}.
\end{eqnarray*}

In similar manner to that for defining the evolution equation in \Cref{subsec:thermal_relax}, the equation describing  the volume fraction variation under the pressure equilibrium takes the following form:
%\begin{equation}\label{eq:dalpdt_md}
%    \dudx{\alpha_1}{t} = \frac{ \left( G_1 \mathcal{E}_1 / \alpha_1 -  G_2 \mathcal{E}_2 / \alpha_2 \right) + \left( p_1\mathcal{C}_1 / \left(\alpha_1 \rho_1 \right) - p_2\mathcal{C}_2 / \left(\alpha_2 \rho_2\right)  \right)  }{\left( G_1/\alpha_1 + G_2/\alpha_2 \right) p_I + \left( p_1/\alpha_1 + p_2 / \alpha_2 \right)},
%\end{equation}
\begin{equation}\label{eq:dalpdt_md}
    \dudx{\alpha_1}{t} = \frac{ \left( \frac{G_1 \mathcal{E}_1 }{ \alpha_1} -  \frac{G_2 \mathcal{E}_2 }{ \alpha_2} \right) + \left( \frac{p_1\mathcal{C}_1 }{\alpha_1 \rho_1 } - \frac{p_2\mathcal{C}_2 }{\alpha_2 \rho_2}  \right)  }{\left( \frac{G_1}{\alpha_1} + \frac{G_2}{\alpha_2} \right) p_I + \left( \frac{p_1}{\alpha_1} + \frac{p_2}{\alpha_2} \right)},
\end{equation}
where $G_k = \gamma_k - 1$.

{\color{black}Note that the diffusion term in \cref{eq:dalpdt_md} is to maintain the obtained pressure equilibrium. This does not contradicts \cref{eq:final_modeld}, but is a result of the sequence in solving the split sub-systems. Similar numerical strategy is adopted in solving phase transition problems\cite{ZEIN20102964}.}

The involved spatial derivatives are approximated with the central difference scheme.

\section{Numerical results}
\label{sec:num_res}
In this section we present some numerical results to validate the proposed model and numerical methods. Moreover, with the aid of the proposed method we investigate the impact of thermal relaxation on RT instability development.

\subsection{The mass diffusion problem}
\label{sec:pure_diff_sec1}
\paragraph{The pure diffusion problem}
Let us consider a pure diffusion problem as in \cite{Thornber2018,Kokkinakis2015}. The two components are characterized by the polytropic EOS with the adiabatic coefficients $\gamma_1 = 2.0$ and $\gamma_2 = 1.4$ and densities $\rho_1 = 20.0$ and $\rho_2 = 1.0$, respectively.
In the computational domain $[0,1]$ the fluids are initially in temperature and pressure equilibrium which means $ \left( \gamma_k - 1 \right) \rho_k C_{vk} = \text{const} $. This relation gives a constraint for prescribing the heat capacities $C_{vk}$.

The initial mixture density and partial density are given as follows:
%\[ \rho = \frac{1}{2}\left( \rho_1 +  \rho_2 \right)  -  \frac{1}{2}\left( \rho_1 -  \rho_2 \right) \text{erf}\left(z\right),  \;\; \rho y_1 = \frac{1}{2} \rho_1  -  \frac{1}{2} \rho_1 \text{erf}\left(z\right),  \]
\begin{align}\label{eq:pure_diff_rho}
    &\rho = \frac{1}{2}\left( \rho_1 +  \rho_2 \right)  -  \frac{1}{2}\left( \rho_1 -  \rho_2 \right) \text{erf}\left(z\right),  \nonumber
    \\ &\rho y_1 = \frac{1}{2} \rho_1  -  \frac{1}{2} \rho_1 \text{erf}\left(z\right), 
\end{align}
where
\[z = \frac{x-x_0}{\sqrt{4Dt + h_0^2}}.\]

In the present test we use $t = 0.0$, $x_0 = 0.5$ and $h_0 = 0.02$ for prescribing the initial condition.

With the pressure as large as $p = 1\times10^5$, the Mach number is so small that the compressibility effect can be neglected. In this case given the mixture density profile one can derive the {\color{black}mass weighted mean velocity}
\[u = - \frac{D}{\rho} \dudx{\rho}{x}.\]
Moreover, the analytical solution for density to this pure diffusion problem is given by \cref{eq:pure_diff_rho} \cite{Thornber2018,Livescu2013}.

\begin{figure}[htbp]
\centering
\subfloat{\includegraphics[width=0.38\textwidth]{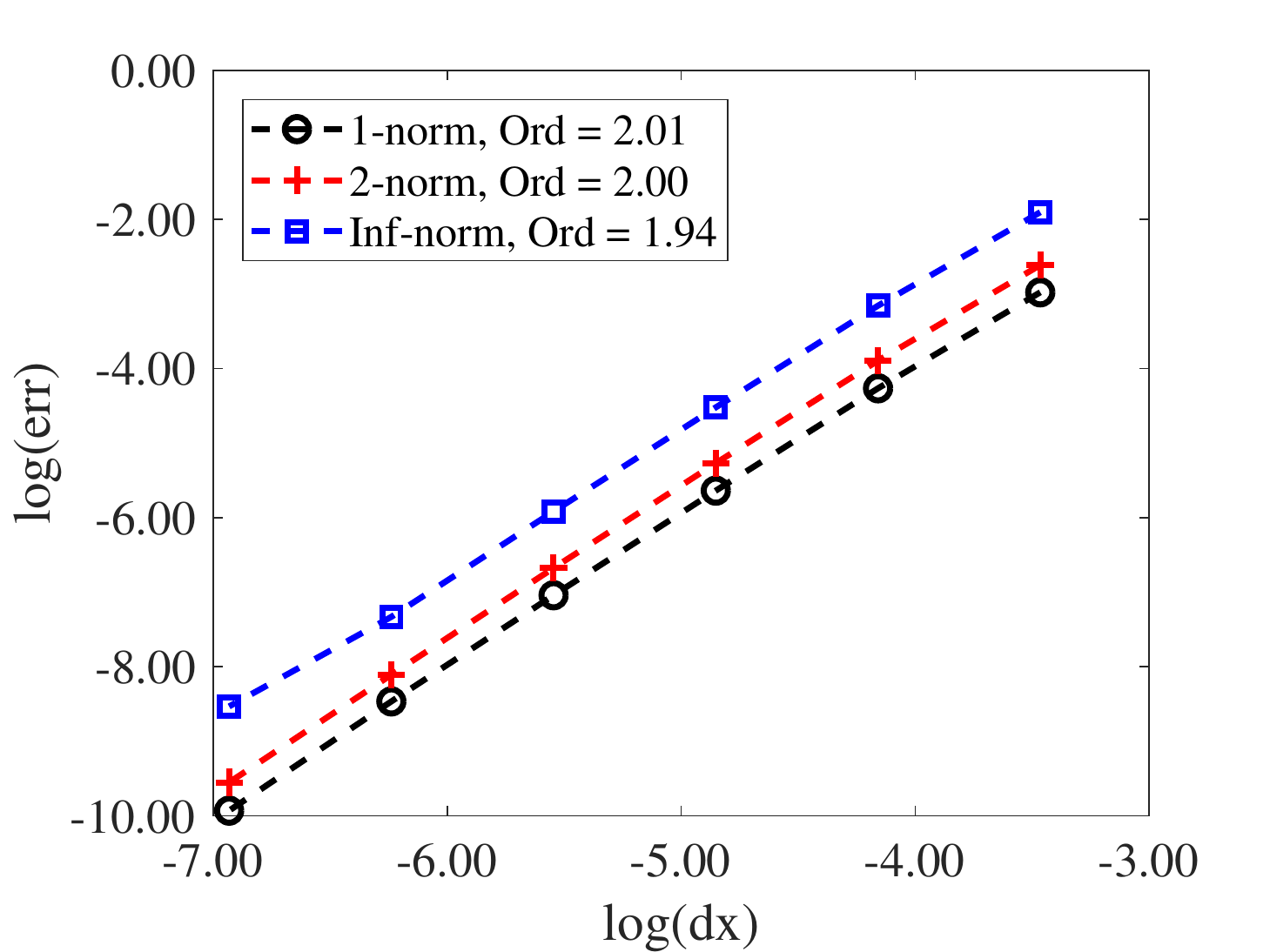}}\\
\subfloat{\includegraphics[width=0.38\textwidth]{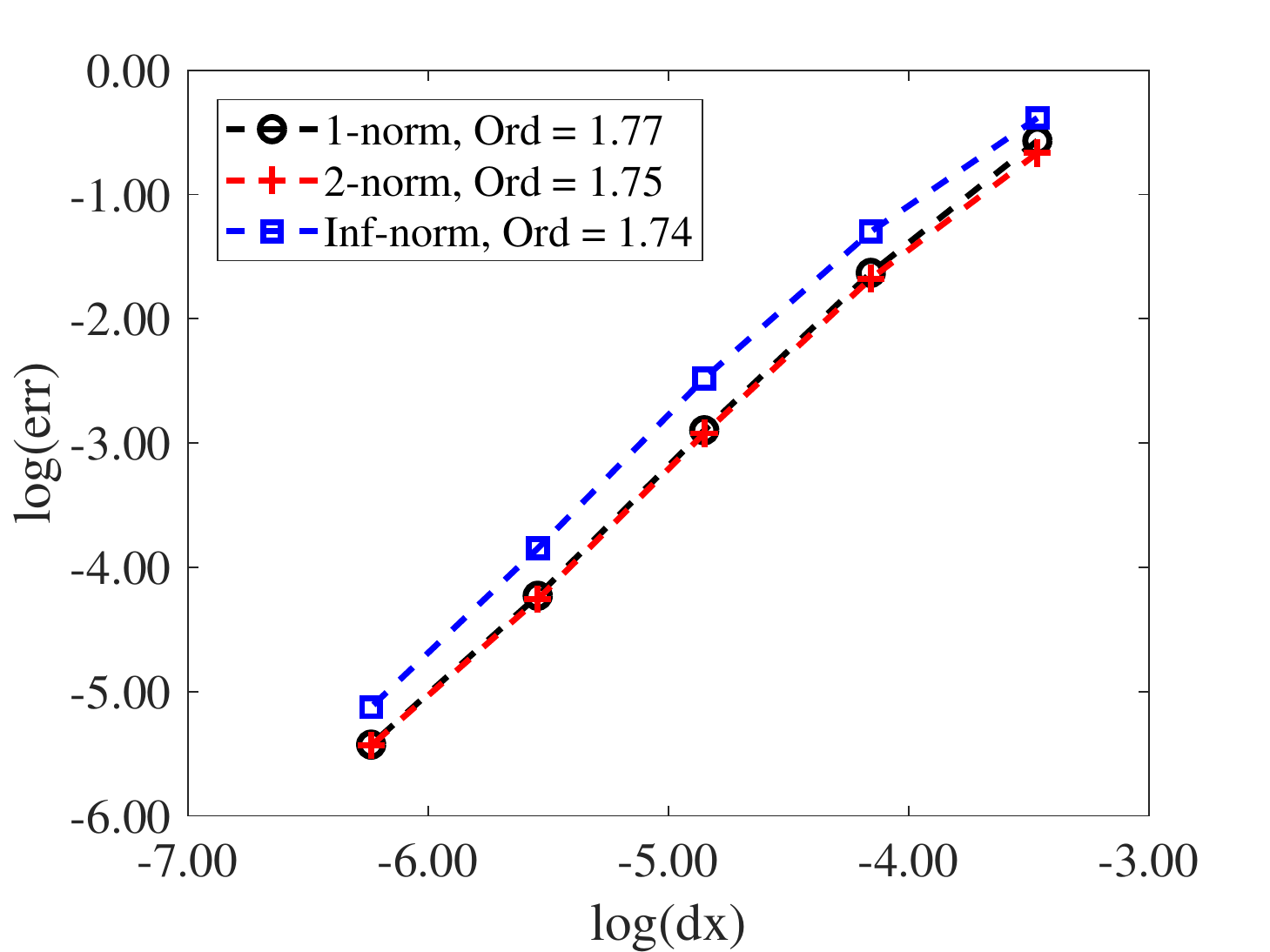}}
\caption{The convergence rate for the mass diffusion problem. {\color{black}Upper}: the pure {\color{black}diffusion} problem, {\color{black}lower}: the advection-diffusion problem. }
\label{fig:convgRate}
\end{figure}

\begin{figure*}
\subfloat[density]{\includegraphics[width=0.38\textwidth]{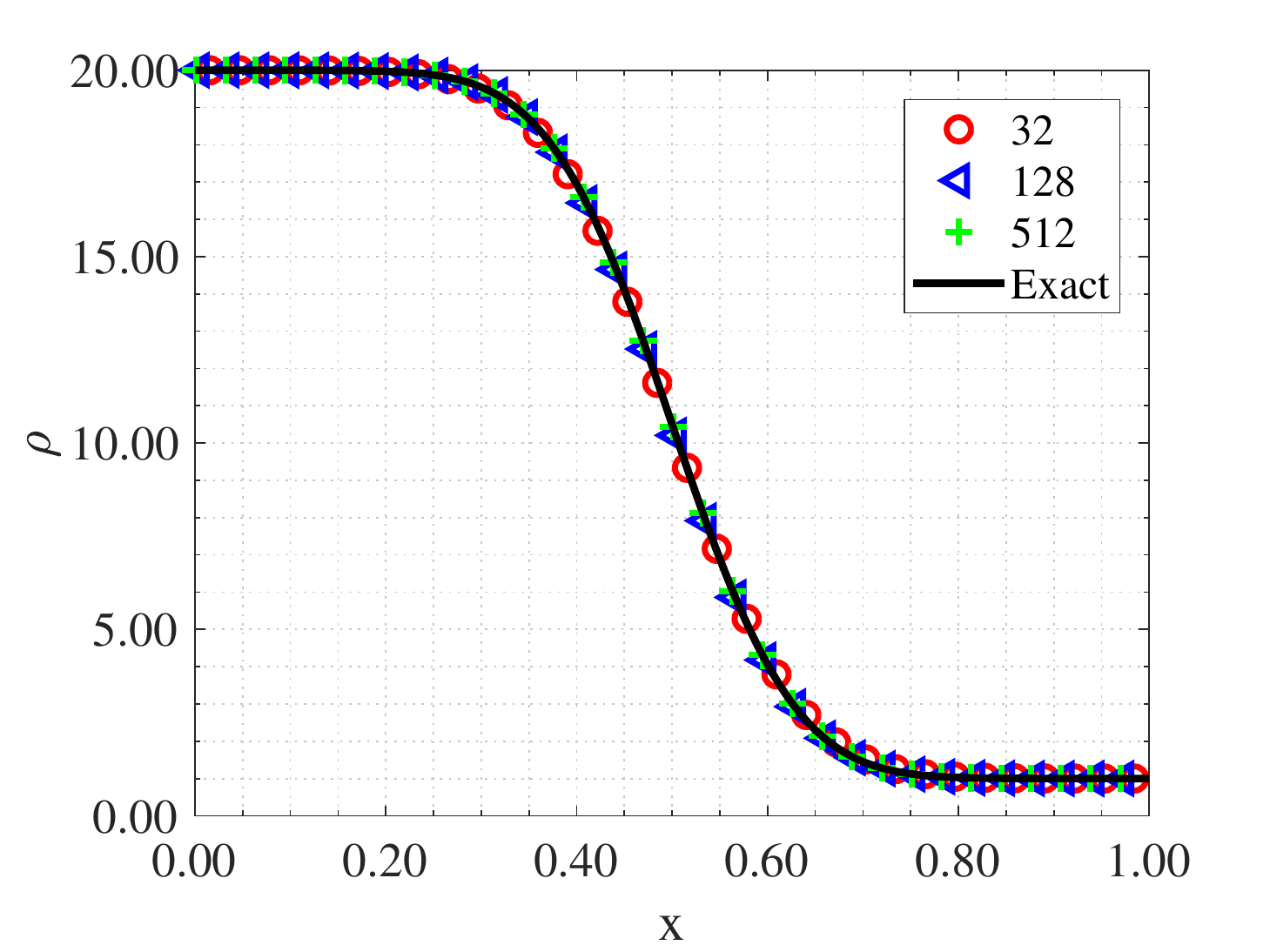}}
\subfloat[velocity]{\includegraphics[width=0.38\textwidth]{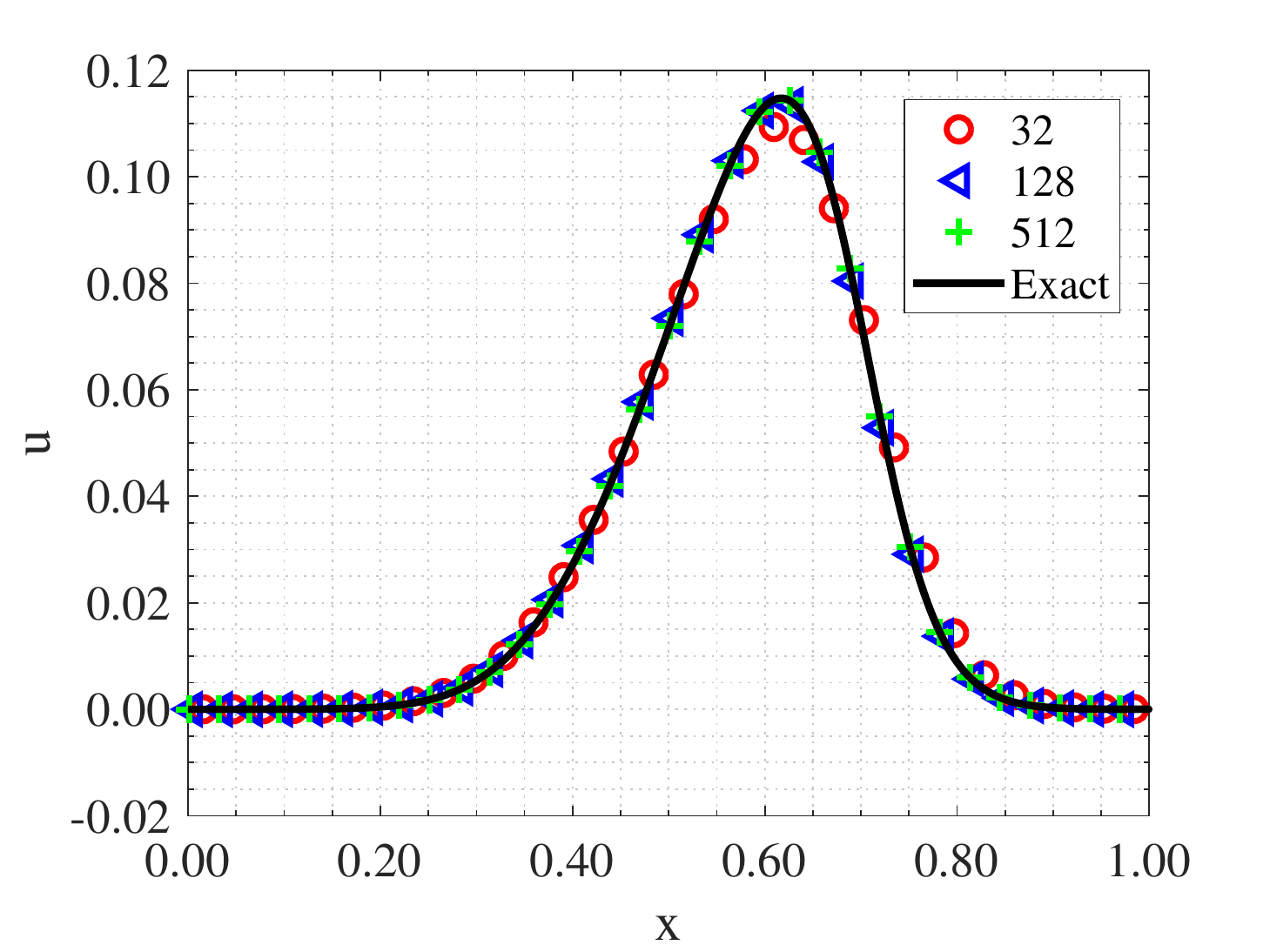}}\\
\subfloat[pressure]{\includegraphics[width=0.38\textwidth]{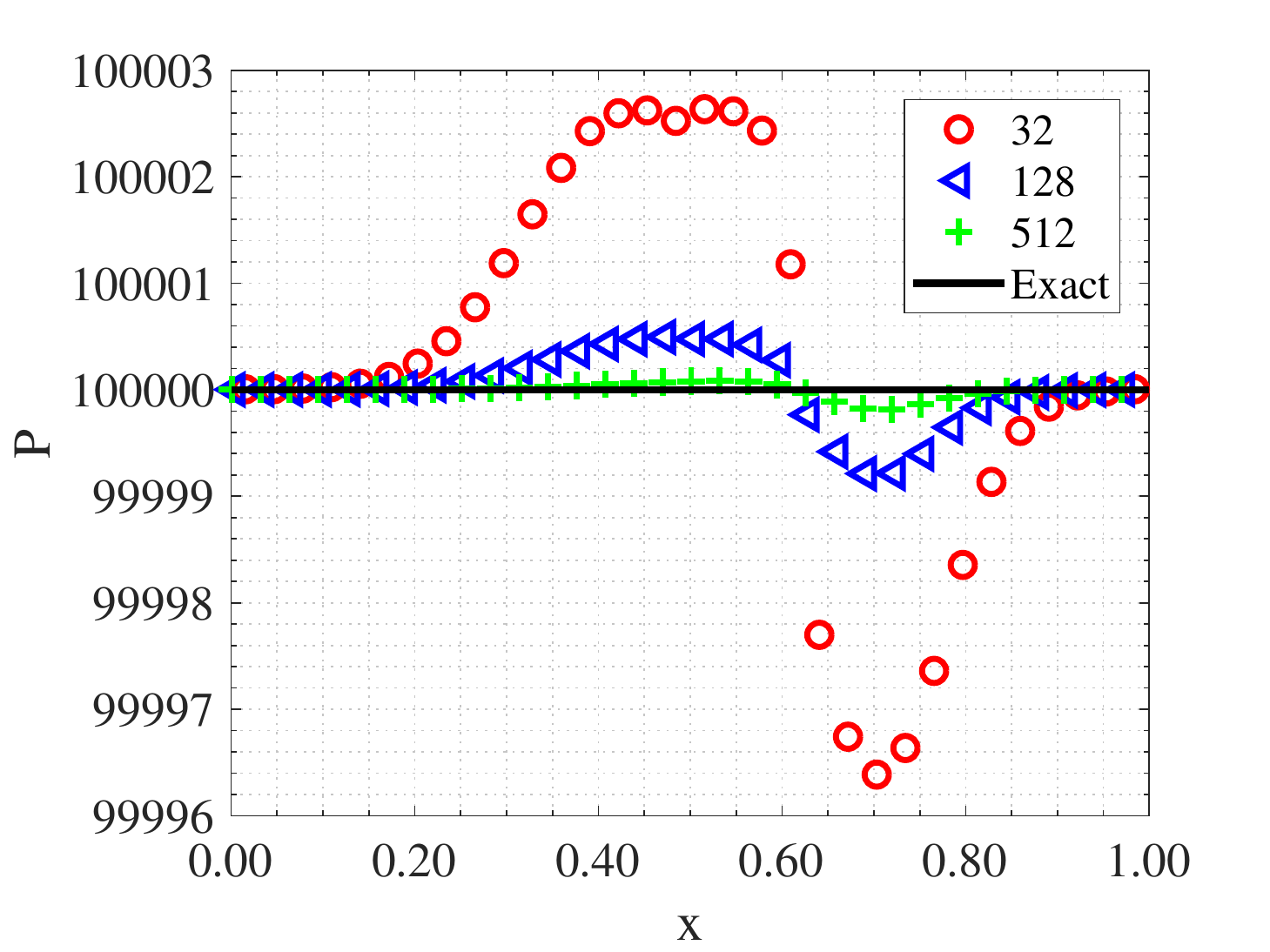}}
\subfloat[temperature]{\includegraphics[width=0.38\textwidth]{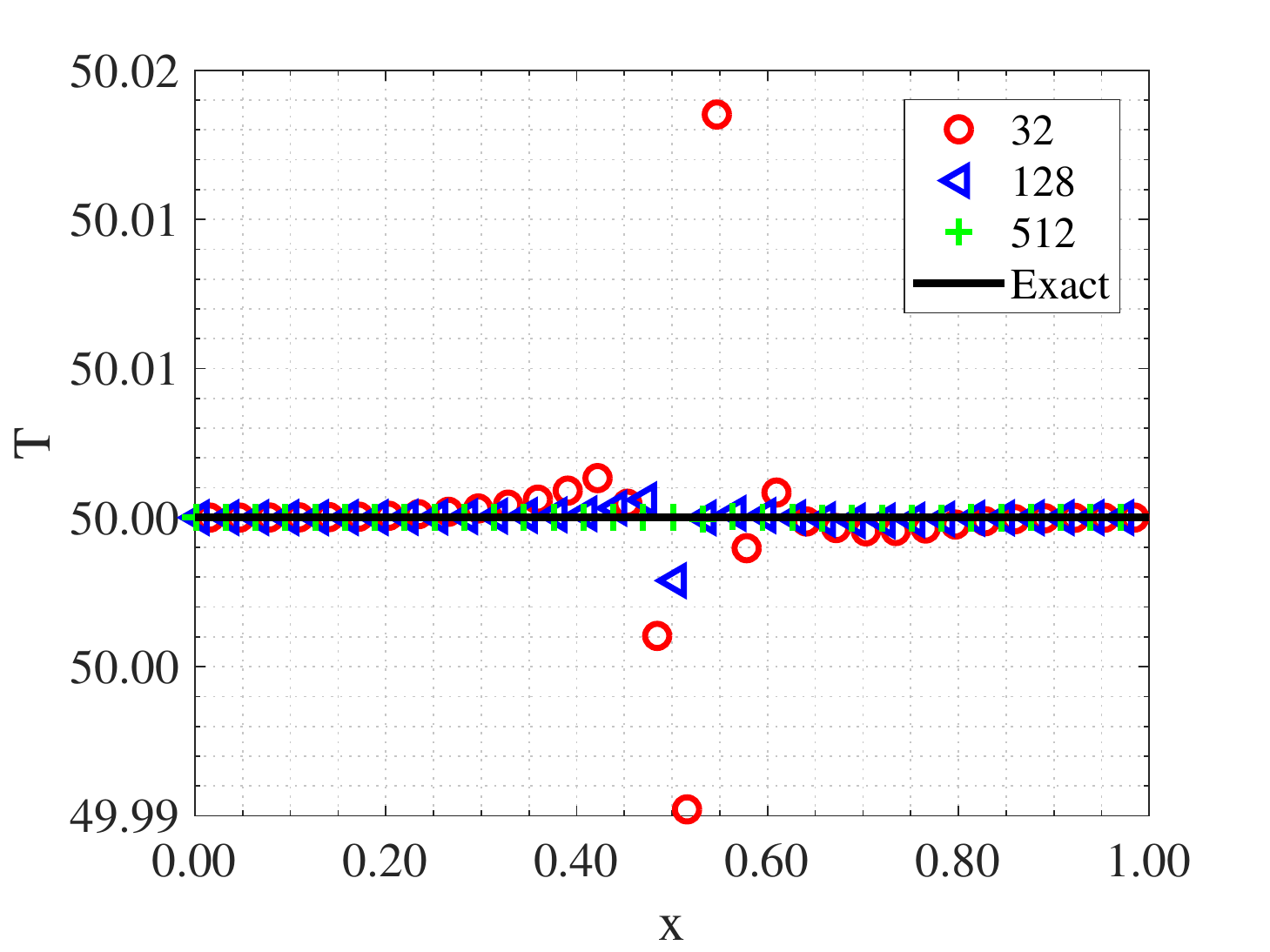}}\\
\subfloat[Mass fraction]{\includegraphics[width=0.38\textwidth]{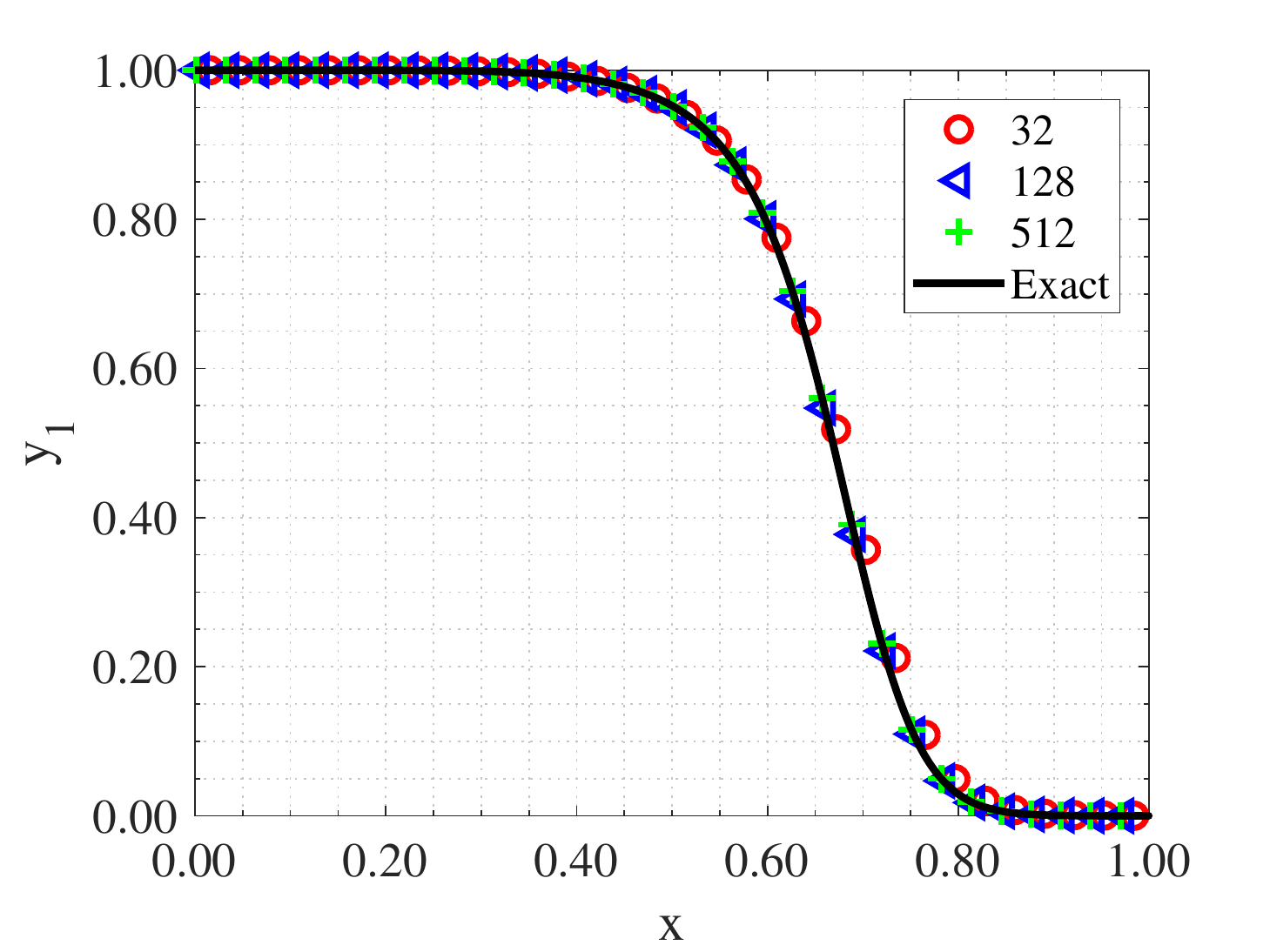}}
\subfloat[Volume fraction]{\includegraphics[width=0.38\textwidth]{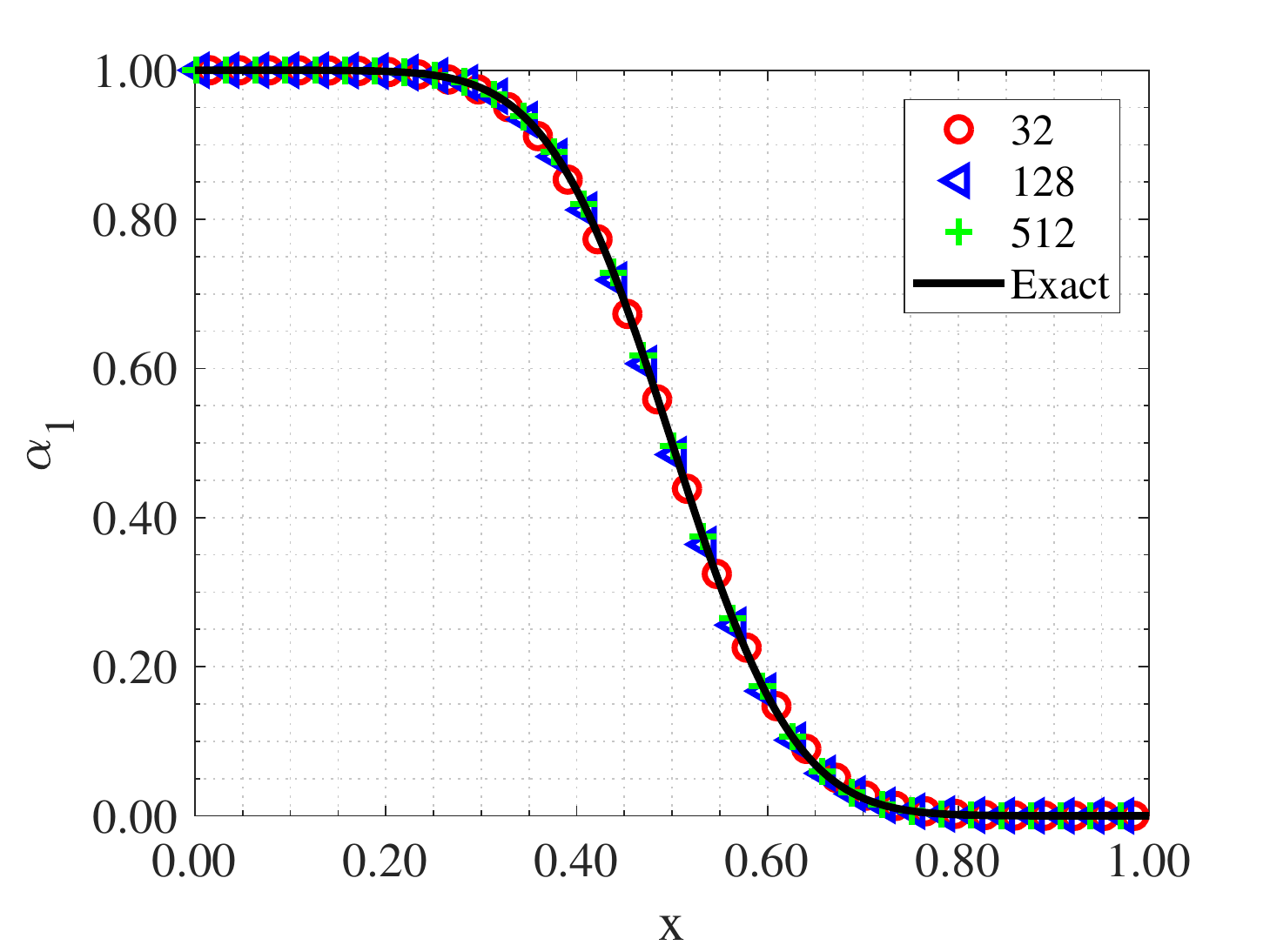}}
\caption{{\color{black}The  numerical results in the pure diffusion problem for density, velocity, pressure, temperature, mass fraction and volume fraction with the refinement of grid. }}
\label{fig:convgVar}
\end{figure*}

Computations are performed {\color{black}to the time moment $t = 0.5$} on a series of refining grids of 32, 64, 128, 256, 512 and 1024 cells. {\color{black}Reflective boundary conditions are imposed on both sides.} The errors for density are defined as its distance to the analytical solution \cref{eq:pure_diff_rho}. The convergence performance is displayed in \Cref{fig:convgRate}.
One can see that the second order is reached as expected. The corresponding convergence performance for different variables are demonstrated in \Cref{fig:convgVar}.

\paragraph{The advection-diffusion problem}
We continue to consider the advection-diffusion problem where the multi-component fluid is transported by a uniform velocity $u = 4.0$ while diffusing. By choosing a reference moving at the transport velocity, one can see that the analytical solution is still described by   \cref{eq:pure_diff_rho},
only with a transported interface center. The computational domain is enlarged to [0,4] and the interface center is transported to $x=2.5$ at $t=0.5$. Similarly, we obtain the convergence rate for this problem as displayed in \Cref{fig:convgRate}.
The convergence rate is somewhat smaller than that for the pure diffusion problem  since the numerical resolution of the advection part adds to some error.

{\color{black}The comparison between the pure-diffusion and advection-diffusion problems are displayed in  \Cref{fig:convgVarVS}. One can observe that on the coarse grid the numerical results (for the density, velocity and mass/volume fraction) of the advection-diffusion problem suffer from more deviations from  the exact solutions. This can be explained by the fact that extra numerical dissipation is needed in solving the advection part. On the other hand, the introduced numerical dissipation also smooths the pressure/temperature oscillations, as can be seen in  \Cref{fig:convgVarVS}(c-d).}

{\color{black}
\begin{figure*}
\subfloat[density]{\includegraphics[width=0.38\textwidth]{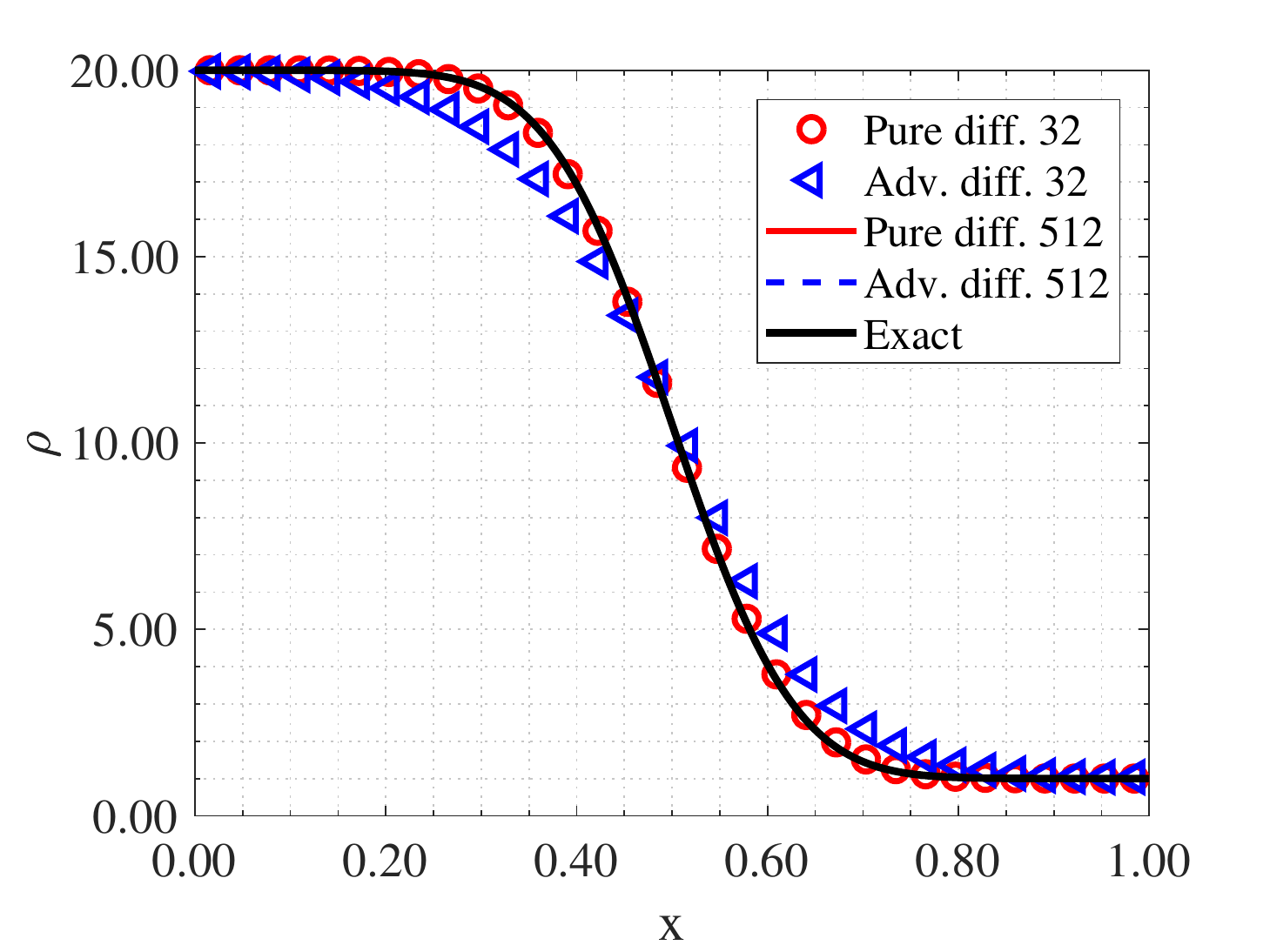}}
\subfloat[velocity]{\includegraphics[width=0.38\textwidth]{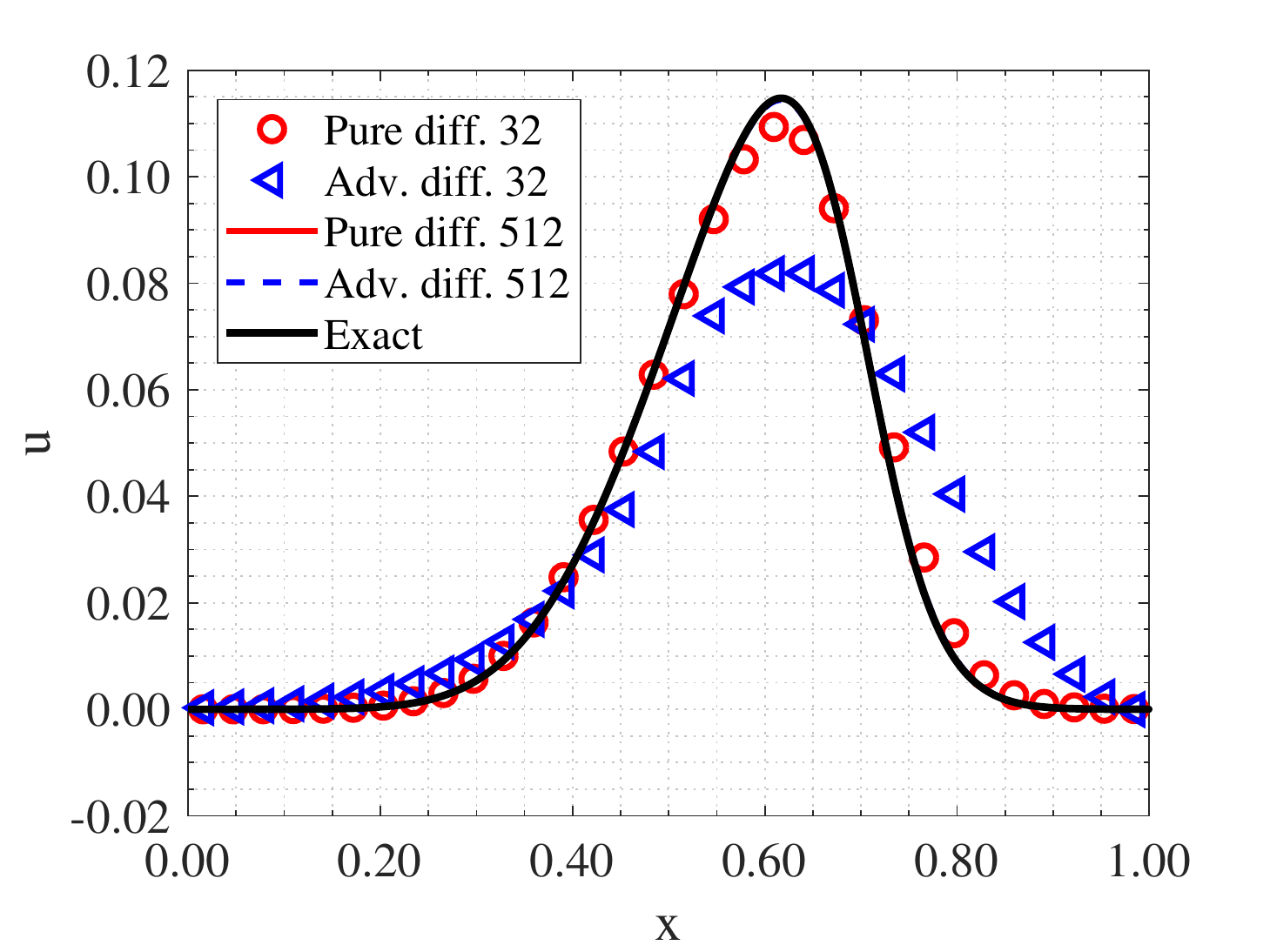}}\\
\subfloat[pressure]{\includegraphics[width=0.38\textwidth]{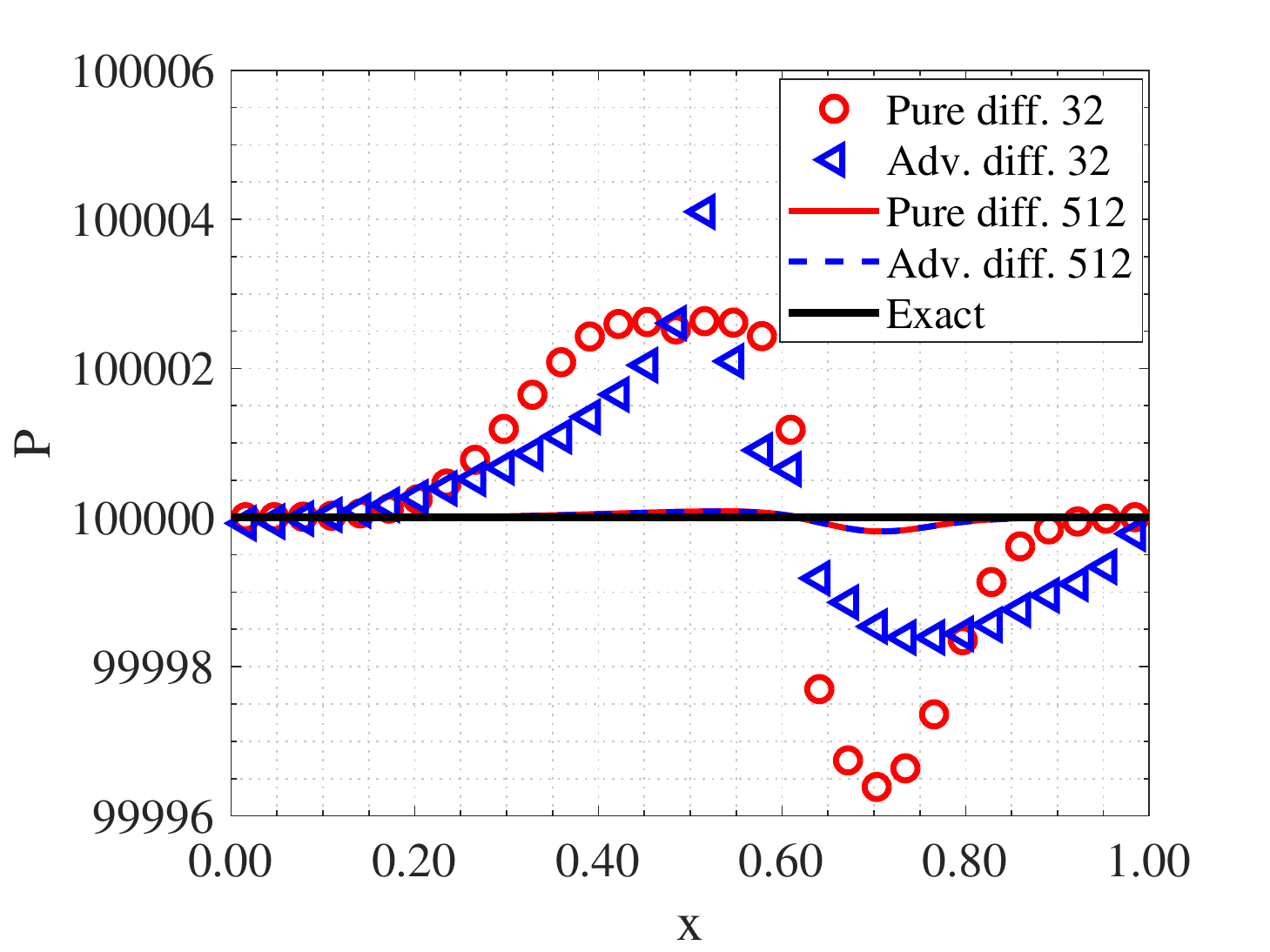}}
\subfloat[temperature]{\includegraphics[width=0.38\textwidth]{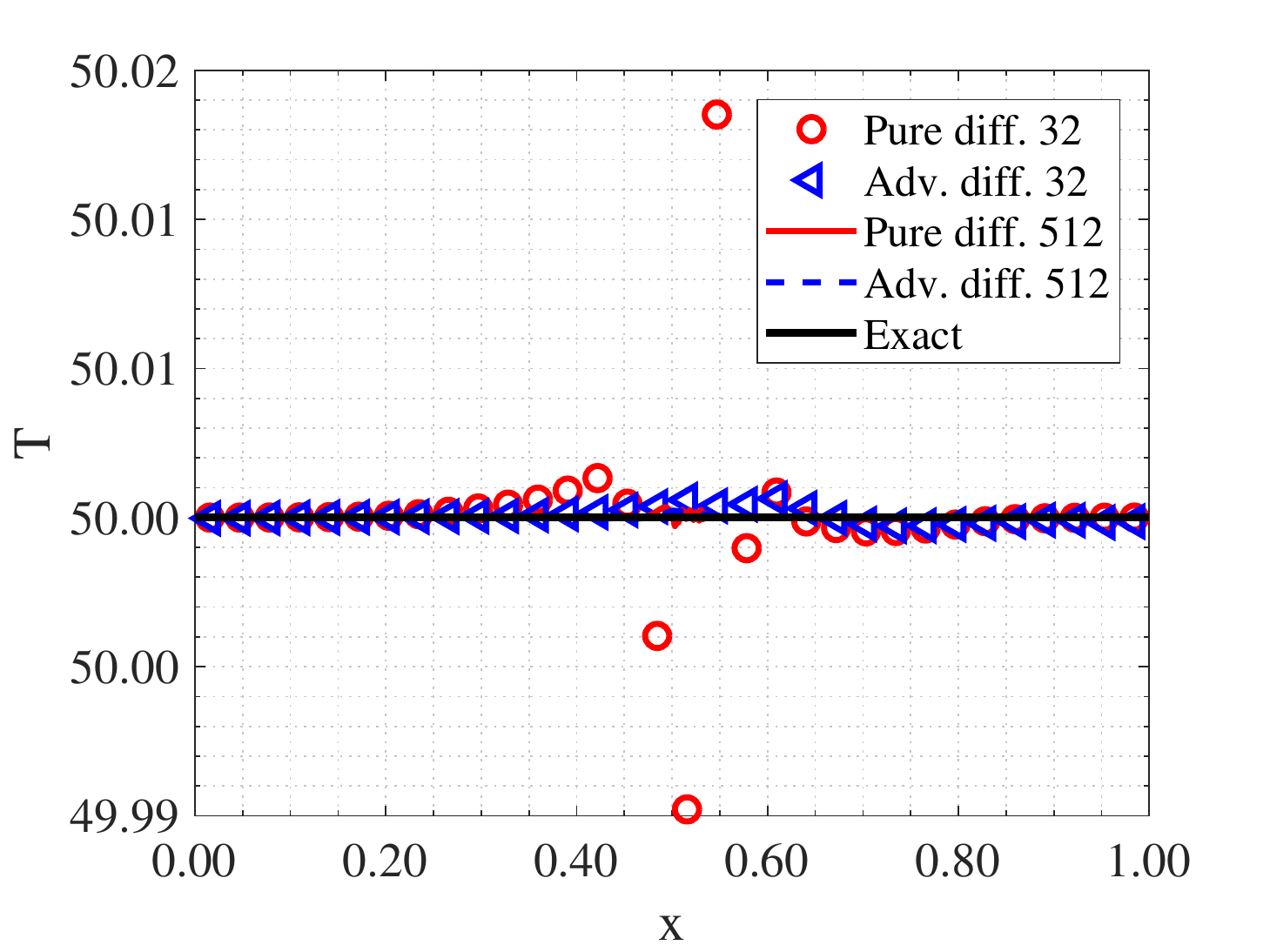}}\\
\subfloat[Mass fraction]{\includegraphics[width=0.38\textwidth]{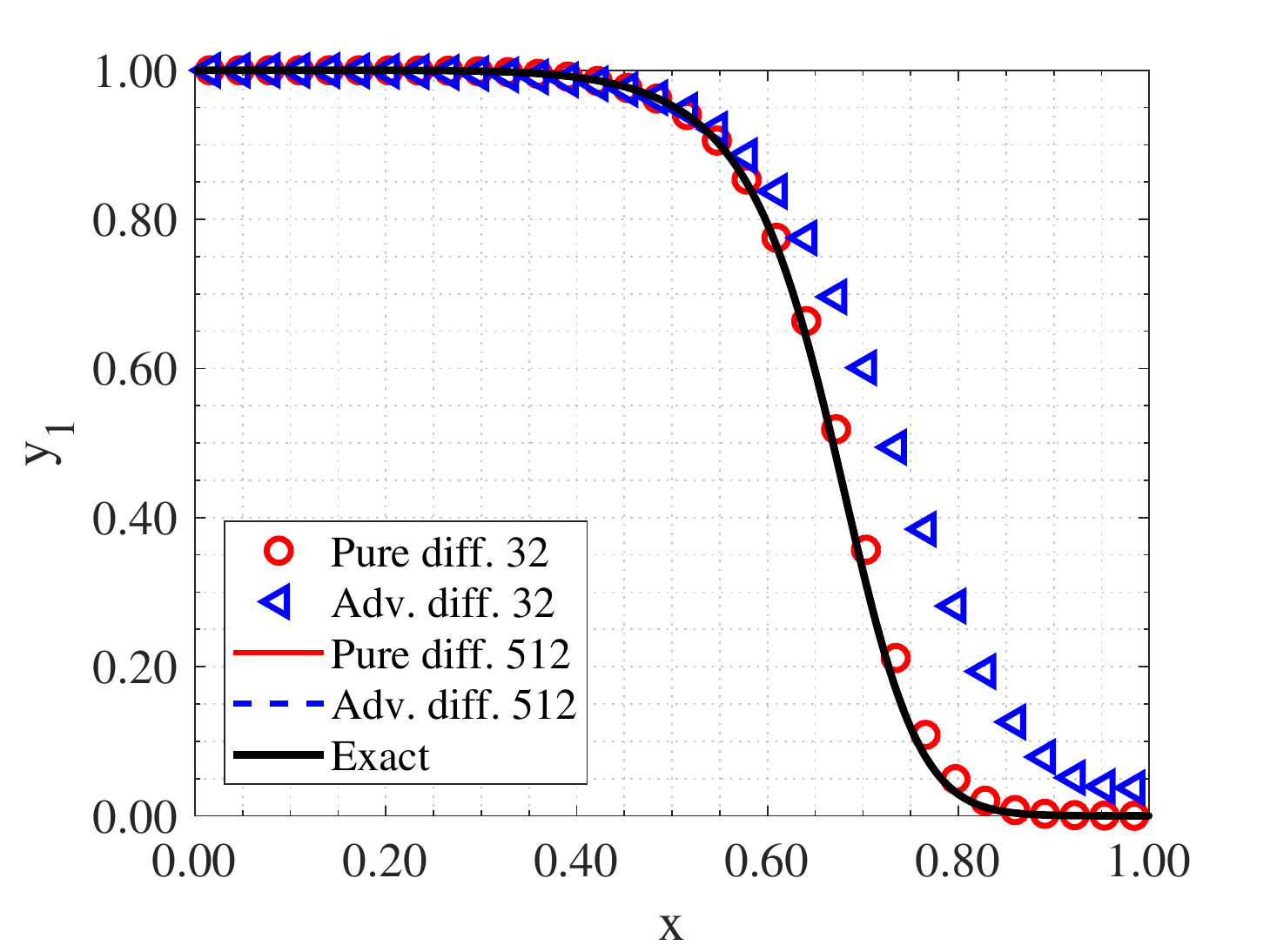}}
\subfloat[Volume fraction]{\includegraphics[width=0.38\textwidth]{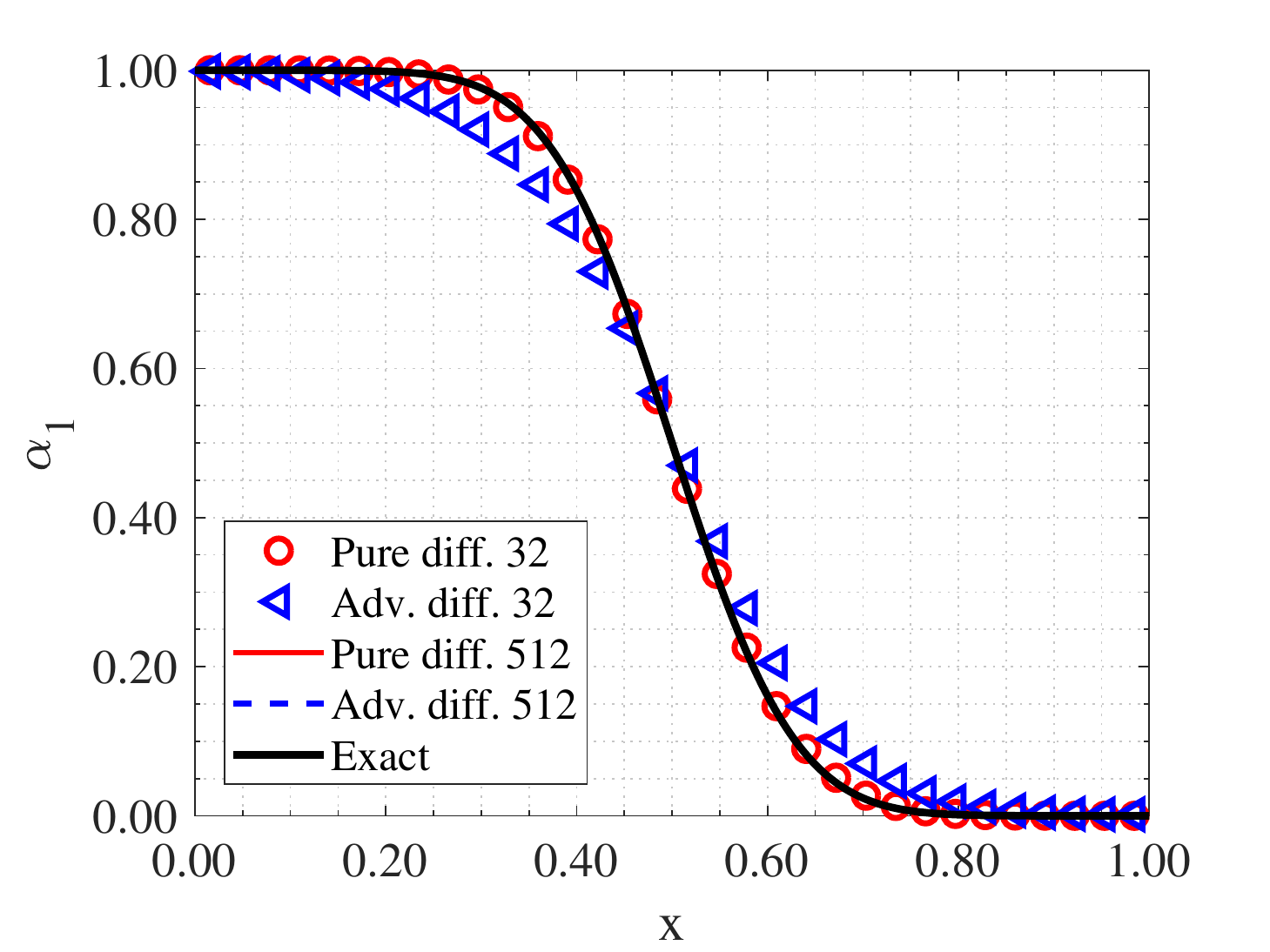}}
\caption{{\color{black}The  numerical results in the pure-diffusion / advection-diffusion problems for density, velocity, pressure, temperature, mass fraction and volume fraction. }}
\label{fig:convgVarVS}
\end{figure*}}

\paragraph{The baro-diffusion and viscous effect}
{\color{black}The initial density and velocity profiles are the same as those in the pure-diffusion problem in \Cref{sec:pure_diff_sec1}.} To show the ability of our method to deal with the baro-diffusion and viscous effect, we consider a problem with considerable compressibility effect under the background pressure $p=100$.  {\color{black}Under such pressure the compressibility results in a non-uniform pressure profile, i.e., non-zero pressure gradient. Thus, the baro-diffusion begins to effect the results.} The kinetic viscosity is taken to be $\nu = 50D$, which is large enough to show the impact of the viscous dissipation. {\color{black}Initial and boundary conditions are the same as the test in \Cref{sec:pure_diff_sec1}.}

\begin{figure}[htbp]
\centering
\subfloat[Density]{\includegraphics[width=0.38\textwidth]{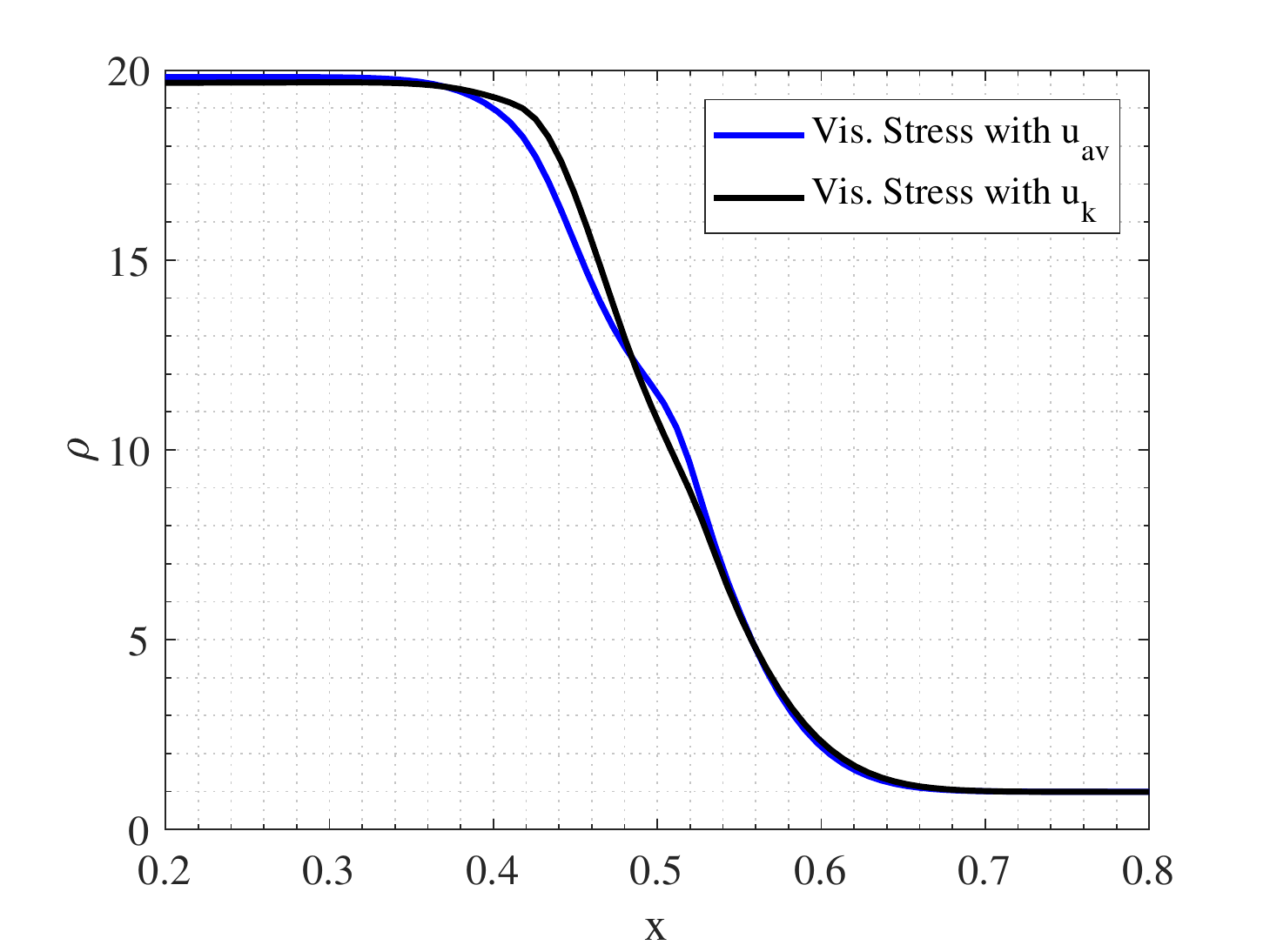}}\\
\subfloat[Mass fraction]{\includegraphics[width=0.38\textwidth]{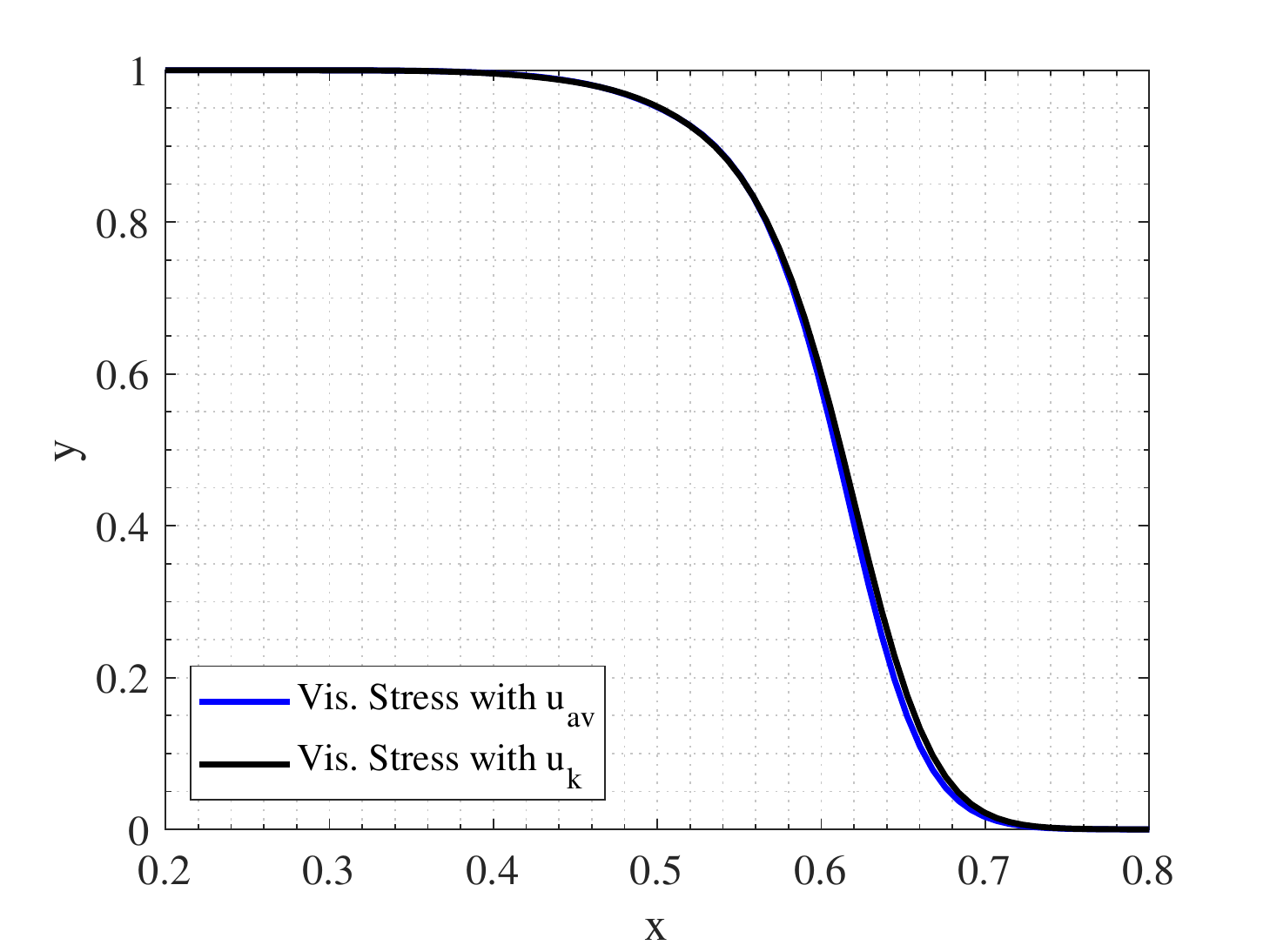}}\\
\subfloat[Pressure]{\includegraphics[width=0.38\textwidth]{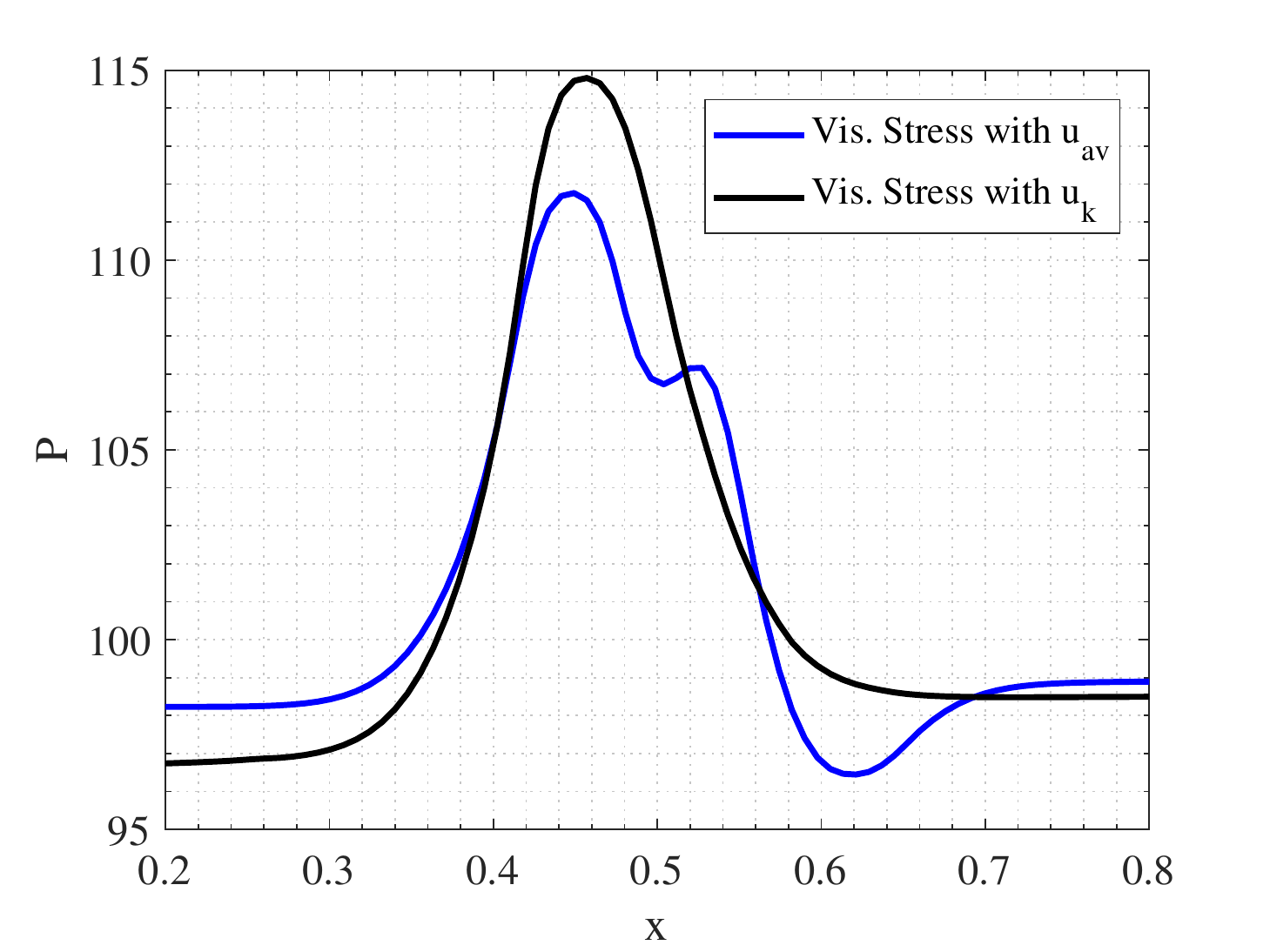}}
\caption{The  numerical results for the mass diffusion problem with viscous effect. ``Vis. stress with $u_k$/$u_{av}$'' $-$ numerical results obtained with the viscous stress being calculated with the component velocity/the mass weighted velocity. }
\label{fig:visdiff}
\end{figure}

As for the viscous part, the conventional way is to determine the viscous stress by using the mass-weighted velocity as in \cite{Cook2009Enthalpy}.
Different from this approach, we use the component velocity to calculate their respective viscous stress. The numerical results obtained by these two approaches are compared in \Cref{fig:visdiff}.
One can see noticeable difference between these numerical results, especially in pressure. New extreme arises in the pressure profile obtained with the $u_{av}$ approach. The difference in pressure results in the corresponding difference in density.

\begin{figure}[htbp]
\centering
\subfloat[Density]{\includegraphics[width=0.38\textwidth]{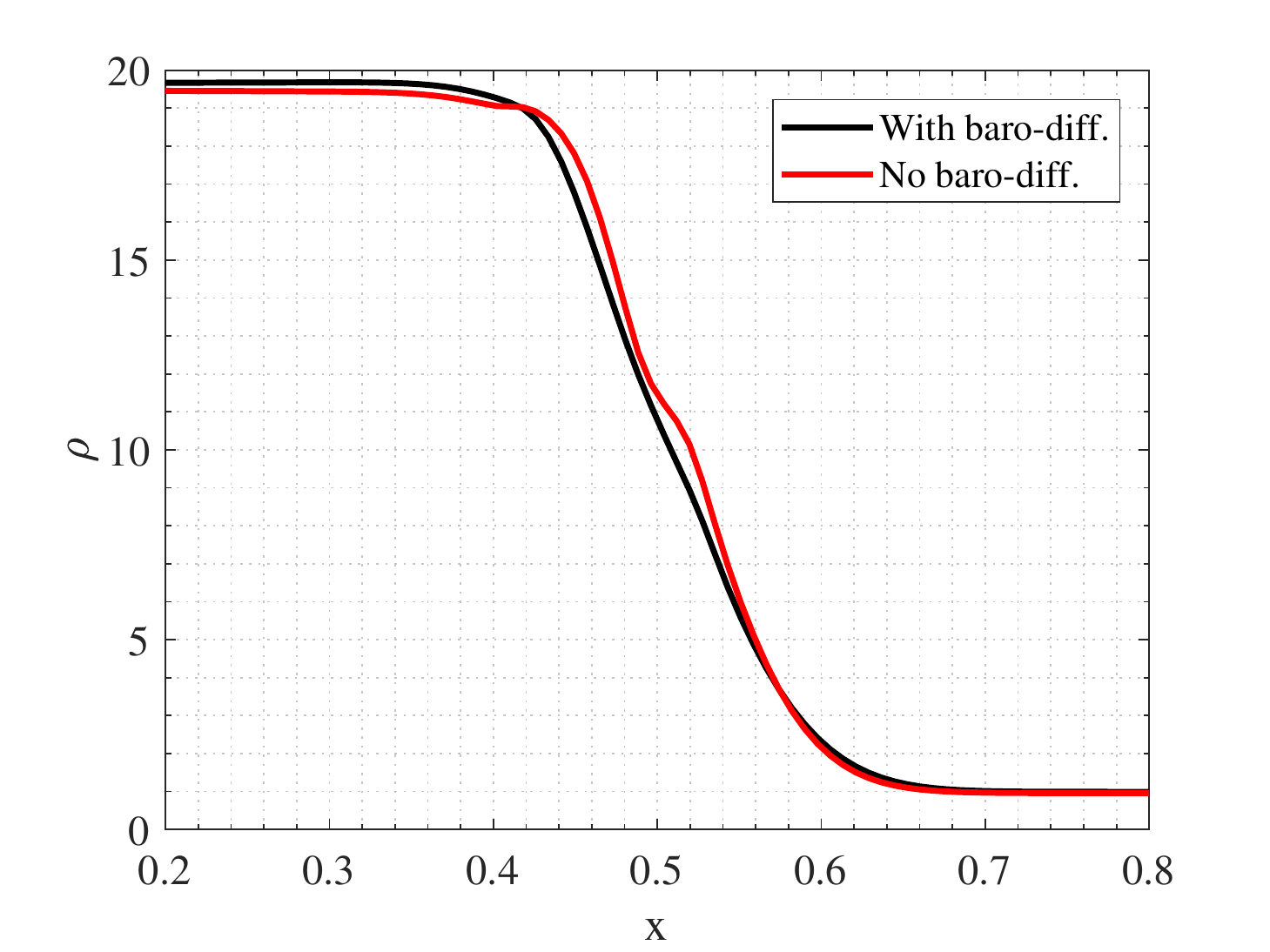}}\\
\subfloat[Mass fraction]{\includegraphics[width=0.38\textwidth]{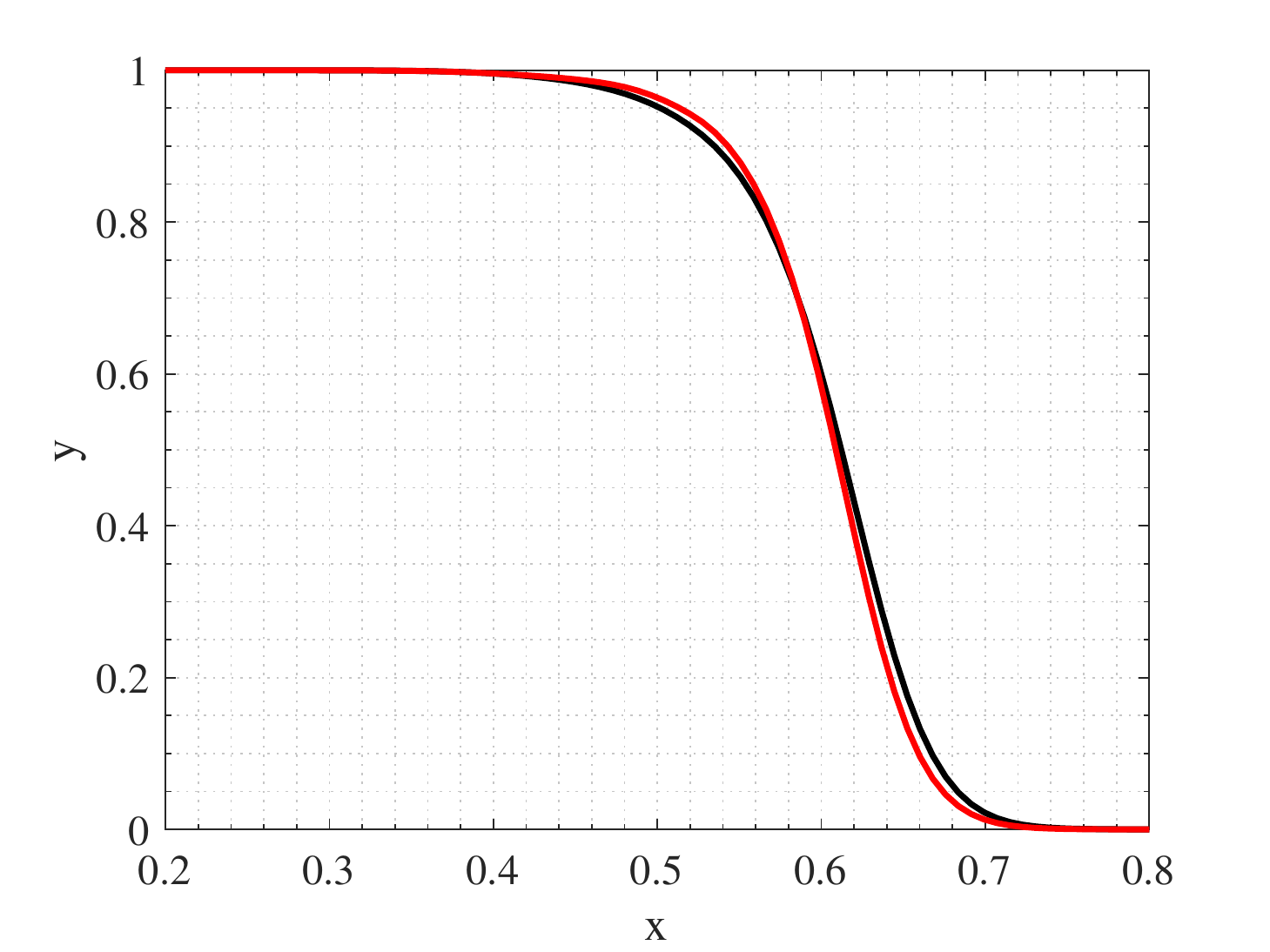}}\\
\subfloat[Pressure]{\includegraphics[width=0.38\textwidth]{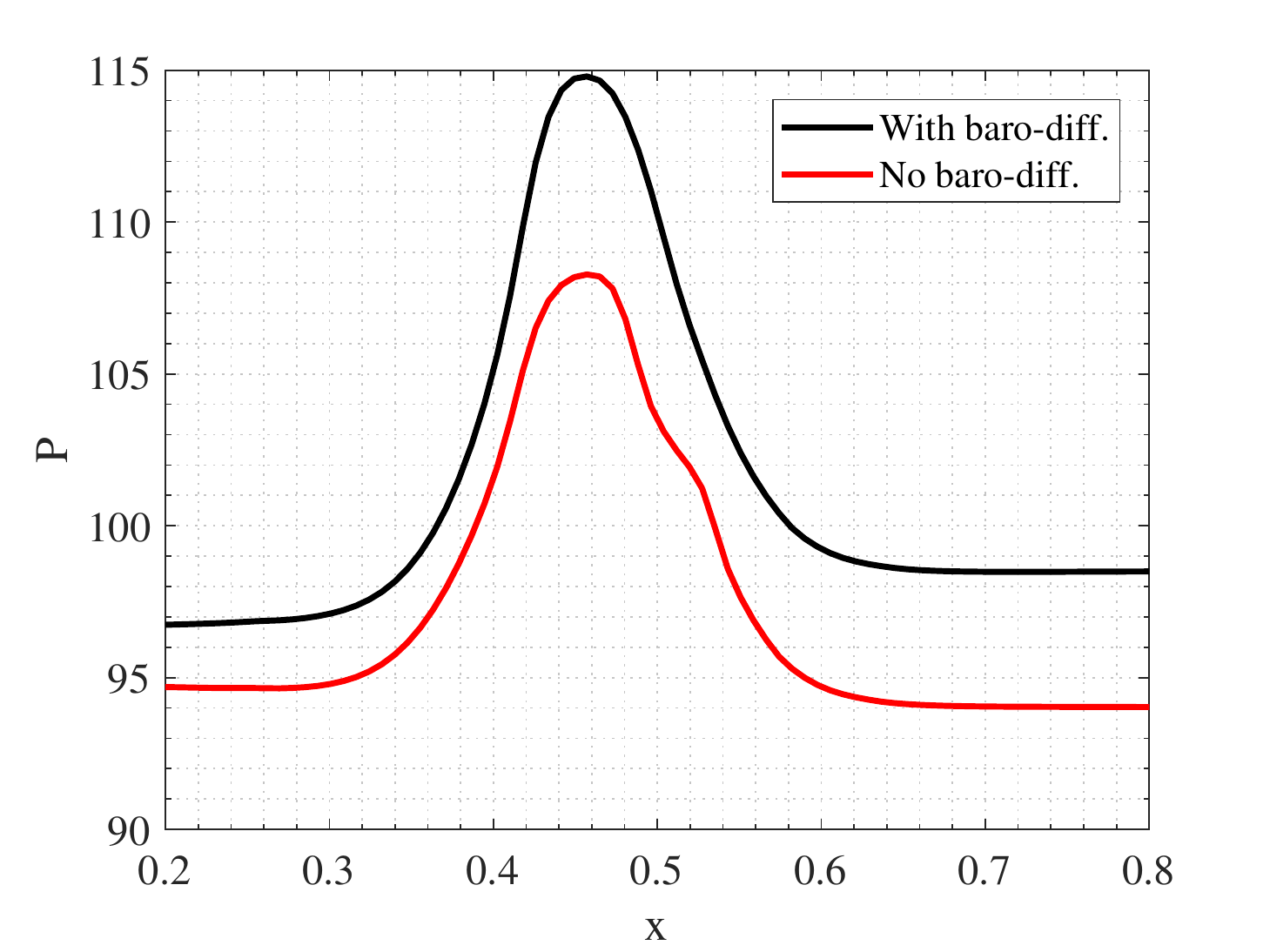}}
\caption{The  numerical results for the mass diffusion problem with/without the baro-diffusion effect.}
\label{fig:pdiff}
\end{figure}

We assume that the two components are Carbon (C) and Deuterium (D), respectively. According to \cref{eq:Dpk}, the baro-diffusion coefficient for Carbon is negative ($D_{p1} < 0$) and that for the Deuterium is positive ($D_{p2} > 0$).
As can be seen from \Cref{fig:pdiff}, in the neighborhood of the diffuse front, the pressure gradient is along $-x$.
Thus, for Carbon the baro-diffusion flux $- \rho D D_{p1} \nabla\text{log}p < 0$, which is opposite to the mass fraction gradient driven flux $- \rho D \nabla y_1 > 0$. Therefore, in \cref{fig:pdiff}, one can observe that the Carbon mass fraction is less diffused with the baro-diffusion effect being included.

%\begin{figure}[htbp]
%\centering
%\subfloat[density]{\includegraphics[width=0.5\textwidth]{./FIGS/dens_pdiffvis.eps}}
%\subfloat[velocity]{\includegraphics[width=0.5\textwidth]{./FIGS/vel_pdiffvis.eps}}\\
%\subfloat[pressure]{\includegraphics[width=0.5\textwidth]{./FIGS/pres_pdiffvis.eps}}
%\subfloat[Mass fraction]{\includegraphics[width=0.5\textwidth]{./FIGS/MF_pdiffvis.eps}}
%\caption{The  numerical results for the mass diffusion problem with viscous and baro-diffusion effect. ``With pdiff., $u_k$'' $-$ numerical results obtained with the pressure diffusion (baro$-$diffusion)
%and the viscous stress being calculated with the component velocity. The other denotations are formulated in a similar way.}
%\label{fig:pdiffvis}
%\end{figure}

{\color{black}
\subsection{The shock passage through a mixing zone}
In this test we demonstrate the capability of the proposed model (\ref{eq:final_model}) to deal with temperature separation phenomenon when the shock travels through the mixing zone. The mixture consists of two polytropic components with adiabatic coefficients $\gamma = 2.0$ and $\gamma = 5/3$. The heat capacity is calculated by $C_v = {N_0 k_b (1 + Z)}/({\gamma A}),$
where $N_0$ is the Avogadro constant, $k_b$ is the Boltzmann constant, $A$ and $Z$ are the atomic weight and number, respectively.  Again, we assume that the two components are Carbon ($A = 12{\text{g/mol}}, \; Z = 6$) and  Deuterium ($A = 2{\text{g/mol}}, \; Z = 1$).

The one-dimensional computational domain is of length $L=$40.96$\mu$m. The initial mixture density is characterized by \cref{eq:pure_diff_rho} with $x_0 = 0.6 L$, $h_0 = L / 10$ and $D = 0.02$cm$^2$/$\mu$s.  
A leftward shock of Mach number 5 hits the mixing interface. The temperature disequilibrium in the initial postshock zone is neglected since the concentration of the second component is negligibly small ($y_2 = 1\times10^{-6}$). 
The pre-shock mixture is in temperature and pressure equilibrium with uniform profile $p = 5000$Mbar and $T = 100$MK.  The component densities can be determined as  $\rho_1 = 2.0617$g/cm$^3$ and $\rho_2 = 1.2027$g/cm$^3$ via the EOSs. The initial profiles for the component temperatures, volume fraction and mass fraction are demonstrated in \Cref{fig:shock_through_mixing}(a). We first perform computation with the physical temperature relaxation rate determined with the Coloumb collosion frequency \cite{Richardson2019}. At the time moment $t = 6\times10^{-3}$ns the shock travels through the mixing interface, temperature disequilibrium arises in postshock zone (\Cref{fig:shock_through_mixing}(b)). The temperature difference can be as large as  525MK. We then increase the physical temperature relaxation rate by 100 times and perform the same computation. The corresponding results are displayed in \Cref{fig:shock_through_mixing}(c). It can be observed that the temperature disequilibrium is reduced. For comparison purpose we also present the numerical results with the temperature equilibrium model (or $\eta \to \infty$) in \Cref{fig:shock_through_mixing}(d). The equilibrium temperature lies between the component temperatures in \Cref{fig:shock_through_mixing}(b-c). The deviation between the mass fraction and the volume fraction is more obvious in \Cref{fig:shock_through_mixing}(d). This is because the temperature relaxation leads to the variation of the volume fraction while it has no impact on the mass fraction, as analysed in \Cref{subsec:thermal_relax}. Comparison between the temperature-equilibrium and the temperature-disequilibrium model indicates that the commonly used temperature-equilibrium model maybe inadequate for evaluating the temperature relaxation effect in mixing topology evolution. 

\begin{figure*}
\centering
\subfloat[Initial data]{\includegraphics[width=0.38\textwidth]{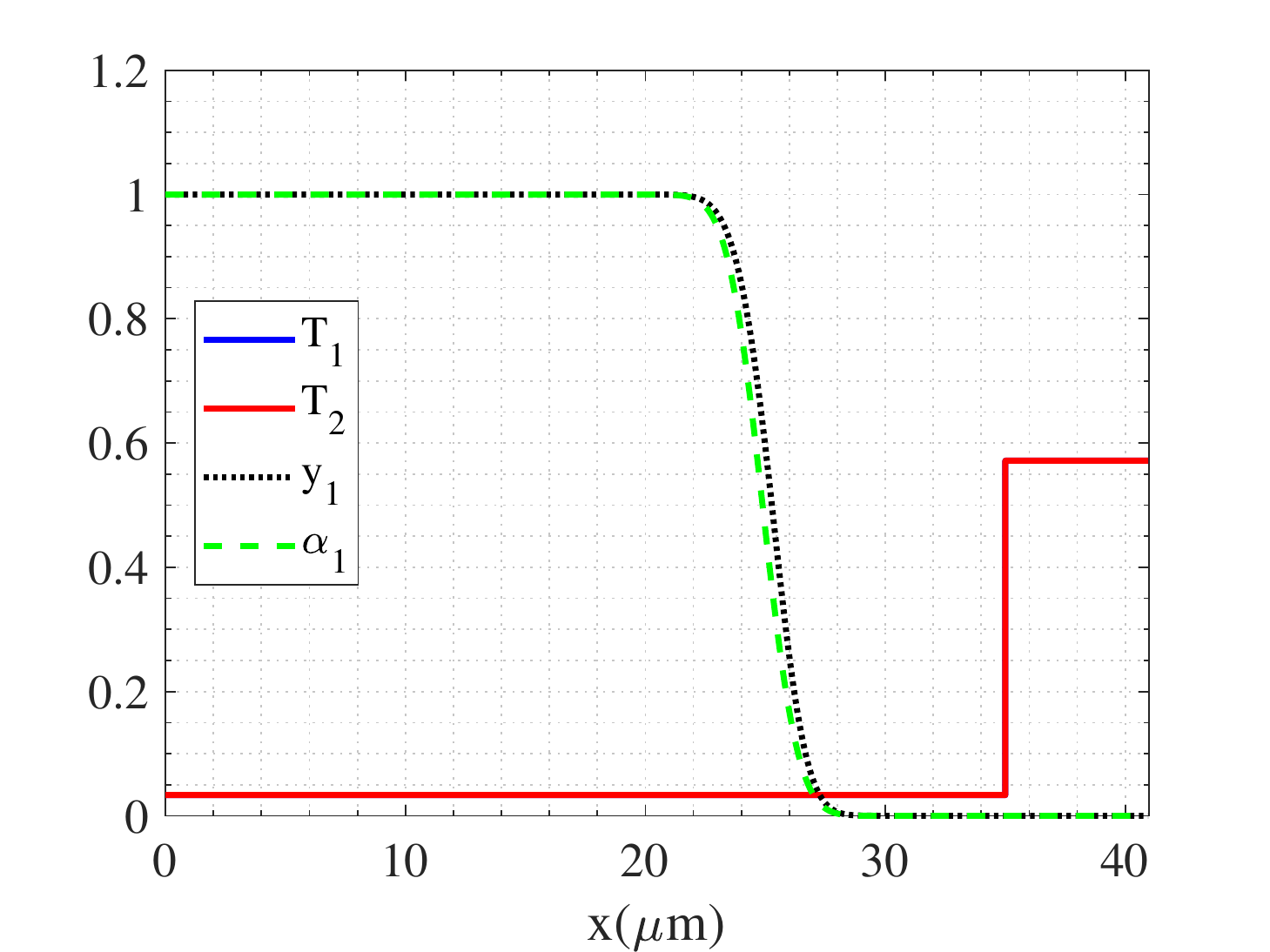}} 
\subfloat[Physical $\eta$]{\includegraphics[width=0.38\textwidth]{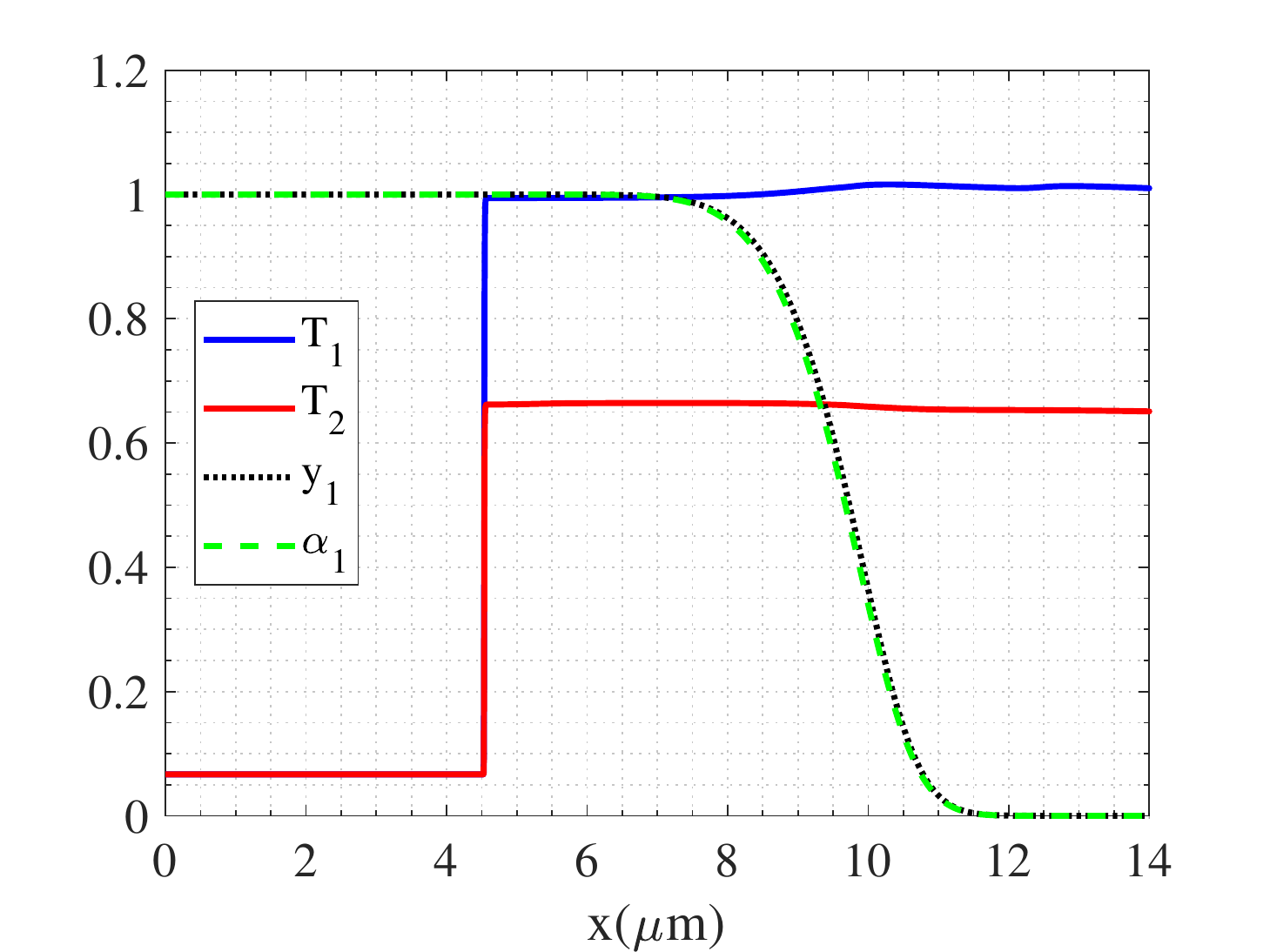}}\\
\subfloat[Physical $\eta$ enlarged by 100 times]{\includegraphics[width=0.38\textwidth]{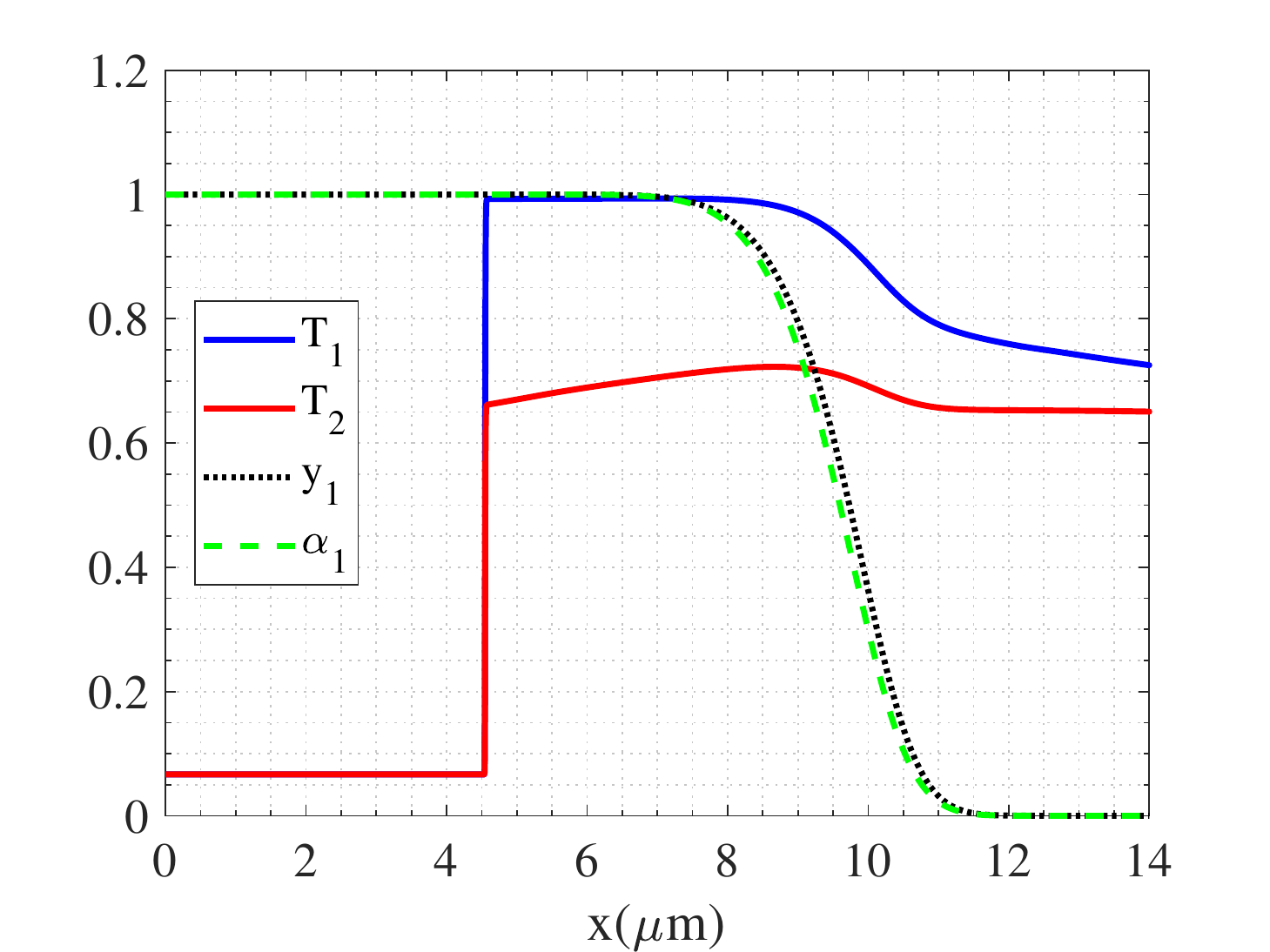}}
\subfloat[Infinite $\eta$]{\includegraphics[width=0.38\textwidth]{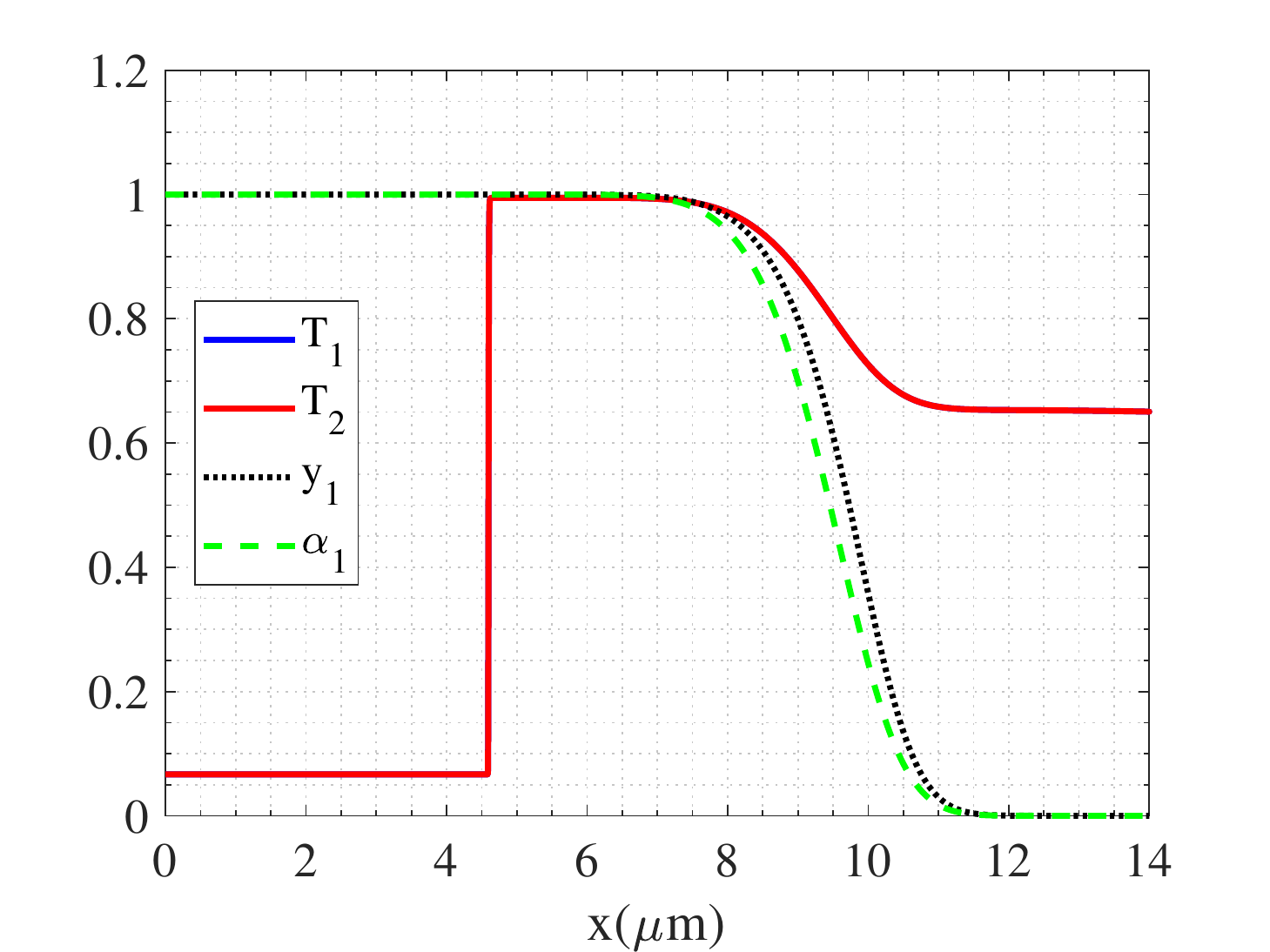}}
\caption{{\color{black}The variable (component temperatures, the volume fraction and the mass fraction) distributions after the shock travels through the mixing zone. Non-dimensional temperatures displayed are  $T_k/(1500{\text{MK}})$. (a) The initial profile, (b) the numerical results at $t = 6\times10^{-3}$ns with physical relaxation rate $\eta_{Phys}$, (c) the numerical results with $100\eta_{Phys}$, (d) the numerical results with $\eta \to \infty$}.}
\label{fig:shock_through_mixing}
\end{figure*}
}

\subsection{The RT instability problem under thermal relaxation}
\begin{figure}[htbp]
\centering
\subfloat{\includegraphics[width=0.4\textwidth]{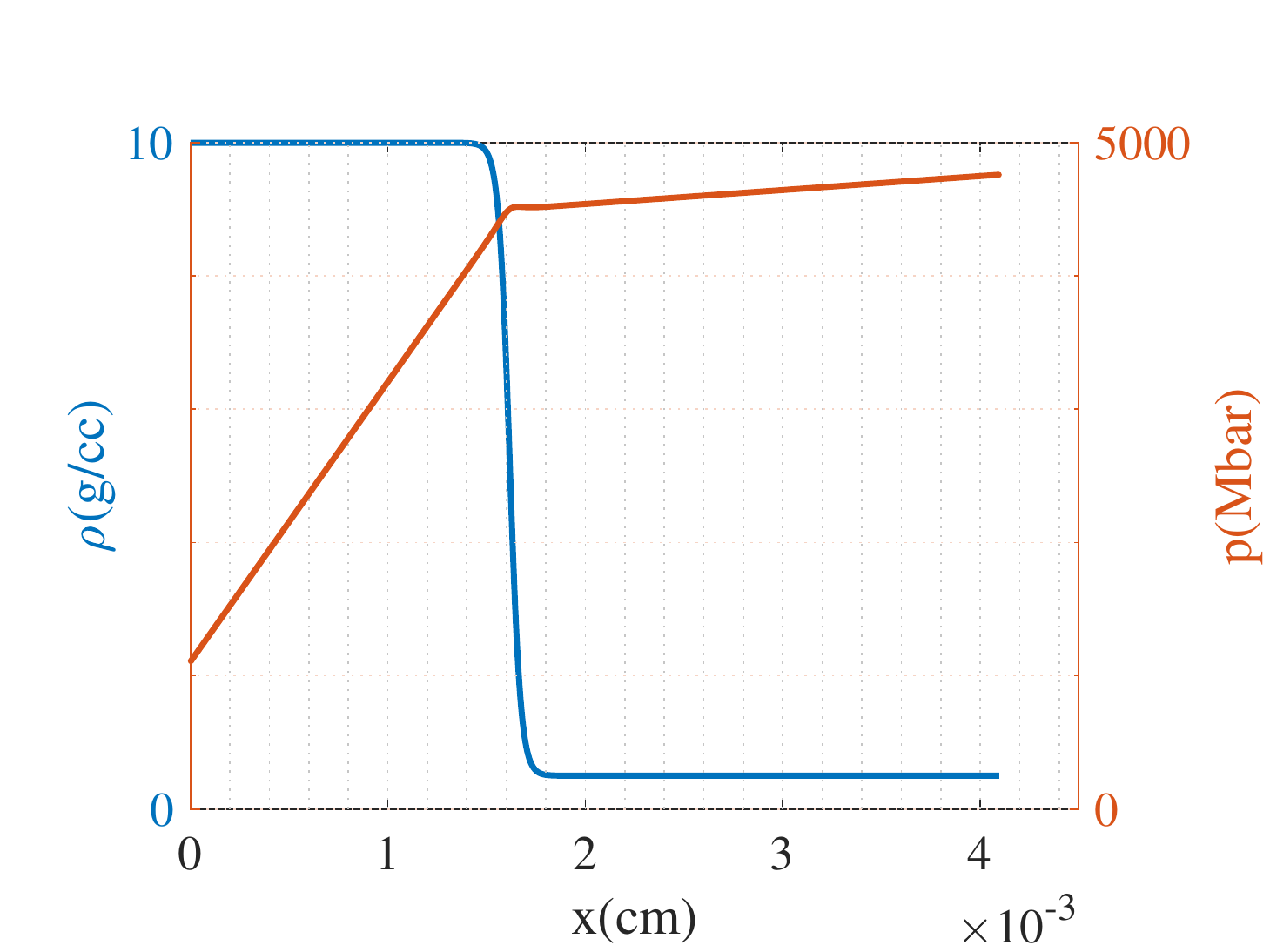}}\\
\subfloat{\includegraphics[width=0.368\textwidth]{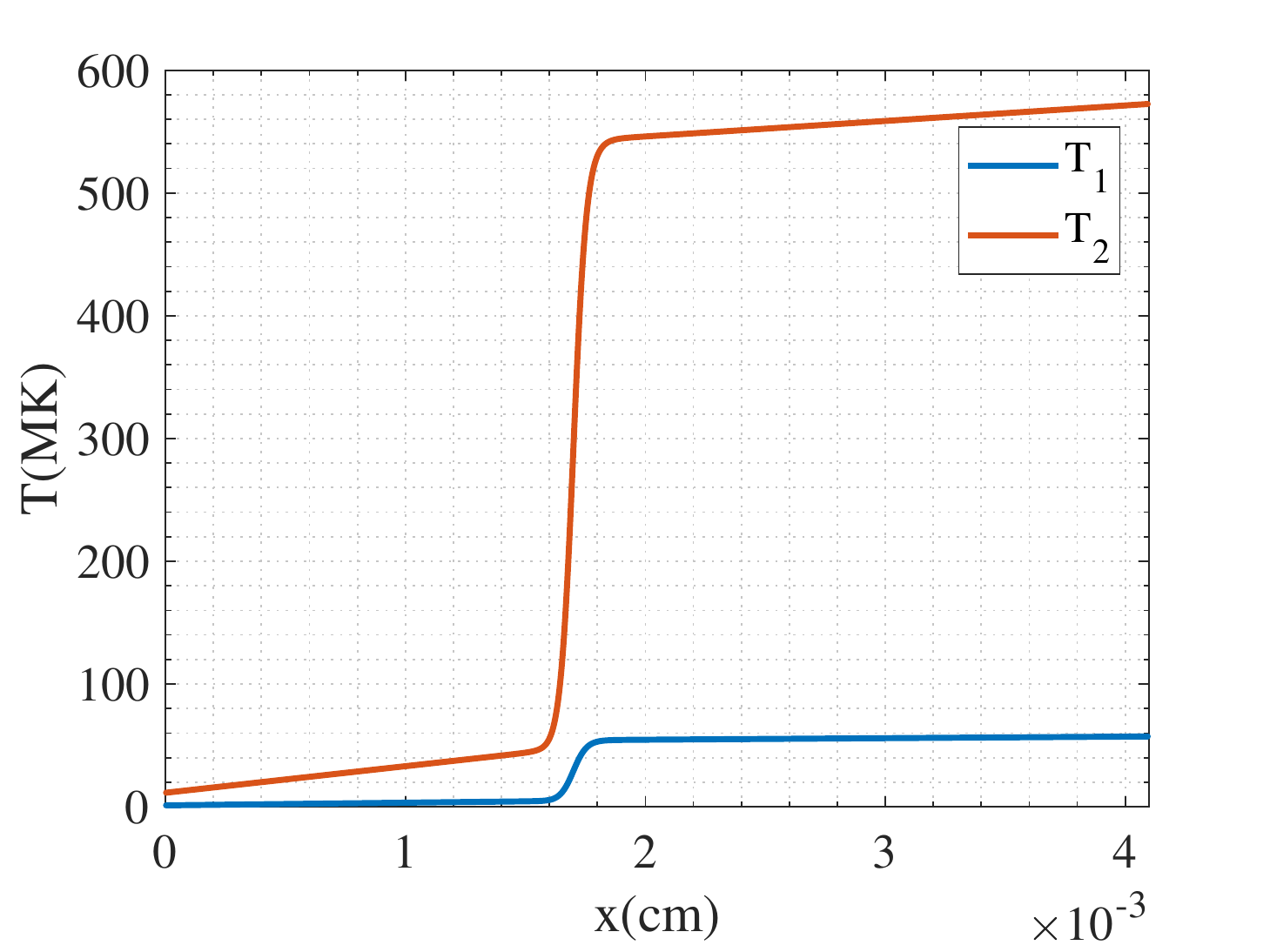}}
\caption{The initial condition for the RT instability problem. {\color{black}Top}: mixture density and pressure, {\color{black}bottom}: component temperature.}
\label{fig:initialRT}
\end{figure}

In this section we consider a planar RT instability problem in a computational domain $\left(x,y\right)\in [0\mu\text{m}, 10.24\mu\text{m}] \times [0\mu\text{m}, 40.96\mu\text{m}]$. The initial condition along $y=L_y/2 = 5.12$cm is demonstrated in  \Cref{fig:initialRT}. The acceleration is set to be $2.1\times10^5$cm$/\mu$s$^2$, which is equal to that of a realistic implosion during the deceleration stage on the Omega facility \cite{vold2021plasma}. The temperature at the interface zone is also within the temperature range in ICF implosion. 

Note that we assume a smeared interface, which maybe a result of various mixing mechanisms such as molecular diffusion or turbulence. {\color{black}The mixing interface center is perturbed with a cosine profile $x_0 = 0.4L_x - 0.03L_y \text{cos}(2\pi y / L_y)$.} Inside the smeared interface the component temperatures relax towards a equilibrium one.
Such thermal relaxation mechanism has a significant impact on the RT instability development, as to be shown below by the following direct simulations.
The relaxation rate is determined with the formula in the NRL plasma formulary\cite{Richardson2019}.

\begin{figure*}[htbp]
\centering
%\subfloat{\includegraphics[width=0.5\textwidth]{./FIGS/ConvgDens1.png}}
%\subfloat{\includegraphics[width=0.4\textwidth]{./FIGS/YYConvg1.png}}
{\includegraphics[width=0.9\textwidth]{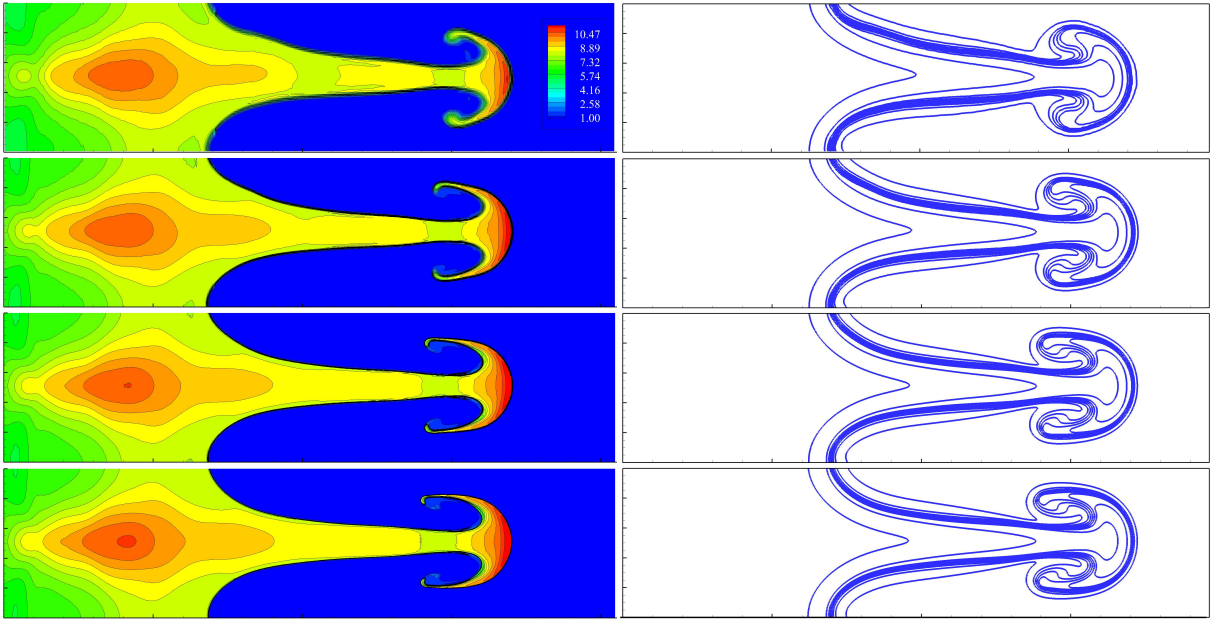}}
\caption{The distribution of the mixture density $\rho$ (Left) and mass fraction $y_1$ (right) on series of refining grids (From top to bottom : 320$\times$80, \; 640$\times$160, \; 1280$\times$320,\; 2560$\times$640 ). 10 uniform contours from 0.01 to 0.99 for $y_1$ is display on the right.}
\label{fig:Convg2D}
\end{figure*}

\begin{figure}[htbp]
\centering
\subfloat{\includegraphics[width=0.4\textwidth]{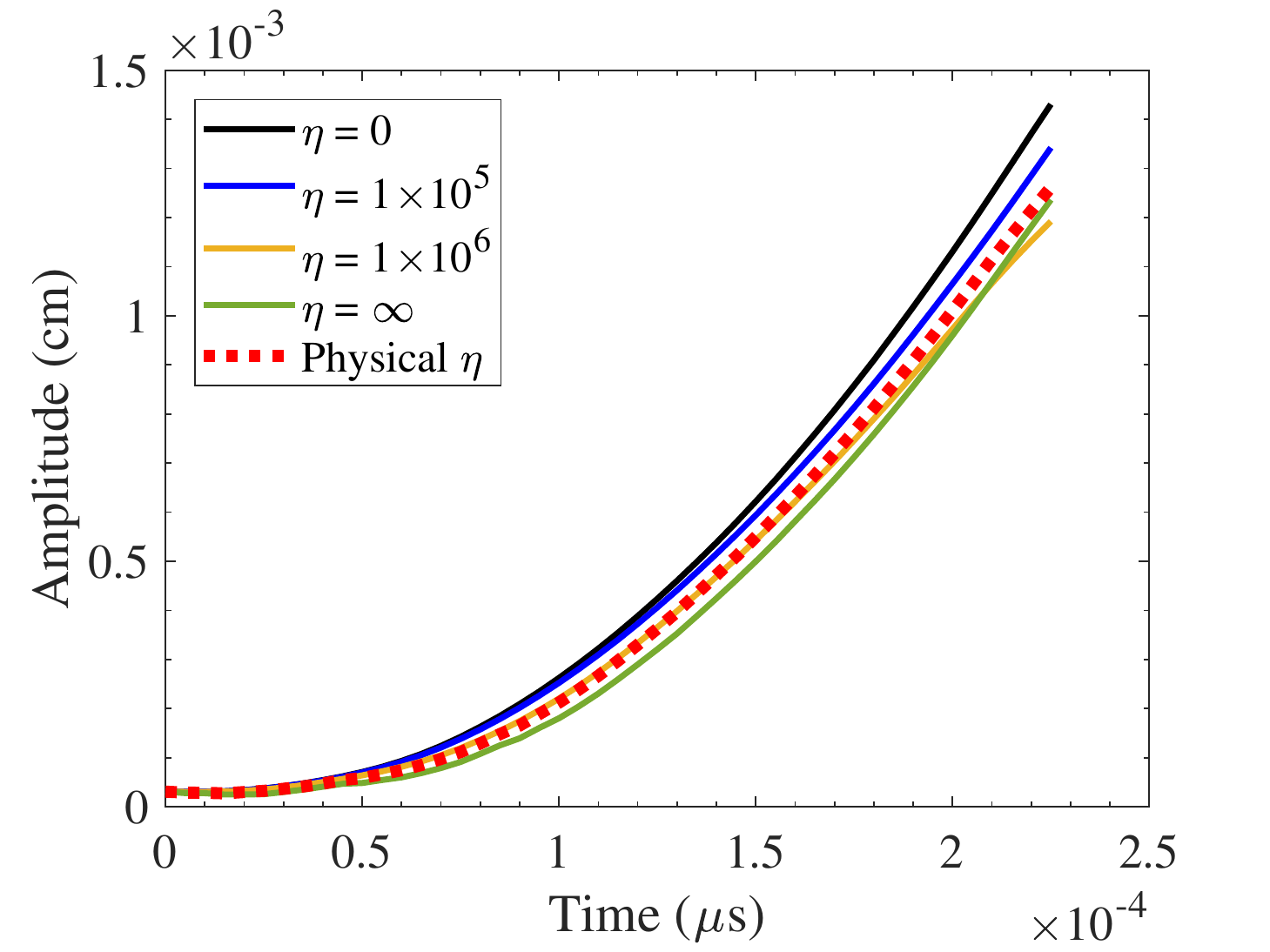}}\\
\subfloat{\includegraphics[width=0.4\textwidth]{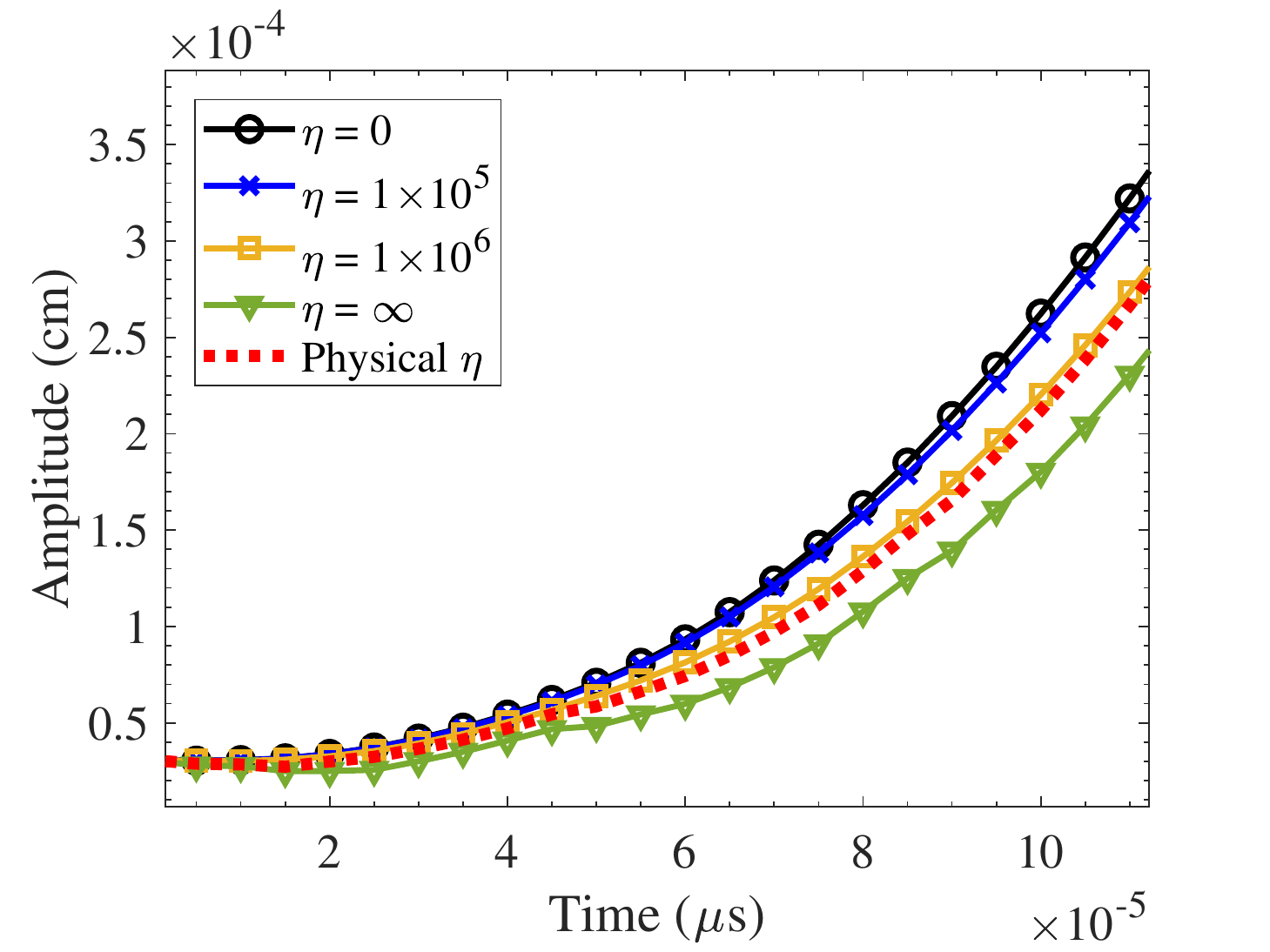}}\\
\subfloat{\includegraphics[width=0.4\textwidth]{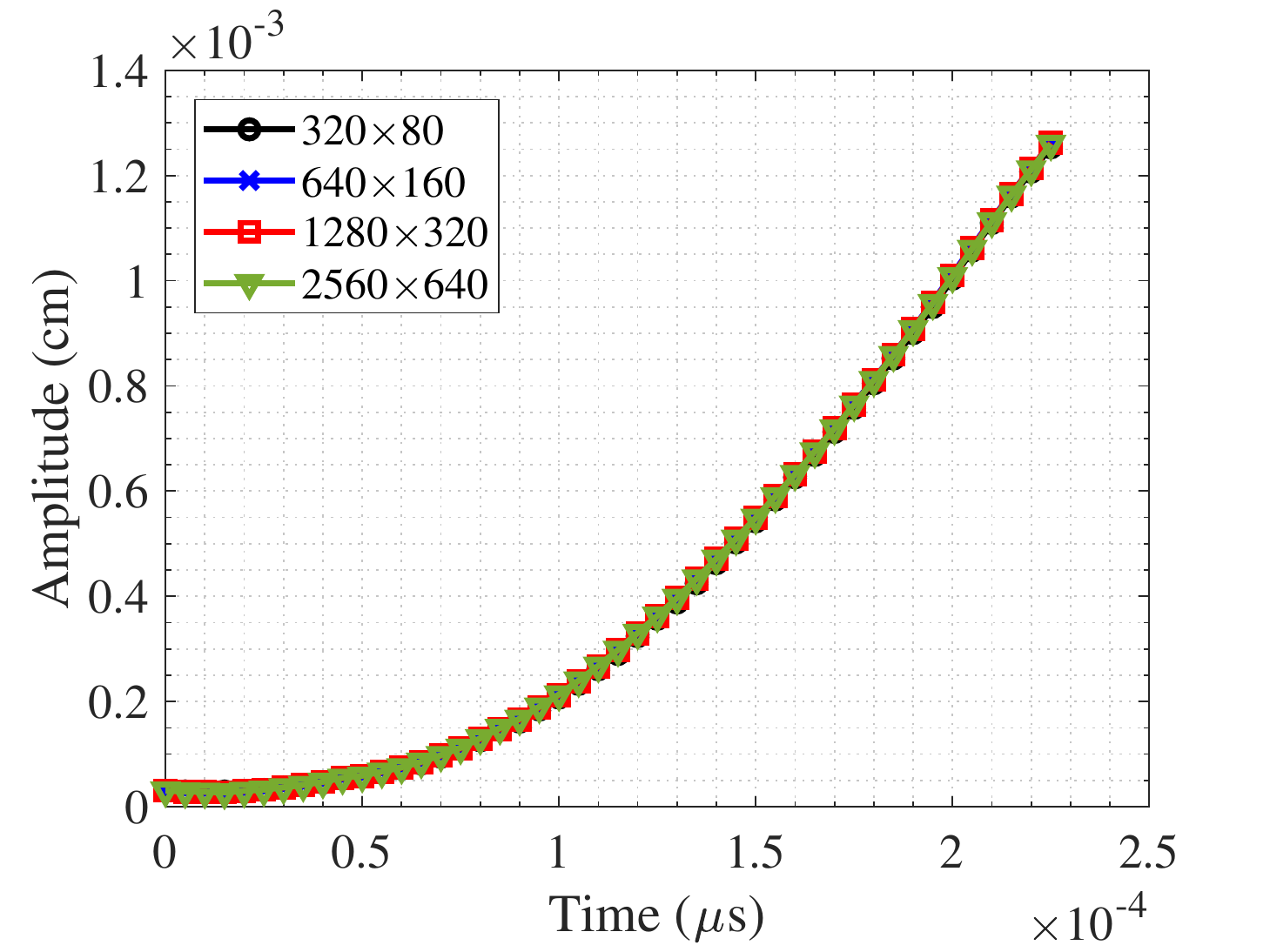}}
\caption{The evolution of the mixing length with time. Top and middle sub-figures are the results corresponding to different relaxation rates $\eta$. The bottom sub-figure shows the grid independence of the numerical results.}
\label{fig:LmixConvg}
\end{figure}

To ensure grid-independence, we first compare the numerical results obtained on a series of refining grids from 320$\times$80 cells to 2560$\times$640 cells. The corresponding results are displayed in  \Cref{fig:Convg2D}.
One can see that the numerical results tend to converge with physical diffusions being included. The mixing length evolution with time displayed in \Cref{fig:LmixConvg} also confirms the convergence.

Then we investigate the sensitivity of the mixing length evolution to the relaxation rate $\eta$ on the $1280\times320$ grid.
The relaxation rate $\eta$ is taken to be $1\times10^5$, $1\times10^6$ and $\infty$. The case $\eta = \infty$ correspond to the case where component temperatures relax instantaneously, i.e., temperature equilibrium.

\begin{figure*}[htbp]
\centering
%\subfloat{\includegraphics[width=0.5\textwidth]{./FIGS/DENS_ETA.png}}
%\subfloat{\includegraphics[width=0.5\textwidth]{./FIGS/YY_ETA.png}}
{\includegraphics[width=0.9\textwidth]{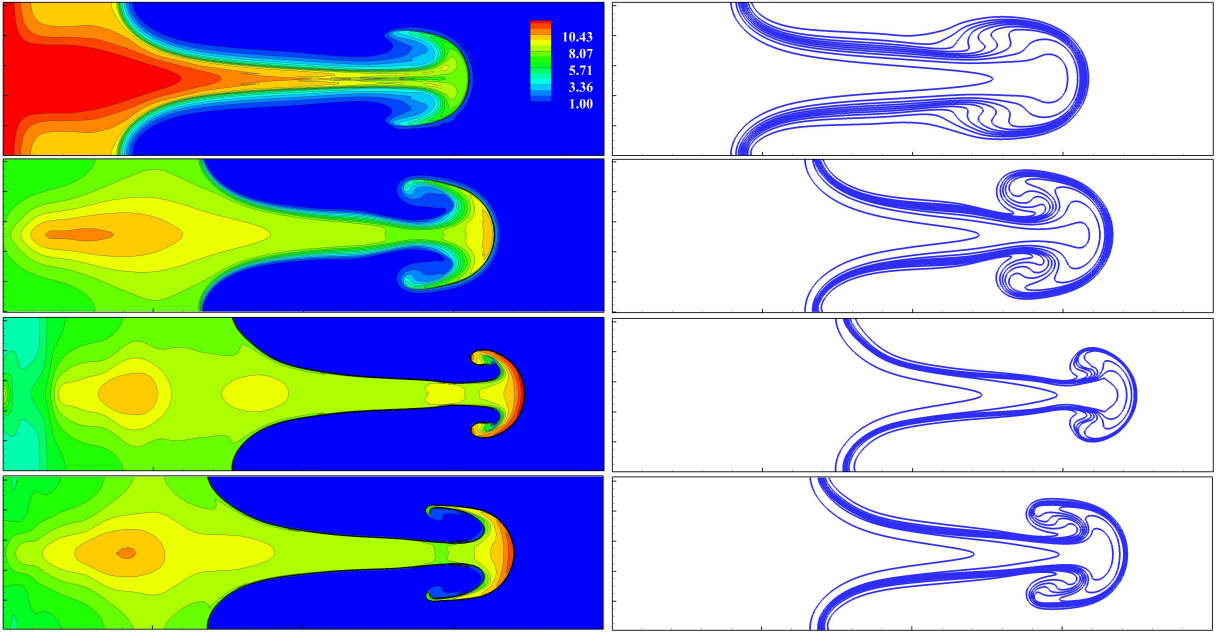}}
\caption{The distribution of the mixture density $\rho$ (Left) and mass fraction $y_1$ (right) with different relaxation rates $\eta$ at the time moment $t = 0.2$ns. From top to bottom : $\eta = 0, \; 1\times10^6, \; \infty,$ and $\eta$ determined by physical model (last row). 10 uniform contours from 0.01 to 0.99 for $y_1$ is display on the right.}
\label{fig:dens_y_eta}
\end{figure*}

The density and mass fraction distributions at 0.2ns with different relaxation rates are shown in \Cref{fig:dens_y_eta}. The evolution of the mixing length (bubble-to-spike distance) with time is demonstrated in \Cref{fig:LmixConvg}.
It can be seen that the thermal relaxation tend to suppress the growth of the mixing length. The main mechanism here lies in the unsteady acceleration of the interface caused by the thermal relaxation.
The details of the physical mechanism will be dealt with in a separate paper and here we focus on the model itself.

Based on the simulation results, we can see that under the ICF deceleration condition the temperature equilibrium or disequilibrium assumptions lead to underestimation or overestimation of the mixing length evolution.
The temperature relaxation timescale is comparable to the ICF-concerned timescale, and the temperature separation is diminishing at a finite rate.
This effect should be considered for the accurate evaluation of the mixing in ICF.

%\begin{figure}[htbp]
%\centering
%{\includegraphics[width=0.75\textwidth]{./FIGS/rhoyConvg.png}}
%\caption{The distribution of the partial density $\rho y_1$ on different grids consisting of 640$\times$100 cells (top) and 1280$\times$200 cells (bottom).}
%\label{fig:rhoyConvg}
%\end{figure}

%\begin{figure}[htbp]
%\centering
%{\includegraphics[width=0.5\textwidth]{./FIGS/yyy.png}}
%\caption{The distribution of the mass fraction $y_1$ with different relaxation rates $\eta$ at the time moment $t = 0.2$ns. From top to bottom : $\eta = 0, \; 1\times10^5, \; 1\times10^6, \; 1\times10^7$. 10 uniform contours from 0.01 to 0.99 for $y_1$ is display on the right.}
%\label{fig:yyy}
%\end{figure}

%
%
%\begin{figure}[htbp]
%\centering
%\subfloat[]{\includegraphics[width=0.5\textwidth]{./FIGS/Lmixetas_VS_time.eps}}
%\subfloat[]{\includegraphics[width=0.5\textwidth]{./FIGS/dLetas_VS_time.eps}}
%\caption{The evolution of mixing length with time under different relaxation rates.}
%\label{fig:Lmixetas_VS_time}
%\end{figure}

%\begin{figure}[htbp]
%\centering
%{\includegraphics[width=0.5\textwidth]{./FIGS/dLetas_VS_time.eps}}
%\caption{The evolution of relative mixing length with time under different relaxation rates.}
%\label{fig:dLetas_VS_time}
%\end{figure}

%\begin{figure}[htbp]
%\centering
%{\includegraphics[width=0.5\textwidth]{./FIGS/dLMax_VS_etas.eps}}
%\caption{The dependence of maximum relative mixing length on the relaxation rates.}
%\label{fig:dLMax_VS_etas}
%\end{figure}
%
%
%
%\begin{figure}[htbp]
%\centering
%{\includegraphics[width=0.5\textwidth]{./FIGS/dLetas_D_ND_VS_time.eps}}
%\caption{The evolution of relative mixing length with time with/without mass diffusivity.}
%\label{fig:dLetas_D_ND_VS_time}
%\end{figure}
%
%
%\begin{figure}[htbp]
%\centering
%{\includegraphics[width=0.5\textwidth]{./FIGS/LmixCvs_VS_time.eps}}
%\caption{The evolution of relative mixing length with time under different initial temperature disequilibria.}
%\label{fig:LmixCvs_VS_time}
%\end{figure}
%
%
%\begin{figure}[htbp]
%\centering
%{\includegraphics[width=0.5\textwidth]{./FIGS/disL3MORM_VS_dT1dT2.eps}}
%\caption{The dependence on initial thermal disequilibrium coefficient of the $L^{\infty}$ distance between the equilibrium and disequilibrium mixing length.}
%\label{fig:disL3MORM_VS_dT1dT2}
%\end{figure}

\section*{Conclusion}
\label{sec:conclusion}
In the present paper we have presented a temperature disequilibrium diffuse-interface model for compressible multicomponent flows with interphase heat transfer and diffusions (including viscous, heat conduction and mass diffusion).
The model is reduced from the BN model in the limit of small Knudsen number ($Kn <<1$) and consists of six equations including phase density equations, mixture momentum equation, phase internal energy equations and the volume fraction equation.
Velocity difference is closed by the mass diffusion laws, thus the velocity of each component is available. The viscous stress is determined by component velocities rather than mixture velocity in literature.
Moreover, the model has included the effect of finite thermal relaxation. We have described second-order numerical methods for solving the advection-diffusion part of the proposed model.
As for the thermal relaxation, we propose method to keep the pressure equilibrium.
Being equipped with this model and its solution methods, we have considered a RT instability problem under ICF deceleration condition with finite thermal relaxation rate.
We have performed a parametric study on the dependence of the mixing length development on the relaxation rate.
Direct numerical simulations demonstrate that for the RT instability at an interface between the high-density low-temperature component and the low-density high-temperature component, the thermal relaxation tend to suppress the development of the instability.
Further details of this mechanism will appear in our future publications.

\section*{Acknowledgement}
The present study is supported by National Natural Science Foundation of China (grant numbers 12205022,11975053). 

\section*{Appendix}
{\color{black}Here we provide the proof of the entropy inequality \cref{eq:entropy_ineq}. Inserting the pressure relaxation  \cref{eq:relaxations} into \cref{eq:entropy_ineq} and performing some reformulations, one obtains
\begin{align}
    &\alpha_1\rho_1 \frac{\text{D}_1 s_1}{\text{D} t} + \alpha_2\rho_2 \frac{\text{D}_2 s_2}{\text{D} t}  + \nabla\dpr\left( \frac{\vc{q}_1}{T_1} \right) + \nabla\dpr\left( \frac{\vc{q}_2}{T_2} \right) - \frac{\mathcal{I}_1}{T_1} - \frac{\mathcal{I}_2}{T_2}\nonumber
    \\&  =\frac{\alpha_2 \varsigma (p_1 - p_2)^2 }{T_1} + \frac{\alpha_1 \varsigma (p_2 - p_1)^2 }{T_2}+ \frac{\eta(T_2-T_1)^2}{T_1 T_2} \nonumber\\
    &  + \frac{\mathcal{S}_1}{T_1} + \frac{\mathcal{S}_2}{T_2} + \vc{q}_1\dpr\nabla\left(\frac{1}{T_1}\right) + \vc{q}_2\dpr\nabla\left(\frac{1}{T_2}\right).
\end{align}
The first three terms on the right hand side are obviously non-negative. The term $\mathcal{S}_k = \alpha_k \overline{\overline{\tau}}_k:\overline{\overline{D}}_k$ represents the kinetic energy dissipation due to viscous friction and is non-negative as long as the viscosity coefficient $\mu_k \geq 0$.
The non-negativity of the terms $\vc{q}_k\dpr\nabla\left(\frac{1}{T_k}\right)$ are ensured by the 
Fourier's law $\vc{q}_k = - \alpha_k \lambda_k \nabla T_k$.
Thus entropy does not decrease in the absence of external heat flux and energy source, which does not contradicts the second law of thermodynamics.
}

\bibliography{apssamp}% Produces the bibliography via BibTeX.

\begin{thebibliography}{46}%
\makeatletter
\providecommand \@ifxundefined [1]{%
 \@ifx{#1\undefined}
}%
\providecommand \@ifnum [1]{%
 \ifnum #1\expandafter \@firstoftwo
 \else \expandafter \@secondoftwo
 \fi
}%
\providecommand \@ifx [1]{%
 \ifx #1\expandafter \@firstoftwo
 \else \expandafter \@secondoftwo
 \fi
}%
\providecommand \natexlab [1]{#1}%
\providecommand \enquote  [1]{``#1''}%
\providecommand \bibnamefont  [1]{#1}%
\providecommand \bibfnamefont [1]{#1}%
\providecommand \citenamefont [1]{#1}%
\providecommand \href@noop [0]{\@secondoftwo}%
\providecommand \href [0]{\begingroup \@sanitize@url \@href}%
\providecommand \@href[1]{\@@startlink{#1}\@@href}%
\providecommand \@@href[1]{\endgroup#1\@@endlink}%
\providecommand \@sanitize@url [0]{\catcode `\\12\catcode `\$12\catcode
  `\&12\catcode `\#12\catcode `\^12\catcode `\_12\catcode `\%12\relax}%
\providecommand \@@startlink[1]{}%
\providecommand \@@endlink[0]{}%
\providecommand \url  [0]{\begingroup\@sanitize@url \@url }%
\providecommand \@url [1]{\endgroup\@href {#1}{\urlprefix }}%
\providecommand \urlprefix  [0]{URL }%
\providecommand \Eprint [0]{\href }%
\providecommand \doibase [0]{https://doi.org/}%
\providecommand \selectlanguage [0]{\@gobble}%
\providecommand \bibinfo  [0]{\@secondoftwo}%
\providecommand \bibfield  [0]{\@secondoftwo}%
\providecommand \translation [1]{[#1]}%
\providecommand \BibitemOpen [0]{}%
\providecommand \bibitemStop [0]{}%
\providecommand \bibitemNoStop [0]{.\EOS\space}%
\providecommand \EOS [0]{\spacefactor3000\relax}%
\providecommand \BibitemShut  [1]{\csname bibitem#1\endcsname}%
\let\auto@bib@innerbib\@empty
%</preamble>
\bibitem [{\citenamefont {Zylstra}\ \emph {et~al.}(2018)\citenamefont
  {Zylstra}, \citenamefont {Hoffman}, \citenamefont {Herrmann}, \citenamefont
  {Schmitt}, \citenamefont {Kim}, \citenamefont {Meaney}, \citenamefont
  {Leatherland}, \citenamefont {Gales}, \citenamefont {Forrest}, \citenamefont
  {Glebov} \emph {et~al.}}]{zylstra2018diffusion}%
  \BibitemOpen
  \bibfield  {author} {\bibinfo {author} {\bibfnamefont {A.}~\bibnamefont
  {Zylstra}}, \bibinfo {author} {\bibfnamefont {N.~M.}\ \bibnamefont
  {Hoffman}}, \bibinfo {author} {\bibfnamefont {H.~W.}\ \bibnamefont
  {Herrmann}}, \bibinfo {author} {\bibfnamefont {M.}~\bibnamefont {Schmitt}},
  \bibinfo {author} {\bibfnamefont {Y.}~\bibnamefont {Kim}}, \bibinfo {author}
  {\bibfnamefont {K.}~\bibnamefont {Meaney}}, \bibinfo {author} {\bibfnamefont
  {A.}~\bibnamefont {Leatherland}}, \bibinfo {author} {\bibfnamefont
  {S.}~\bibnamefont {Gales}}, \bibinfo {author} {\bibfnamefont
  {C.}~\bibnamefont {Forrest}}, \bibinfo {author} {\bibfnamefont {V.~Y.}\
  \bibnamefont {Glebov}}, \emph {et~al.},\ }\bibfield  {title} {\bibinfo
  {title} {Diffusion-dominated mixing in moderate convergence implosions},\
  }\href@noop {} {\bibfield  {journal} {\bibinfo  {journal} {Physical Review
  E}\ }\textbf {\bibinfo {volume} {97}},\ \bibinfo {pages} {061201} (\bibinfo
  {year} {2018})}\BibitemShut {NoStop}%
\bibitem [{\citenamefont {Wilson}\ \emph {et~al.}(1011)\citenamefont {Wilson},
  \citenamefont {Ebey}, \citenamefont {Sangster}, \citenamefont {Shmayda},
  \citenamefont {Glebov},\ and\ \citenamefont {Lerche}}]{Wilson2011}%
  \BibitemOpen
  \bibfield  {author} {\bibinfo {author} {\bibfnamefont {D.~C.}\ \bibnamefont
  {Wilson}}, \bibinfo {author} {\bibfnamefont {P.~S.}\ \bibnamefont {Ebey}},
  \bibinfo {author} {\bibfnamefont {T.~C.}\ \bibnamefont {Sangster}}, \bibinfo
  {author} {\bibfnamefont {W.~T.}\ \bibnamefont {Shmayda}}, \bibinfo {author}
  {\bibfnamefont {V.~Y.}\ \bibnamefont {Glebov}},\ and\ \bibinfo {author}
  {\bibfnamefont {R.~A.}\ \bibnamefont {Lerche}},\ }\bibfield  {title}
  {\bibinfo {title} {Atomic mix in directly driven inertial confinement
  implosions},\ }\href@noop {} {\bibfield  {journal} {\bibinfo  {journal}
  {Physics of Plasmas}\ }\textbf {\bibinfo {volume} {18}},\ \bibinfo {pages}
  {5} (\bibinfo {year} {1011})}\BibitemShut {NoStop}%
\bibitem [{\citenamefont {Dimonte}\ and\ \citenamefont
  {Tipton}(2006)}]{Dimonte2006}%
  \BibitemOpen
  \bibfield  {author} {\bibinfo {author} {\bibfnamefont {G.}~\bibnamefont
  {Dimonte}}\ and\ \bibinfo {author} {\bibfnamefont {R.}~\bibnamefont
  {Tipton}},\ }\bibfield  {title} {\bibinfo {title} {K-l turbulence model for
  the self-similar growth of the {Rayleigh}-{Taylor} and {Richtmyer}-{Meshkov}
  instabilities},\ }\href@noop {} {\bibfield  {journal} {\bibinfo  {journal}
  {Physics of Fluids}\ }\textbf {\bibinfo {volume} {18}},\ \bibinfo {pages}
  {085101} (\bibinfo {year} {2006})}\BibitemShut {NoStop}%
\bibitem [{\citenamefont {Schilling}(2021)}]{schilling2021self}%
  \BibitemOpen
  \bibfield  {author} {\bibinfo {author} {\bibfnamefont {O.}~\bibnamefont
  {Schilling}},\ }\bibfield  {title} {\bibinfo {title} {Self-similar
  reynolds-averaged mechanical--scalar turbulence models for
  {Rayleigh}-{Taylor}, {Richtmyer}-{Meshkov}, and {Kelvin}-{Helmholtz}
  instability-induced mixing in the small atwood number limit},\ }\href@noop {}
  {\bibfield  {journal} {\bibinfo  {journal} {Physics of Fluids}\ }\textbf
  {\bibinfo {volume} {33}},\ \bibinfo {pages} {085129} (\bibinfo {year}
  {2021})}\BibitemShut {NoStop}%
\bibitem [{\citenamefont {Besnard}\ \emph {et~al.}(1992)\citenamefont
  {Besnard}, \citenamefont {Harlow}, \citenamefont {Rauenzahn},\ and\
  \citenamefont {Zemach}}]{Besnard1992}%
  \BibitemOpen
  \bibfield  {author} {\bibinfo {author} {\bibfnamefont {D.}~\bibnamefont
  {Besnard}}, \bibinfo {author} {\bibfnamefont {F.}~\bibnamefont {Harlow}},
  \bibinfo {author} {\bibfnamefont {R.~M.}\ \bibnamefont {Rauenzahn}},\ and\
  \bibinfo {author} {\bibfnamefont {C.}~\bibnamefont {Zemach}},\ }\bibfield
  {title} {\bibinfo {title} {Turbulence transport equations for
  variable-density turbulence and their relationship to two-field models},\
  }\href@noop {} {\bibfield  {journal} {\bibinfo  {journal} {LAUR-12303}\ }
  (\bibinfo {year} {1992})}\BibitemShut {NoStop}%
\bibitem [{\citenamefont {Grinstein}\ \emph {et~al.}(2021)\citenamefont
  {Grinstein}, \citenamefont {Saenz},\ and\ \citenamefont
  {Germano}}]{grinstein2021coarse}%
  \BibitemOpen
  \bibfield  {author} {\bibinfo {author} {\bibfnamefont {F.~F.}\ \bibnamefont
  {Grinstein}}, \bibinfo {author} {\bibfnamefont {J.~A.}\ \bibnamefont
  {Saenz}},\ and\ \bibinfo {author} {\bibfnamefont {M.}~\bibnamefont
  {Germano}},\ }\bibfield  {title} {\bibinfo {title} {Coarse grained
  simulations of shock-driven turbulent material mixing},\ }\href@noop {}
  {\bibfield  {journal} {\bibinfo  {journal} {Physics of Fluids}\ }\textbf
  {\bibinfo {volume} {33}},\ \bibinfo {pages} {035131} (\bibinfo {year}
  {2021})}\BibitemShut {NoStop}%
\bibitem [{\citenamefont {Olson}\ \emph {et~al.}(2020)\citenamefont {Olson},
  \citenamefont {Murphy}, \citenamefont {Haines}, \citenamefont {Douglas},
  \citenamefont {Albright}, \citenamefont {Gunderson}, \citenamefont {Kim},
  \citenamefont {Cardenas}, \citenamefont {Hamilton},\ and\ \citenamefont
  {Randolph}}]{olson2020development}%
  \BibitemOpen
  \bibfield  {author} {\bibinfo {author} {\bibfnamefont {R.~E.}\ \bibnamefont
  {Olson}}, \bibinfo {author} {\bibfnamefont {T.~J.}\ \bibnamefont {Murphy}},
  \bibinfo {author} {\bibfnamefont {B.~M.}\ \bibnamefont {Haines}}, \bibinfo
  {author} {\bibfnamefont {M.~R.}\ \bibnamefont {Douglas}}, \bibinfo {author}
  {\bibfnamefont {B.~J.}\ \bibnamefont {Albright}}, \bibinfo {author}
  {\bibfnamefont {M.~A.}\ \bibnamefont {Gunderson}}, \bibinfo {author}
  {\bibfnamefont {Y.}~\bibnamefont {Kim}}, \bibinfo {author} {\bibfnamefont
  {T.}~\bibnamefont {Cardenas}}, \bibinfo {author} {\bibfnamefont {C.~E.}\
  \bibnamefont {Hamilton}},\ and\ \bibinfo {author} {\bibfnamefont {R.~B.}\
  \bibnamefont {Randolph}},\ }\bibfield  {title} {\bibinfo {title} {Development
  of the marble experimental platform at the national ignition facility},\
  }\href@noop {} {\bibfield  {journal} {\bibinfo  {journal} {Physics of
  Plasmas}\ }\textbf {\bibinfo {volume} {27}},\ \bibinfo {pages} {102703}
  (\bibinfo {year} {2020})}\BibitemShut {NoStop}%
\bibitem [{\citenamefont {Murphy}\ \emph {et~al.}(2021)\citenamefont {Murphy},
  \citenamefont {Albright}, \citenamefont {Douglas}, \citenamefont {Cardenas},
  \citenamefont {Cooley}, \citenamefont {Day}, \citenamefont {Denissen},
  \citenamefont {Gore}, \citenamefont {Gunderson}, \citenamefont {Haack} \emph
  {et~al.}}]{murphy2021results}%
  \BibitemOpen
  \bibfield  {author} {\bibinfo {author} {\bibfnamefont {T.~J.}\ \bibnamefont
  {Murphy}}, \bibinfo {author} {\bibfnamefont {B.}~\bibnamefont {Albright}},
  \bibinfo {author} {\bibfnamefont {M.}~\bibnamefont {Douglas}}, \bibinfo
  {author} {\bibfnamefont {T.}~\bibnamefont {Cardenas}}, \bibinfo {author}
  {\bibfnamefont {J.}~\bibnamefont {Cooley}}, \bibinfo {author} {\bibfnamefont
  {T.}~\bibnamefont {Day}}, \bibinfo {author} {\bibfnamefont {N.}~\bibnamefont
  {Denissen}}, \bibinfo {author} {\bibfnamefont {R.}~\bibnamefont {Gore}},
  \bibinfo {author} {\bibfnamefont {M.}~\bibnamefont {Gunderson}}, \bibinfo
  {author} {\bibfnamefont {J.}~\bibnamefont {Haack}}, \emph {et~al.},\
  }\bibfield  {title} {\bibinfo {title} {Results from single-shock marble
  experiments studying thermonuclear burn in the presence of heterogeneous mix
  on the national ignition facility},\ }\href@noop {} {\bibfield  {journal}
  {\bibinfo  {journal} {High Energy Density Physics}\ }\textbf {\bibinfo
  {volume} {38}},\ \bibinfo {pages} {100929} (\bibinfo {year}
  {2021})}\BibitemShut {NoStop}%
\bibitem [{\citenamefont {Haines}\ \emph {et~al.}(2020)\citenamefont {Haines},
  \citenamefont {Shah}, \citenamefont {Smidt},\ and\ \citenamefont
  {Abright}}]{haines2020nc}%
  \BibitemOpen
  \bibfield  {author} {\bibinfo {author} {\bibfnamefont {B.~M.}\ \bibnamefont
  {Haines}}, \bibinfo {author} {\bibfnamefont {R.}~\bibnamefont {Shah}},
  \bibinfo {author} {\bibfnamefont {J.}~\bibnamefont {Smidt}},\ and\ \bibinfo
  {author} {\bibfnamefont {B.}~\bibnamefont {Abright}},\ }\bibfield  {title}
  {\bibinfo {title} {Observation of persistent species temperature separation
  in inertial confinement fusion mixtures},\ }\href@noop {} {\bibfield
  {journal} {\bibinfo  {journal} {Nature communications}\ }\textbf {\bibinfo
  {volume} {11}},\ \bibinfo {pages} {1} (\bibinfo {year} {2020})}\BibitemShut
  {NoStop}%
\bibitem [{\citenamefont {Llor}(2005)}]{llor2005statistical}%
  \BibitemOpen
  \bibfield  {author} {\bibinfo {author} {\bibfnamefont {A.}~\bibnamefont
  {Llor}},\ }\href@noop {} {\emph {\bibinfo {title} {Statistical Hydrodynamic
  Models for Developed Mixing Instability Flows: Analytical" 0D" Evaluation
  Criteria, and Comparison of Single-and Two-Phase Flow Approaches}}},\ Vol.\
  \bibinfo {volume} {681}\ (\bibinfo  {publisher} {Springer Science \& Business
  Media},\ \bibinfo {year} {2005})\BibitemShut {NoStop}%
\bibitem [{\citenamefont {Llor}\ and\ \citenamefont
  {Bailly}(2003)}]{llor2003new}%
  \BibitemOpen
  \bibfield  {author} {\bibinfo {author} {\bibfnamefont {A.}~\bibnamefont
  {Llor}}\ and\ \bibinfo {author} {\bibfnamefont {P.}~\bibnamefont {Bailly}},\
  }\bibfield  {title} {\bibinfo {title} {A new turbulent two-field concept for
  modeling {Rayleigh}-{Taylor}, {Richtmyer}-{Meshkov}, and {Kelvin}-{Helmholtz}
  mixing layers},\ }\href@noop {} {\bibfield  {journal} {\bibinfo  {journal}
  {Laser and Particle Beams}\ }\textbf {\bibinfo {volume} {21}},\ \bibinfo
  {pages} {311} (\bibinfo {year} {2003})}\BibitemShut {NoStop}%
\bibitem [{\citenamefont {Youngs}(1994)}]{youngs1994numerical}%
  \BibitemOpen
  \bibfield  {author} {\bibinfo {author} {\bibfnamefont {D.~L.}\ \bibnamefont
  {Youngs}},\ }\bibfield  {title} {\bibinfo {title} {Numerical simulation of
  mixing by {Rayleigh}-{Taylor} and {Richtmyer}-{Meshkov} instabilities},\
  }\href@noop {} {\bibfield  {journal} {\bibinfo  {journal} {Laser and Particle
  Beams}\ }\textbf {\bibinfo {volume} {12}},\ \bibinfo {pages} {725} (\bibinfo
  {year} {1994})}\BibitemShut {NoStop}%
\bibitem [{\citenamefont {Saurel}\ \emph {et~al.}(2012)\citenamefont {Saurel},
  \citenamefont {Huber}, \citenamefont {Jourdan}, \citenamefont {Lap{\'e}bie},\
  and\ \citenamefont {Munier}}]{saurel2012modelling}%
  \BibitemOpen
  \bibfield  {author} {\bibinfo {author} {\bibfnamefont {R.}~\bibnamefont
  {Saurel}}, \bibinfo {author} {\bibfnamefont {G.}~\bibnamefont {Huber}},
  \bibinfo {author} {\bibfnamefont {G.}~\bibnamefont {Jourdan}}, \bibinfo
  {author} {\bibfnamefont {E.}~\bibnamefont {Lap{\'e}bie}},\ and\ \bibinfo
  {author} {\bibfnamefont {L.}~\bibnamefont {Munier}},\ }\bibfield  {title}
  {\bibinfo {title} {Modelling spherical explosions with turbulent mixing and
  post-detonation},\ }\href@noop {} {\bibfield  {journal} {\bibinfo  {journal}
  {Physics of Fluids}\ }\textbf {\bibinfo {volume} {24}},\ \bibinfo {pages}
  {115101} (\bibinfo {year} {2012})}\BibitemShut {NoStop}%
\bibitem [{BAE(1986)}]{BAER1986861}%
  \BibitemOpen
  \bibfield  {title} {\bibinfo {title} {A two-phase mixture theory for the
  deflagration-to-detonation transition ({DDT}) in reactive granular
  materials},\ }\href@noop {} {\bibfield  {journal} {\bibinfo  {journal}
  {International Journal of Multiphase Flow}\ }\textbf {\bibinfo {volume}
  {12}},\ \bibinfo {pages} {861 } (\bibinfo {year} {1986})}\BibitemShut
  {NoStop}%
\bibitem [{\citenamefont {Kapila}\ \emph {et~al.}(2001)\citenamefont {Kapila},
  \citenamefont {Menikoff}, \citenamefont {Bdzil}, \citenamefont {Son},\ and\
  \citenamefont {Stewart}}]{kapila2001two}%
  \BibitemOpen
  \bibfield  {author} {\bibinfo {author} {\bibfnamefont {A.}~\bibnamefont
  {Kapila}}, \bibinfo {author} {\bibfnamefont {R.}~\bibnamefont {Menikoff}},
  \bibinfo {author} {\bibfnamefont {J.}~\bibnamefont {Bdzil}}, \bibinfo
  {author} {\bibfnamefont {S.}~\bibnamefont {Son}},\ and\ \bibinfo {author}
  {\bibfnamefont {D.~S.}\ \bibnamefont {Stewart}},\ }\bibfield  {title}
  {\bibinfo {title} {Two-phase modeling of deflagration-to-detonation
  transition in granular materials: Reduced equations},\ }\href@noop {}
  {\bibfield  {journal} {\bibinfo  {journal} {Physics of Fluids}\ }\textbf
  {\bibinfo {volume} {13}},\ \bibinfo {pages} {3002} (\bibinfo {year}
  {2001})}\BibitemShut {NoStop}%
\bibitem [{\citenamefont {Lund}(2012)}]{lund2012hierarchy}%
  \BibitemOpen
  \bibfield  {author} {\bibinfo {author} {\bibfnamefont {H.}~\bibnamefont
  {Lund}},\ }\bibfield  {title} {\bibinfo {title} {A hierarchy of relaxation
  models for two-phase flow},\ }\href@noop {} {\bibfield  {journal} {\bibinfo
  {journal} {SIAM Journal on Applied Mathematics}\ }\textbf {\bibinfo {volume}
  {72}},\ \bibinfo {pages} {1713} (\bibinfo {year} {2012})}\BibitemShut
  {NoStop}%
\bibitem [{\citenamefont {Saurel}\ \emph
  {et~al.}(2009{\natexlab{a}})\citenamefont {Saurel}, \citenamefont
  {Petitpas},\ and\ \citenamefont {Berry}}]{Saurel2009}%
  \BibitemOpen
  \bibfield  {author} {\bibinfo {author} {\bibfnamefont {R.}~\bibnamefont
  {Saurel}}, \bibinfo {author} {\bibfnamefont {F.}~\bibnamefont {Petitpas}},\
  and\ \bibinfo {author} {\bibfnamefont {R.}~\bibnamefont {Berry}},\ }\bibfield
   {title} {\bibinfo {title} {Simple and efficient relaxation methods for
  interfaces separating compressible fluids, cavitating flows and shocks in
  multiphase mixtures},\ }\href@noop {} {\bibfield  {journal} {\bibinfo
  {journal} {Journal of Computational Physics}\ }\textbf {\bibinfo {volume}
  {228}},\ \bibinfo {pages} {1678} (\bibinfo {year}
  {2009}{\natexlab{a}})}\BibitemShut {NoStop}%
\bibitem [{\citenamefont {Saurel}\ and\ \citenamefont
  {Abgrall}(1999)}]{saurel1999multiphase}%
  \BibitemOpen
  \bibfield  {author} {\bibinfo {author} {\bibfnamefont {R.}~\bibnamefont
  {Saurel}}\ and\ \bibinfo {author} {\bibfnamefont {R.}~\bibnamefont
  {Abgrall}},\ }\bibfield  {title} {\bibinfo {title} {A multiphase godunov
  method for compressible multifluid and multiphase flows},\ }\href@noop {}
  {\bibfield  {journal} {\bibinfo  {journal} {Journal of Computational
  Physics}\ }\textbf {\bibinfo {volume} {150}},\ \bibinfo {pages} {425}
  (\bibinfo {year} {1999})}\BibitemShut {NoStop}%
\bibitem [{\citenamefont {Petitpas}\ and\ \citenamefont
  {Le~Martelot}(2014)}]{petitpas2014}%
  \BibitemOpen
  \bibfield  {author} {\bibinfo {author} {\bibfnamefont {F.}~\bibnamefont
  {Petitpas}}\ and\ \bibinfo {author} {\bibfnamefont {S.}~\bibnamefont
  {Le~Martelot}},\ }\bibfield  {title} {\bibinfo {title} {A discrete method to
  treat heat conduction in compressible two-phase flows},\ }\href@noop {}
  {\bibfield  {journal} {\bibinfo  {journal} {Comput. Therm. Sci.}\ }\textbf
  {\bibinfo {volume} {6}},\ \bibinfo {pages} {251} (\bibinfo {year}
  {2014})}\BibitemShut {NoStop}%
\bibitem [{\citenamefont {Perigaud}\ and\ \citenamefont
  {Saurel}(2005)}]{perigaud2005compressible}%
  \BibitemOpen
  \bibfield  {author} {\bibinfo {author} {\bibfnamefont {G.}~\bibnamefont
  {Perigaud}}\ and\ \bibinfo {author} {\bibfnamefont {R.}~\bibnamefont
  {Saurel}},\ }\bibfield  {title} {\bibinfo {title} {A compressible flow model
  with capillary effects},\ }\href@noop {} {\bibfield  {journal} {\bibinfo
  {journal} {Journal of Computational Physics}\ }\textbf {\bibinfo {volume}
  {209}},\ \bibinfo {pages} {139} (\bibinfo {year} {2005})}\BibitemShut
  {NoStop}%
\bibitem [{\citenamefont {Drew}(1983)}]{drew1983mathematical}%
  \BibitemOpen
  \bibfield  {author} {\bibinfo {author} {\bibfnamefont {D.~A.}\ \bibnamefont
  {Drew}},\ }\bibfield  {title} {\bibinfo {title} {Mathematical modeling of
  two-phase flow},\ }\href@noop {} {\bibfield  {journal} {\bibinfo  {journal}
  {Annual Review of Fluid Mechanics}\ }\textbf {\bibinfo {volume} {15}},\
  \bibinfo {pages} {261} (\bibinfo {year} {1983})}\BibitemShut {NoStop}%
\bibitem [{\citenamefont {Chinnayya}\ \emph {et~al.}(2004)\citenamefont
  {Chinnayya}, \citenamefont {Daniel},\ and\ \citenamefont
  {Saurel}}]{CHINNAYYA2004490}%
  \BibitemOpen
  \bibfield  {author} {\bibinfo {author} {\bibfnamefont {A.}~\bibnamefont
  {Chinnayya}}, \bibinfo {author} {\bibfnamefont {E.}~\bibnamefont {Daniel}},\
  and\ \bibinfo {author} {\bibfnamefont {R.}~\bibnamefont {Saurel}},\
  }\bibfield  {title} {\bibinfo {title} {Modelling detonation waves in
  heterogeneous energetic materials},\ }\href
  {https://doi.org/https://doi.org/10.1016/j.jcp.2003.11.015} {\bibfield
  {journal} {\bibinfo  {journal} {Journal of Computational Physics}\ }\textbf
  {\bibinfo {volume} {196}},\ \bibinfo {pages} {490} (\bibinfo {year}
  {2004})}\BibitemShut {NoStop}%
\bibitem [{\citenamefont {Abgrall}\ and\ \citenamefont
  {Saurel}(2003)}]{ABGRALL2003361}%
  \BibitemOpen
  \bibfield  {author} {\bibinfo {author} {\bibfnamefont {R.}~\bibnamefont
  {Abgrall}}\ and\ \bibinfo {author} {\bibfnamefont {R.}~\bibnamefont
  {Saurel}},\ }\bibfield  {title} {\bibinfo {title} {Discrete equations for
  physical and numerical compressible multiphase mixtures},\ }\href
  {https://doi.org/https://doi.org/10.1016/S0021-9991(03)00011-1} {\bibfield
  {journal} {\bibinfo  {journal} {Journal of Computational Physics}\ }\textbf
  {\bibinfo {volume} {186}},\ \bibinfo {pages} {361} (\bibinfo {year}
  {2003})}\BibitemShut {NoStop}%
\bibitem [{\citenamefont {Hantke}\ \emph {et~al.}(2021)\citenamefont {Hantke},
  \citenamefont {M{\"u}ller},\ and\ \citenamefont
  {Grabowsky}}]{hantke2021news}%
  \BibitemOpen
  \bibfield  {author} {\bibinfo {author} {\bibfnamefont {M.}~\bibnamefont
  {Hantke}}, \bibinfo {author} {\bibfnamefont {S.}~\bibnamefont {M{\"u}ller}},\
  and\ \bibinfo {author} {\bibfnamefont {L.}~\bibnamefont {Grabowsky}},\
  }\bibfield  {title} {\bibinfo {title} {News on baer--nunziato-type model at
  pressure equilibrium},\ }\href@noop {} {\bibfield  {journal} {\bibinfo
  {journal} {Continuum Mechanics and Thermodynamics}\ }\textbf {\bibinfo
  {volume} {33}},\ \bibinfo {pages} {767} (\bibinfo {year} {2021})}\BibitemShut
  {NoStop}%
\bibitem [{\citenamefont {Saurel}\ and\ \citenamefont
  {Pantano}(2018)}]{saurel2018diffuse}%
  \BibitemOpen
  \bibfield  {author} {\bibinfo {author} {\bibfnamefont {R.}~\bibnamefont
  {Saurel}}\ and\ \bibinfo {author} {\bibfnamefont {C.}~\bibnamefont
  {Pantano}},\ }\bibfield  {title} {\bibinfo {title} {Diffuse-interface
  capturing methods for compressible two-phase flows},\ }\href@noop {}
  {\bibfield  {journal} {\bibinfo  {journal} {Annual Review of Fluid
  Mechanics}\ }\textbf {\bibinfo {volume} {50}},\ \bibinfo {pages} {105}
  (\bibinfo {year} {2018})}\BibitemShut {NoStop}%
\bibitem [{\citenamefont {Murrone}\ and\ \citenamefont
  {Guillard}(2005)}]{murrone2005five}%
  \BibitemOpen
  \bibfield  {author} {\bibinfo {author} {\bibfnamefont {A.}~\bibnamefont
  {Murrone}}\ and\ \bibinfo {author} {\bibfnamefont {H.}~\bibnamefont
  {Guillard}},\ }\bibfield  {title} {\bibinfo {title} {A five equation reduced
  model for compressible two phase flow problems},\ }\href@noop {} {\bibfield
  {journal} {\bibinfo  {journal} {Journal of Computational Physics}\ }\textbf
  {\bibinfo {volume} {202}},\ \bibinfo {pages} {664} (\bibinfo {year}
  {2005})}\BibitemShut {NoStop}%
\bibitem [{\citenamefont {Kagan}\ and\ \citenamefont
  {Xianzhu}(2014)}]{kagan2014}%
  \BibitemOpen
  \bibfield  {author} {\bibinfo {author} {\bibfnamefont {G.}~\bibnamefont
  {Kagan}}\ and\ \bibinfo {author} {\bibfnamefont {T.}~\bibnamefont
  {Xianzhu}},\ }\bibfield  {title} {\bibinfo {title} {Thermo-diffusion in
  inertially confined plasmas},\ }\href@noop {} {\bibfield  {journal} {\bibinfo
   {journal} {Pyscial Letters A}\ }\textbf {\bibinfo {volume} {378}},\ \bibinfo
  {pages} {1531 } (\bibinfo {year} {2014})}\BibitemShut {NoStop}%
\bibitem [{\citenamefont {Saurel}\ \emph
  {et~al.}(2009{\natexlab{b}})\citenamefont {Saurel}, \citenamefont
  {Petitpas},\ and\ \citenamefont {Berry}}]{saurel2009simple}%
  \BibitemOpen
  \bibfield  {author} {\bibinfo {author} {\bibfnamefont {R.}~\bibnamefont
  {Saurel}}, \bibinfo {author} {\bibfnamefont {F.}~\bibnamefont {Petitpas}},\
  and\ \bibinfo {author} {\bibfnamefont {R.}~\bibnamefont {Berry}},\ }\bibfield
   {title} {\bibinfo {title} {Simple and efficient relaxation methods for
  interfaces separating compressible fluids, cavitating flows and shocks in
  multiphase mixtures},\ }\href@noop {} {\bibfield  {journal} {\bibinfo
  {journal} {Journal of Computational Physics}\ }\textbf {\bibinfo {volume}
  {228}},\ \bibinfo {pages} {1678} (\bibinfo {year}
  {2009}{\natexlab{b}})}\BibitemShut {NoStop}%
\bibitem [{\citenamefont {Zhang}\ \emph {et~al.}(2022)\citenamefont {Zhang},
  \citenamefont {Menshov}, \citenamefont {Lifeng},\ and\ \citenamefont
  {Zhijun}}]{zhang2022a}%
  \BibitemOpen
  \bibfield  {author} {\bibinfo {author} {\bibfnamefont {C.}~\bibnamefont
  {Zhang}}, \bibinfo {author} {\bibfnamefont {I.}~\bibnamefont {Menshov}},
  \bibinfo {author} {\bibfnamefont {W.}~\bibnamefont {Lifeng}},\ and\ \bibinfo
  {author} {\bibfnamefont {S.}~\bibnamefont {Zhijun}},\ }\bibfield  {title}
  {\bibinfo {title} {Diffuse interface relaxation model for two-phase
  compressible flows with diffusion processes},\ }\href@noop {} {\bibfield
  {journal} {\bibinfo  {journal} {Journal of Computational Physics}\ ,\
  \bibinfo {pages} {111356}} (\bibinfo {year} {2022})}\BibitemShut {NoStop}%
\bibitem [{\citenamefont {Cook}(2008)}]{Cook2009Enthalpy}%
  \BibitemOpen
  \bibfield  {author} {\bibinfo {author} {\bibfnamefont {A.}~\bibnamefont
  {Cook}},\ }\bibfield  {title} {\bibinfo {title} {Enthalpy diffusion in
  multicomponent flows},\ }\href@noop {} {\bibfield  {journal} {\bibinfo
  {journal} {Physics of Fluids}\ }\textbf {\bibinfo {volume} {21}} (\bibinfo
  {year} {2008})}\BibitemShut {NoStop}%
\bibitem [{\citenamefont {Allaire}\ \emph {et~al.}(2002)\citenamefont
  {Allaire}, \citenamefont {Clerc},\ and\ \citenamefont
  {Kokh}}]{allaire2002five}%
  \BibitemOpen
  \bibfield  {author} {\bibinfo {author} {\bibfnamefont {G.}~\bibnamefont
  {Allaire}}, \bibinfo {author} {\bibfnamefont {S.}~\bibnamefont {Clerc}},\
  and\ \bibinfo {author} {\bibfnamefont {S.}~\bibnamefont {Kokh}},\ }\bibfield
  {title} {\bibinfo {title} {A five-equation model for the simulation of
  interfaces between compressible fluids},\ }\href@noop {} {\bibfield
  {journal} {\bibinfo  {journal} {Journal of Computational Physics}\ }\textbf
  {\bibinfo {volume} {181}},\ \bibinfo {pages} {577} (\bibinfo {year}
  {2002})}\BibitemShut {NoStop}%
\bibitem [{\citenamefont {Johnsen}\ and\ \citenamefont
  {Ham}(2012)}]{JOHNSEN20125705}%
  \BibitemOpen
  \bibfield  {author} {\bibinfo {author} {\bibfnamefont {E.}~\bibnamefont
  {Johnsen}}\ and\ \bibinfo {author} {\bibfnamefont {F.}~\bibnamefont {Ham}},\
  }\bibfield  {title} {\bibinfo {title} {Preventing numerical errors generated
  by interface-capturing schemes in compressible multi-material flows},\ }\href
  {https://doi.org/https://doi.org/10.1016/j.jcp.2012.04.048} {\bibfield
  {journal} {\bibinfo  {journal} {Journal of Computational Physics}\ }\textbf
  {\bibinfo {volume} {231}},\ \bibinfo {pages} {5705} (\bibinfo {year}
  {2012})}\BibitemShut {NoStop}%
\bibitem [{\citenamefont {Williams}(2019)}]{williams2019fully}%
  \BibitemOpen
  \bibfield  {author} {\bibinfo {author} {\bibfnamefont {R.~J.}\ \bibnamefont
  {Williams}},\ }\bibfield  {title} {\bibinfo {title} {Fully-conservative
  contact-capturing schemes for multi-material advection},\ }\href@noop {}
  {\bibfield  {journal} {\bibinfo  {journal} {Journal of Computational
  Physics}\ }\textbf {\bibinfo {volume} {398}},\ \bibinfo {pages} {108809}
  (\bibinfo {year} {2019})}\BibitemShut {NoStop}%
\bibitem [{\citenamefont {Spitzer~Jr}\ and\ \citenamefont
  {H{\"a}rm}(1953)}]{Spitzer1953}%
  \BibitemOpen
  \bibfield  {author} {\bibinfo {author} {\bibfnamefont {L.}~\bibnamefont
  {Spitzer~Jr}}\ and\ \bibinfo {author} {\bibfnamefont {R.}~\bibnamefont
  {H{\"a}rm}},\ }\bibfield  {title} {\bibinfo {title} {Transport phenomena in a
  completely ionized gas},\ }\href@noop {} {\bibfield  {journal} {\bibinfo
  {journal} {Physical Review}\ }\textbf {\bibinfo {volume} {89}},\ \bibinfo
  {pages} {977} (\bibinfo {year} {1953})}\BibitemShut {NoStop}%
\bibitem [{\citenamefont {Cl{\'e}rouin}\ \emph {et~al.}(1998)\citenamefont
  {Cl{\'e}rouin}, \citenamefont {Cherfi},\ and\ \citenamefont
  {Z{\'e}rah}}]{clerouin1998viscosity}%
  \BibitemOpen
  \bibfield  {author} {\bibinfo {author} {\bibfnamefont {J.}~\bibnamefont
  {Cl{\'e}rouin}}, \bibinfo {author} {\bibfnamefont {M.}~\bibnamefont
  {Cherfi}},\ and\ \bibinfo {author} {\bibfnamefont {G.}~\bibnamefont
  {Z{\'e}rah}},\ }\bibfield  {title} {\bibinfo {title} {The viscosity of dense
  plasmas mixtures},\ }\href@noop {} {\bibfield  {journal} {\bibinfo  {journal}
  {EPL (Europhysics Letters)}\ }\textbf {\bibinfo {volume} {42}},\ \bibinfo
  {pages} {37} (\bibinfo {year} {1998})}\BibitemShut {NoStop}%
\bibitem [{\citenamefont {Paquette}\ \emph {et~al.}(1986)\citenamefont
  {Paquette}, \citenamefont {Pelletier}, \citenamefont {Fontaine},\ and\
  \citenamefont {Michaud}}]{paquette1986diffusion}%
  \BibitemOpen
  \bibfield  {author} {\bibinfo {author} {\bibfnamefont {C.}~\bibnamefont
  {Paquette}}, \bibinfo {author} {\bibfnamefont {C.}~\bibnamefont {Pelletier}},
  \bibinfo {author} {\bibfnamefont {G.}~\bibnamefont {Fontaine}},\ and\
  \bibinfo {author} {\bibfnamefont {G.}~\bibnamefont {Michaud}},\ }\bibfield
  {title} {\bibinfo {title} {Diffusion coefficients for stellar plasmas},\
  }\href@noop {} {\bibfield  {journal} {\bibinfo  {journal} {The Astrophysical
  Journal Supplement Series}\ }\textbf {\bibinfo {volume} {61}},\ \bibinfo
  {pages} {177} (\bibinfo {year} {1986})}\BibitemShut {NoStop}%
\bibitem [{\citenamefont {Kagan}\ and\ \citenamefont
  {Baalrud}(2018)}]{kagan2018}%
  \BibitemOpen
  \bibfield  {author} {\bibinfo {author} {\bibfnamefont {G.}~\bibnamefont
  {Kagan}}\ and\ \bibinfo {author} {\bibfnamefont {S.}~\bibnamefont
  {Baalrud}},\ }\bibfield  {title} {\bibinfo {title} {Transport formulas for
  multi-component plasmas within the effective potential theory framework},\
  }\href@noop {} {\bibfield  {journal} {\bibinfo  {journal} {arXiv}\ ,\
  \bibinfo {pages} {1611.09872v2}} (\bibinfo {year} {2018})}\BibitemShut
  {NoStop}%
\bibitem [{\citenamefont {Simakov}\ and\ \citenamefont
  {Molvig}(2016{\natexlab{a}})}]{simakov2016a}%
  \BibitemOpen
  \bibfield  {author} {\bibinfo {author} {\bibfnamefont {A.}~\bibnamefont
  {Simakov}}\ and\ \bibinfo {author} {\bibfnamefont {K.}~\bibnamefont
  {Molvig}},\ }\bibfield  {title} {\bibinfo {title} {Hydrodynamic description
  of an unmagnetized plasma with multiple ion species. {I}. general
  formulation},\ }\href@noop {} {\bibfield  {journal} {\bibinfo  {journal}
  {Phys. Plasmas}\ }\textbf {\bibinfo {volume} {23}},\ \bibinfo {pages}
  {032115} (\bibinfo {year} {2016}{\natexlab{a}})}\BibitemShut {NoStop}%
\bibitem [{\citenamefont {Balashov}\ and\ \citenamefont
  {Savenkov}(2018)}]{molvig2014}%
  \BibitemOpen
  \bibfield  {author} {\bibinfo {author} {\bibfnamefont {V.}~\bibnamefont
  {Balashov}}\ and\ \bibinfo {author} {\bibfnamefont {E.}~\bibnamefont
  {Savenkov}},\ }\bibfield  {title} {\bibinfo {title} {Classical transport
  equations for burning gas-metal plasmas},\ }\href@noop {} {\bibfield
  {journal} {\bibinfo  {journal} {Journal of Applied Mechanics \& Technical
  Physics}\ }\textbf {\bibinfo {volume} {59}},\ \bibinfo {pages} {434}
  (\bibinfo {year} {2018})}\BibitemShut {NoStop}%
\bibitem [{\citenamefont {Simakov}\ and\ \citenamefont
  {Molvig}(2016{\natexlab{b}})}]{simakov2016hydrodynamic}%
  \BibitemOpen
  \bibfield  {author} {\bibinfo {author} {\bibfnamefont {A.~N.}\ \bibnamefont
  {Simakov}}\ and\ \bibinfo {author} {\bibfnamefont {K.}~\bibnamefont
  {Molvig}},\ }\bibfield  {title} {\bibinfo {title} {Hydrodynamic description
  of an unmagnetized plasma with multiple ion species. i. general
  formulation},\ }\href@noop {} {\bibfield  {journal} {\bibinfo  {journal}
  {Physics of Plasmas}\ }\textbf {\bibinfo {volume} {23}},\ \bibinfo {pages}
  {032115} (\bibinfo {year} {2016}{\natexlab{b}})}\BibitemShut {NoStop}%
\bibitem [{ZEI(2010)}]{ZEIN20102964}%
  \BibitemOpen
  \bibfield  {title} {\bibinfo {title} {Modeling phase transition for
  compressible two-phase flows applied to metastable liquids},\ }\href@noop {}
  {\bibfield  {journal} {\bibinfo  {journal} {Journal of Computational
  Physics}\ }\textbf {\bibinfo {volume} {229}},\ \bibinfo {pages} {2964}
  (\bibinfo {year} {2010})}\BibitemShut {NoStop}%
\bibitem [{\citenamefont {Thornber}\ \emph {et~al.}(2018)\citenamefont
  {Thornber}, \citenamefont {Groom},\ and\ \citenamefont
  {Youngs}}]{Thornber2018}%
  \BibitemOpen
  \bibfield  {author} {\bibinfo {author} {\bibfnamefont {B.}~\bibnamefont
  {Thornber}}, \bibinfo {author} {\bibfnamefont {M.}~\bibnamefont {Groom}},\
  and\ \bibinfo {author} {\bibfnamefont {D.}~\bibnamefont {Youngs}},\
  }\bibfield  {title} {\bibinfo {title} {A five-equation model for the
  simulation of miscible and viscous compressible fluids},\ }\href@noop {}
  {\bibfield  {journal} {\bibinfo  {journal} {Journal of Computational
  Physics}\ }\textbf {\bibinfo {volume} {372}} (\bibinfo {year}
  {2018})}\BibitemShut {NoStop}%
\bibitem [{\citenamefont {Kokkinakis}\ \emph {et~al.}(2015)\citenamefont
  {Kokkinakis}, \citenamefont {Drikakis}, \citenamefont {Youngs},\ and\
  \citenamefont {Williams}}]{Kokkinakis2015}%
  \BibitemOpen
  \bibfield  {author} {\bibinfo {author} {\bibfnamefont {I.}~\bibnamefont
  {Kokkinakis}}, \bibinfo {author} {\bibfnamefont {D.}~\bibnamefont
  {Drikakis}}, \bibinfo {author} {\bibfnamefont {D.}~\bibnamefont {Youngs}},\
  and\ \bibinfo {author} {\bibfnamefont {R.}~\bibnamefont {Williams}},\
  }\bibfield  {title} {\bibinfo {title} {Two-equation and multi-fluid
  turbulence models for {Rayleigh}--{Taylor} mixing},\ }\href@noop {}
  {\bibfield  {journal} {\bibinfo  {journal} {International Journal of Heat and
  Fluid Flow}\ }\textbf {\bibinfo {volume} {56}},\ \bibinfo {pages} {233}
  (\bibinfo {year} {2015})}\BibitemShut {NoStop}%
\bibitem [{\citenamefont {Livescu}(2013)}]{Livescu2013}%
  \BibitemOpen
  \bibfield  {author} {\bibinfo {author} {\bibfnamefont {D.}~\bibnamefont
  {Livescu}},\ }\bibfield  {title} {\bibinfo {title} {A multiphase model with
  internal degrees of freedom: application to shock{-}bubble interaction},\
  }\href@noop {} {\bibfield  {journal} {\bibinfo  {journal} {Philosophical
  transactions. Series A, Mathematical, physical, and engineering sciences}\
  }\textbf {\bibinfo {volume} {371}},\ \bibinfo {pages} {283} (\bibinfo {year}
  {2013})}\BibitemShut {NoStop}%
\bibitem [{\citenamefont {Richardson}(2019)}]{Richardson2019}%
  \BibitemOpen
  \bibfield  {author} {\bibinfo {author} {\bibfnamefont {A.}~\bibnamefont
  {Richardson}},\ }\href@noop {} {\emph {\bibinfo {title} {2019 NRL plasma
  formulary}}}\ (\bibinfo {year} {2019})\BibitemShut {NoStop}%
\bibitem [{\citenamefont {Vold}\ \emph {et~al.}(2021)\citenamefont {Vold},
  \citenamefont {Yin},\ and\ \citenamefont {Albright}}]{vold2021plasma}%
  \BibitemOpen
  \bibfield  {author} {\bibinfo {author} {\bibfnamefont {E.}~\bibnamefont
  {Vold}}, \bibinfo {author} {\bibfnamefont {L.}~\bibnamefont {Yin}},\ and\
  \bibinfo {author} {\bibfnamefont {B.}~\bibnamefont {Albright}},\ }\bibfield
  {title} {\bibinfo {title} {Plasma transport simulations of
  {Rayleigh}--{Taylor} instability in near-{ICF} deceleration regimes},\
  }\href@noop {} {\bibfield  {journal} {\bibinfo  {journal} {Physics of
  Plasmas}\ }\textbf {\bibinfo {volume} {28}},\ \bibinfo {pages} {092709}
  (\bibinfo {year} {2021})}\BibitemShut {NoStop}%
\end{thebibliography}%

\end{document}